\documentclass[aps,prl,twocolumn,superscriptaddress,longbibliography,nofootinbib]{revtex4-2}
\usepackage{amsmath,amssymb,bm,graphicx,booktabs,xcolor}
\usepackage{placeins}
\usepackage[colorlinks=true,linkcolor=blue!60!black,citecolor=blue!60!black,urlcolor=blue!60!black]{hyperref}

\newcommand{\ee}{\mathrm{e}}
\newcommand{\ii}{\mathrm{i}}
\newcommand{\Vb}{\bar V}
\newcommand{\Pf}{\mathrm{Pf}}

\newcommand{\sgn}{\mathrm{sgn}}
\newcommand{\qbg}{q_{bg}}
\newcommand{\Eunit}{\pi\hbar^2n/m}

\begin{document}

\title{Variational wavefunctions for  ideal anyon gas: superfluid and skyrmion Hall crystal}

\author{Ya-Hui Zhang}
\affiliation{William H. Miller III Department of Physics and Astronomy, Johns Hopkins University, Baltimore, Maryland
21218, USA}

\date{\today}

\begin{abstract}
The ground state of an ideal gas of anyons with statistical angle $\theta\in(0,\pi)$ is a long-standing open problem that has gained renewed relevance with the observation of fractional Chern insulators. We construct variational wavefunctions on the sphere that incorporate the exact anyon statistics, using composite-boson, composite-fermion and parton states as starting points for neural-network optimization. For both spinless and SU(2) semions, independent constructions support a paired anyon superfluid whose energy lies well below that of the best crystal states. Liquids and crystals compete more closely for spinless anyons at $\theta=\pi/3$. For $\theta=\pi/3$ anyons with SU(3) flavor, we identify two competing candidates: a skyrmion Hall crystal, whose pinned flavor texture gives an integer Hall response of the anyons, and a chiral color superconductor of six-anyon clusters. This competition offers a possible explanation for the adjacent reentrant integer quantum anomalous Hall and superconducting phases in twisted MoTe$_2$. In this setting, the color superconductor has $f+if$ symmetry but differs from its BCS counterpart by 3 in chiral central charge. We also provide parameter-free analytical wavefunctions for the color-superconductor ansatz, which reach energies comparable to, or even lower than, those of the best optimized states for SU(3) anyons at $\theta=\pi/3$ and for SU(2) semions.
\end{abstract}

\maketitle

%=======================================================================================================================
\paragraph{Introduction.}
In two dimensions, exchanging identical particles can multiply their wavefunction by an arbitrary phase $\ee^{\ii\theta}$ \cite{Leinaas1977,Wilczek1982}, and particles with these anyonic statistics arise as quasiparticles of fractional quantum Hall (FQH) states \cite{Arovas1984,Halperin1984}. The collective behavior of anyons at finite density is a long-standing question \cite{Arovas1985,Nakajima2025} that remains unresolved even in the absence of interactions. For $\theta/\pi=1-1/p$ with integer $p$, the composite-fermion theory built on flux attachment predicts a superfluid \cite{Laughlin1988,Fetter1989,Chen1989,Xie1990,Girvin1990}. For $\theta/\pi=1/p$, the composite-boson theory predicts a vortex lattice \cite{Lundholm2015,Correggi2019,Komargodski2026,Le2026} for large integer $p$, but whether a new anyon superfluid phase exists at smaller $p$ remains open. Recent observations of fractional quantum anomalous Hall (FQAH) states in twisted MoTe$_2$ (tMoTe$_2$) \cite{Cai2023,Zeng2023,Park2023,Xu2023} and rhombohedral graphene \cite{Lu2024} have brought this question for $\theta=\pi/3$ into experimental focus, particularly with the discovery of superconductivity adjacent to reentrant integer quantum anomalous Hall (RIQAH) states upon doping the FQAH phase in tMoTe$_2$ \cite{Xu2025a}. These experiments have motivated proposals for anyon superconductors, fractionalized metals and crystals \cite{Kim2025,Shi2025a,Divic2025,Xu2025b,Zhang2026a,Nosov2026,Shi2025b,Pichler2026,Huang2025,Wang2025,Kuhlenkamp2025,Tuo2025,Seo2026b,Lotric2026b,WangT2026,Fan2026,WangT2026b,Senthil2026,Han2026,Zhang2026b,Mehta2026,Seo2026,Lotric2026,Cui2026}. However, the ground state of even a non-interacting gas of the relevant anyons, with statistics $\theta=\pi/3$ and SU(3) flavor, remains unknown. In this work, for anyons with $\theta/\pi=1/p$, we introduce a color-superconductor ansatz with $N_c=p$ colors, which gives a new class of anyon superconductors, analogous to the color superconductors of dense quark matter \cite{Alford1999,Alford2008}. Semions, with $p=2$, belong to both series, and for them the color construction reaches Laughlin's paired superfluid \cite{Laughlin1988} by a different route.

To compare the candidate phases on a common footing, we construct variational wavefunctions for the ideal anyon gas on the sphere, using composite-boson (CB), composite-fermion (CF) and parton states as starting points for neural-network optimization. Neural-network wavefunctions, or neural quantum states \cite{Carleo2017,Pfau2020,Hermann2020,vonGlehn2023}, have recently been applied to FQH and moir\'e systems \cite{Luo2024,Teng2025,Geier2025,Luo2025,Li2025,Qian2025,Gattu2025,Abouelkomsan2026a,Abouelkomsan2026b,PerezFadon2026a,Graham2026}, as well as to anyons in a harmonic trap and in anyonic atoms \cite{PerezFadon2026b}. In our calculations, states built from a known physical construction often reach lower energies than general neural networks initialized with random parameters. Their structure also connects the optimized wavefunction to a low-energy effective theory, allowing us to identify phases that cannot be distinguished by symmetry-breaking order parameters. We also construct parameter-free analytical color-superconductor wavefunctions for semions and for SU(3) anyons at $\theta=\pi/3$. Their energies are comparable to, and in some cases lower than, those obtained by direct neural-network optimization.

For spinless and SU(2) semions, three independent constructions support the paired anyon superfluid: a CF integer quantum Hall state at $\nu^*=2$, general CF neural networks initialized with random parameters, and a parton construction with an SU(2) color gauge symmetry. The liquid states obtained from these constructions agree in energy and lie clearly below the best crystal state. Pair correlations are stronger in the SU(2) gas than in the spinless gas, making this case particularly relevant to doped chiral spin liquids \cite{Kalmeyer1987,Laughlin1988,Song2021,Song2023,Divic2025,Pichler2026,Kuhlenkamp2025,ChenF2025,Roth2026}. For spinless anyons at $\theta=\pi/3$, the energetic comparison is less conclusive: a crystal and a Fermi liquid of trions, the bound states of three anyons, are too close in energy to determine which is the ground state.

For $\theta=\pi/3$ anyons with SU(3) flavor, the lowest-energy crystal is a \emph{skyrmion Hall crystal}, whose pinned SU(3) flavor texture gives an integer Hall response of the anyons with $\sigma_{xy}=3e_*^2/h$, where $e_*=e/3$. This contributes an additional $\frac13e^2/h$ to the Hall response of the parent fractional Chern insulator (FCI), so that doping the $\nu=2/3$ FCI state of tMoTe$_2$ produces an RIQAH state with $\sigma_{xy}=e^2/h$. The lowest-energy state in our calculations is a color superconductor that lies below this crystal. It arises through color-flavor-locked pairing of the SU(3) holon metal \cite{Zhang2026b,Mehta2026}, which has both an SU(3) color gauge symmetry and an SU(3) flavor symmetry. The other descendants of the holon metal, with U(1)$^2$ \cite{Zhang2026a,Fan2026,Zhang2026b} or Z$_3$ \cite{Senthil2026,Han2026,Zhang2026b} gauge structure, have higher energies, as do general CF neural networks initialized with random orbitals. In tMoTe$_2$, the color superconductor is a charge-$2e$ $f+if$ superconductor whose chiral central charge differs by 3 from that of the BCS $f+if$ state. Chiral superconductivity can also arise from electron pairing driven by repulsive interactions, as proposed through the Kohn--Luttinger mechanism \cite{Xu2025b} and supported by numerical studies \cite{Guerci2025,Guerci2026,Wang2025,Gallegos2026,Pichler2026b}; which of the two is realized in tMoTe$_2$ remains open.

\begin{figure*}[t]
\centering
\includegraphics[width=0.92\textwidth]{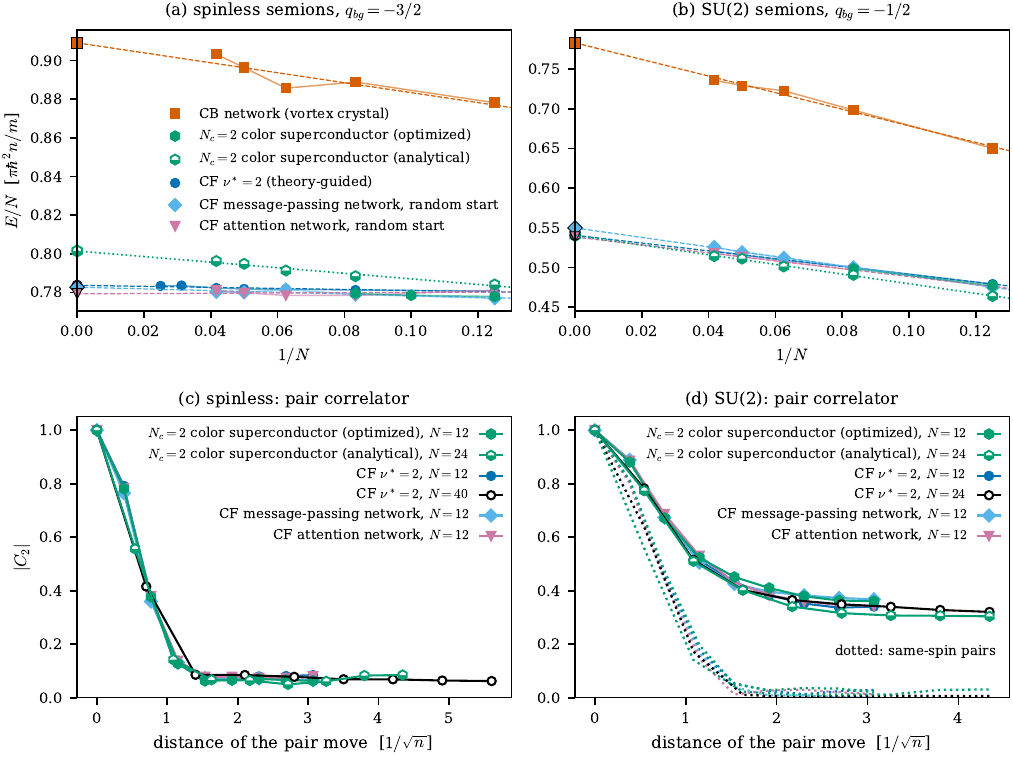}
\caption{We compare spinless (a,c) and SU(2) (b,d) semions on the sphere. Panels (a,b) show $E/N$ in units of $\Eunit$, with dashed lines indicating linear fits for $N\ge8$ and statistical errors smaller than the symbols. Panels (c,d) show the pair correlator $|C_2|$  as a function of the pair displacement; the dotted curves in (d) correspond to same-spin pairs.}
\label{fig:semions}
\end{figure*}

%=======================================================================================================================
\paragraph{Model and wavefunctions.}
We study $\mathcal H=\frac{\hbar^2}{2m}\sum_i[-\ii\nabla_i-\mathbf A(\mathbf r_i)]^2$ for $N$ anyons with statistics $\theta=\alpha\pi$ on a sphere of radius $R=\sqrt N/2$ and density $n=1/\pi$, where $\mathbf A$ is the vector potential of a background monopole carrying $\qbg$ flux quanta. The monopole selects the shift sector \cite{Haldane1983}, and we compare states at fixed $(\qbg,N)$. We express $E/N$ in units of $\Eunit$, for which the free spinless fermion and free boson energies are 1 and 0, respectively. For a gas with $N_f$ flavors and $N/N_f$ anyons per flavor, braiding any two anyons gives $\ee^{2\ii\theta}$, while exchanging two anyons of the same flavor gives $\ee^{\ii\theta}$.

Using planar coordinates $z_j=x_j+\ii y_j$, we write the wavefunction $\Psi=\prod_{i<j}(z_{ij}/|z_{ij}|)^{\alpha_s}\,\Phi$, where $\Phi$ is bosonic for $\alpha_s=\alpha$ (CB gauge) and fermionic for $\alpha_s=\alpha-1$ (CF gauge). In this representation, the Hamiltonian acting on $\Phi$ is
\begin{equation}
\mathcal H_\Phi=\frac{\hbar^2}{2m}\sum_i\bigl[-\ii\nabla_i-\mathbf A(\mathbf r_i)-\mathbf a(\mathbf r_i)\bigr]^2,
\label{eq:hphi}
\end{equation}
where the statistical vector potential is $\mathbf a(\mathbf r_i)=-\alpha_s\sum_{j\neq i}\hat z\times\mathbf r_{ij}/r_{ij}^2$, so that each particle carries $-\alpha_s$ flux quanta and $\nabla\times\mathbf a=-2\pi\alpha_s\rho$. In zero applied field we define the signed Landau-level (LL) filling by $\nu^*=2\pi n/\bar b$, where $\bar b$ is the mean statistical field. For CBs, $\bar b=-2\pi\alpha n$ gives $\nu^*=-1/\alpha$. For CFs, $\bar b=2\pi(1-\alpha)n$ gives $\nu^*=1/(1-\alpha)$.

\begin{figure}[t]
\centering
\includegraphics[width=\columnwidth]{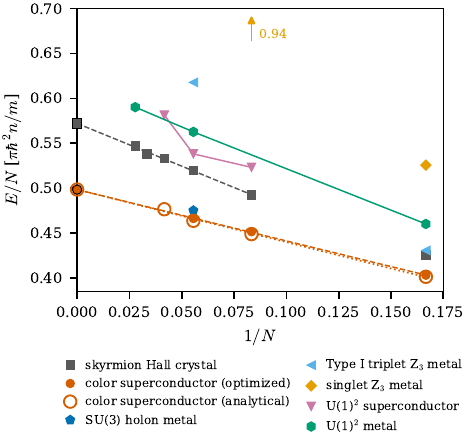}
\caption{We compare the energies of SU(3) anyons at $\theta=\pi/3$ in the $\qbg=-1/3$ sector selected by the Chern--Simons term, with $E/N$ in units of $\Eunit$. The color superconductor, SU(3) holon metal and Type I triplet Z$_3$ metal retain the color sum in Eq.~\eqref{eq:colour}, whereas the U(1)$^2$ states and singlet Z$_3$ metal are color-flavor-locked products. The singlet Z$_3$ energies use the CF ansatz of Eq.~\eqref{eq:z3singlet-cf}. The arrow marks the $N=12$ value above the plotted range. Dashed lines show linear fits and their extrapolated $e_\infty$, using $N\ge12$ for the crystal and $N=6$--18 for the color superconductor. Open circles and the dotted line show the analytical wavefunction of Eq.~\eqref{eq:csc0}, which has no variational parameter, and its linear fit over $N=6$--24.}
\label{fig:su3}
\end{figure}

We organize the variational states into four families, supplementing their underlying structures with a Jastrow factor $\ee^{J}$ and a backflow transformation $z_i\to z'_i$ (SM):  

\textbf{(1)} The basic \emph{CB crystal} ansatz uses $1.2$--$1.7\times10^4$ parameters and takes the form $\Phi_B=\prod_{i<j}|z_{ij}|^{\alpha}\sum_{k=1}^{K}\prod_if^k(i)\,\ee^{J}$ with $K=2$ or 4. Each $f^k(i)$ is a function of $z'_i$ with $N_v\approx\alpha N$ antiholomorphic zeros (antivortices), whose positions depend on the environment of particle $i$ through rotation-invariant network features $h_i$. We initialize this state as a vortex lattice, $f(i)=\prod_{l=1}^{N_v}(\bar z_i-\bar v_l)$. The SU(3) crystal results in Fig.~\ref{fig:su3} and Table~\ref{tab:main} use $K=8$.

\textbf{(2)} The \emph{theory-guided CF states} optimize a prescribed CF wavefunction through backflow and the Jastrow factor, using $3\times10^3$--$1.5\times10^4$ parameters. For semions, $\Phi_F=\det[\varphi_a(i)]\,\ee^{J}$ retains the orbitals of the $\nu^*=2$ integer quantum Hall (IQH) state, with two filled LLs for spinless particles or one LL per spin for SU(2).

\textbf{(3)} The \emph{general CF neural networks}, with $5\times10^3$--$1.8\times10^5$ parameters, instead use $\Phi_F=\sum_{k=1}^{K}\det[v^k_a(i)]\,\ee^{J}$ with $K=2$--64. Their generalized orbitals $v^k_a(i)$ are expanded in CF LL orbitals with coefficients that depend on the full particle configuration, and training begins from random orbitals.

\textbf{(4)} The \emph{SU($N_c$) color states}, with $N_c=1/\alpha$ and $1.1$--$4.4\times10^4$ parameters, are parton states in the CB gauge. We decompose the composite boson of flavor $b$ as $\phi_b=\sum_{a=1}^{N_c}\chi_a\psi_{ab}$, with a U($N_c$) gauge symmetry acting on the color index $a$. Each color of the fermions $\chi_a$ fills the lowest LL of the CB field, while the fermionic holons $\psi_{ab}$ form a metal with $N_c\times N_f$ pockets or its descendants. Projection onto gauge singlets gives an exact sum over color configurations $a=(a_1,\ldots,a_N)$,
\begin{equation}
\Phi_B=\sum_{a_1\ldots a_N}\sgn(a)\,\Phi_\psi(\{z_i,a_i,b_i\})\prod_{c=1}^{N_c}\Vb_{S_c(a)}\;\ee^{J},
\label{eq:colour}
\end{equation}
where $\Vb_S=\prod_{i<j\in S}\bar z_{ij}$, $S_c(a)=\{i:a_i=c\}$ is the set of particles assigned color $c$ in the color configuration, and $\sgn(a)$ is the sign of the permutation that sorts the particles by color.

\textbf{Parameter-free analytical wavefunctions.} Besides these variational states with many parameters, we construct analytical versions of the color states that contain no variational parameters:
\begin{align}
\Psi_{\rm s}&=\prod_{i<j}z_{ij}^{1/2}J_{ij}\sum_a\sgn(a)\,\Pf\big[\epsilon_{a_ia_j}\bar z_{ij}u_{ij}\big]U_a,\label{eq:an-sp}\\
\Psi_{\rm SU(2)}&=\prod_{i<j}z_{ij}^{1/2}J_{ij}\sum_a\sgn(a)\,\Pf\big[\epsilon_{a_ia_j}\epsilon_{s_is_j}u_{ij}\big]U_a,\label{eq:an-su2}\\
\Psi_{\rm SU(3)}&=\prod_{i<j}z_{ij}^{1/3}J_{ij}\sum_a\sgn(a)\,\Pf\big[\sigma_{ij}u_{ij}\big]U_a,\label{eq:csc0}
\end{align}
for spinless semions, SU(2) semions with spins $s_i$ and SU(3) anyons at $\theta=\pi/3$. The factor $U_a$ represents the IQH states of $\chi$, and the holons pair in the $d-id$, spin-singlet $p-ip$ and color-flavor-locked $p-ip$ channels, respectively (SM). Here $u_{ij}=\bar z_{ij}|z_{ij}|^{-2\alpha}$, $U_a=\prod_c\prod_{i<j\in S_c(a)}u_{ij}$, and $\sigma_{ij}=\sum_A\epsilon_{Aa_ia_j}\epsilon_{Ab_ib_j}$ locks the colors of each Cooper pair to its flavors. The radial power in $u_{ij}$ follows from the two-anyon solution.  For SU(3) $\theta=\pi/3$ anyons, the same-color factor in $U_a$ weakly breaks the SU(3) flavor symmetry. An exact flavor singlet would require a projection, but its effect is quite small. The factor $J_{ij}$ cancels the plasma-like correlations of the color sum and adds a phonon factor \cite{Reatto1967}.

%=======================================================================================================================
\paragraph{Semions: paired superfluid.}
For semions, the CF state fills two LLs in the spinless case, $\Phi_F=\det[z_i^k\,|\,\bar z_iz_i^k]$ with $k=0,\ldots,N/2-1$, or one LL per spin in the SU(2) case, $\Phi_F=\Delta(z_\uparrow)\Delta(z_\downarrow)$ with $\Delta(z)=\prod_{i<j}(z_i-z_j)$ over a single spin species. These wavefunctions follow the original theory of anyon superconductivity~\cite{Laughlin1988,Chen1989,Seo2026} and describe a superfluid of semion pairs (SM). The same superfluid arises from color-singlet pairing of the holons in the $N_c=2$ color states, Eqs.~\eqref{eq:an-sp} and \eqref{eq:an-su2}. The optimized $N_c=2$ color superconductor has a fidelity of 0.95--0.99 with the CF $\nu^*=2$ state at $N=8$ and 12, confirming that the two constructions describe the same phase (SM).

Figure~\ref{fig:semions} brings together the numerical evidence for the paired superfluid. In Fig.~\ref{fig:semions}(a,b), the CF $\nu^*=2$ IQH state optimized through its Jastrow factor and coordinate backflow reaches the same liquid phase as general CF neural networks initialized with random orbitals, even though pairing is not imposed on those networks. The $N_c=2$ color superconductor gives nearly the same energy, both when optimized and in the parameter-free form of Eqs.~\eqref{eq:an-sp} and \eqref{eq:an-su2}, and all these liquids approach a common plateau of the pair correlator in Fig.~\ref{fig:semions}(c,d), with substantially stronger correlations for the spin-singlet pairs of SU(2) semions. This agreement among independent constructions, together with their effective theory and the much higher crystal energies in Fig.~\ref{fig:semions}(a,b), strongly suggests that the ground state of the ideal semion gas is a paired superfluid.

%=======================================================================================================================
\paragraph{Spinless anyons at $\theta=\pi/3$.}
At $\theta=\pi/3$, the CF filling is $\nu^*=3/2$, with a fully occupied lowest LL and a half-filled second LL. Attaching two additional flux quanta to the CFs in the second level, as in the composite Fermi liquid (CFL) of electrons at $\nu=1/2$ \cite{Halperin1993}, yields a conventional Fermi liquid of trions, or bound states of three anyons, with no Hall conductance. This liquid competes closely with a crystal: although its energy is slightly lower at the largest size studied (Table~\ref{tab:main}), the difference is comparable to the variational uncertainty and changes sign between shift sectors and system sizes (SM). Our calculations therefore do not determine whether the ground state is a trion liquid or a crystal, nor do they resolve a possible pairing instability of the trion Fermi liquid. General CF networks initialized with random orbitals do not converge to this liquid, but instead settle into non-uniform states of higher energy (SM).

%=======================================================================================================================
\paragraph{$\theta=\frac{\pi}{3}$ anyons with SU(3) flavor: color superconductor.}
The charge-$e/3$ anyons of a doped $C=-1/3$ FCI carry a three-valued flavor enforced by the projected translation symmetry, while their $\pi/3$ statistics are independent of flavor \cite{Shi2025a}. We therefore consider a $\theta=\pi/3$ anyon gas with SU(3) flavor, for which the color states with $N_c=3$ \cite{Zhang2026b} achieve the lowest variational energies.  In the CB representation, the decomposition $\phi_b=\sum_a\chi_a\psi_{ab}$ introduces a U(3) gauge field $\beta$ with U(1) component $\beta_c=\frac13\mathrm{Tr}\,\beta$, such that $\chi$ couples to $a+\beta$ and the holons $\psi$ couple to $-\beta^T$. Each color of $\chi$ fills the lowest LL of the CB field with Chern number $-1$, generating a U(3)$_{-1}$ term for $a+\beta$. Its contribution $-\frac{3}{4\pi}a\,da$, with $X\,dY\equiv\epsilon^{\mu\nu\lambda}X_\mu\partial_\nu Y_\lambda$, cancels the flux-attachment term because $N_c=1/\alpha$, leaving $a$ as a Lagrange multiplier that locks $\beta_c$ to the electromagnetic field $A_c$:
\begin{equation}
\begin{split}
\mathcal L={}&\mathrm{tr}\,\psi^\dagger\Big(\ii D_t+\frac{\mathbf D^2}{2m_\psi}\Big)\psi+\mathcal L_{\rm int}[\psi]\\
&-\frac{1}{4\pi}\mathrm{Tr}\Big(\beta\,d\beta-\frac{2\ii}{3}\beta^3\Big)+\frac{1}{2\pi}a\,d(A_c-3\beta_c).
\end{split}
\label{eq:Lholon}
\end{equation}
Here $\psi=(\psi_{ab})$ is a color$\times$flavor matrix, $D_\mu\psi=(\partial_\mu+\ii\beta^T_\mu)\psi$, and all Chern--Simons levels are integers. Setting $\beta_c=\frac13A_c$ assigns charge $-e/3$ to the holons and separates the U(3) term into the FCI response $-\frac{1}{12\pi}A_c\,dA_c$ and an SU(3)$_{-1}$ gauge field, recovering the parton theory of the doped FCI in Ref.~\cite{Zhang2026b} with the opposite chirality. This Lagrangian describes an SU(3) holon metal with nine Fermi pockets coupled to the SU(3) gauge field.

A pairing instability produces the color superconductor through color-flavor-locked pairing, $\langle\psi_{ab}\psi_{a'b'}\rangle\propto\epsilon_{Aaa'}\epsilon_{Abb'}$, which gaps all nine pockets and completely Higgses the SU(3) gauge field; the resulting charge-$2e$ condensate binds six anyons \cite{Zhang2026b}. It is the lowest-energy state we find, and its analytical wavefunction, Eq.~\eqref{eq:csc0}, matches the energy of the optimized state without any variational parameter, with a high fidelity between the two (SM). The other Higgs descendants of the holon metal all lie higher (Fig.~\ref{fig:su3}): the U(1)$^2$ metal and its paired descendant \cite{Zhang2026a,Fan2026,Zhang2026b}, and the Z$_3$ orthogonal metals \cite{Senthil2026,Han2026,Zhang2026b}. General CF neural networks initialized with random orbitals also stay higher, above even the crystal for $N\ge12$ (SM). The CB network instead produces a skyrmion Hall crystal, about 15\% higher in energy after extrapolation and 12\% higher at $N=18$ (Fig.~\ref{fig:su3}).

In the color superconductor, all $\sim3^N$ colorings in Eq.~\eqref{eq:colour} interfere, giving a hyperdeterminant \cite{Hu2024,Lin2026,Hu2026b}, or its Pfaffian analog, whereas states with color locked to flavor keep a single coloring and reduce to products of parton determinants. Hyperdeterminants are NP-hard to evaluate in general \cite{Gurvits2005}, and an exact polarization identity \cite{Glynn2010,Ishikawa1995} reduces the cost only to $\sim2^N$; this limits the optimized states to $N\le18$ and the analytical wavefunction to $N \le 24$.

In tMoTe$_2$, whose parent FCI has $C=-1/3$, the Chern--Simons term selects $p-ip$ holon pairing \cite{Zhang2026b}, while the physical charge-$2e$ pairs have $-3$ times the angular momentum of a holon pair and therefore have $f+if$ symmetry. The chiral central charge is $c_-=1+\frac32=\frac52$, with 1 contributed by the filled Chern band. A conventional $f+if$ BCS superconductor has the same pairing symmetry but $c_-=1-\frac32=-\frac12$; the color superconductor is thus a distinct phase, and a transition between the two involves six gapless Dirac cones.

%=======================================================================================================================
\paragraph{Competing state: skyrmion Hall crystal.}
For the $\theta=\frac{\pi}{3}$ anyon gas with SU(3) flavor, the lowest-energy crystal is obtained from the CB network with flavor-resolved vortices. The density of each flavor vanishes at its own $N_v=N/3$ vortices, which occupy three interleaved sublattices, while the total density remains nearly uniform. In the CB gauge, the relative flavor amplitudes define a CP$^2$ texture. For the reference crystal, there is a skyrmion number $|Q_{\rm sk}|=N_v$, providing an effective magnetic flux for the anyons. With the skyrmion texture pinned, the anyons form a $C=3$ integer quantum Hall state under this  effective flux, with an additional contribution to Hall conductivity: $\sigma_{xy}=3\,e_*^2/h=\frac13\,e^2/h$. In tMoTe$_2$, the resulting phase is an RIQAH state with $\sigma_{xy}=e^2/h$. Its composite boson description connects it to the commensurate FCI-CDW transition of Ref.~\cite{Song2024}, with doping introducing statistical flux and a vortex texture. Earlier interpretations of the observed RIQAH states \cite{Xu2025a} invoke disorder to localize the doped charge \cite{Nosov2026,Shi2025a,Shi2025b}, while our picture starts from a spontaneously formed skyrmion crystal. Since the closest competing liquid is a chiral color superconductor of the same anyons, the adjacent RIQAH and superconducting phases observed in tMoTe$_2$ \cite{Xu2025a} may correspond to these two competing states of the anyon gas.

%=======================================================================================================================
\paragraph{Conclusion.}
We have constructed variational wavefunctions for the ideal anyon gas that incorporate the exact statistics and combine composite-boson, composite-fermion and parton structures with neural-network correlations. For semions, the convergence of independent liquid constructions and their energetic advantage over crystals support the paired anyon superfluid as the ground-state phase. For spinless anyons at $\theta=\pi/3$, the trion Fermi liquid and crystal remain too close in energy to distinguish. For SU(3) anyons at the same statistical angle, the case relevant to twisted MoTe$_2$, the color superconductor has the lowest variational energy and competes with a skyrmion Hall crystal. This competition suggests an interpretation of the adjacent experimental phases in tMoTe$_2$: the RIQAH state may be a pinned skyrmion Hall crystal of doped anyons, while the observed superconductivity may arise from their color superconductor. Methodologically, the color-superconductor ansatz provides a unified framework for $\theta=\pi/2$ and $\pi/3$, with parameter-free analytical wavefunctions at both angles. These wavefunctions take the form of hyperdeterminants, or their Pfaffian analogs, and go beyond the conventional products of parton wavefunctions; we hope that this work motivates further exploration of this class of wavefunctions.

\paragraph{Note added.} While finalizing this paper, we became aware of a preprint \cite{Khalifa2026} that studies the
semion gas on a torus with continuum neural wavefunctions.

\paragraph{Acknowledgment.} I conceived the various variational wavefunctions and then used Claude Opus 5.5 to implement the variational Monte Carlo calculations. I wrote the manuscript with Claude's assistance in
editing the language, and I take full responsibility for its content. This work is supported by the National Science
Foundation under Grant No.~DMR-2237031.

\bibliography{refs}

%=======================================================================================================================
\clearpage
\onecolumngrid
\begin{center}
\textbf{\large Supplemental Material for\\[2pt]
``Variational wavefunctions for  ideal anyon gas: superfluid and skyrmion Hall crystal''}
\end{center}
\setcounter{equation}{0}
\setcounter{figure}{0}
\setcounter{table}{0}
\setcounter{section}{0}
\renewcommand{\theequation}{S\arabic{equation}}
\renewcommand{\thefigure}{S\arabic{figure}}
\renewcommand{\thetable}{S\arabic{table}}
\renewcommand{\thesection}{S\arabic{section}}
\renewcommand{\thesubsection}{\thesection.\arabic{subsection}}
\makeatletter\renewcommand{\p@subsection}{}\makeatother   % revtex prefixes subsection labels with the section
\setcounter{secnumdepth}{2}

\tableofcontents

%=======================================================================================================================
\section{Variational wavefunctions}\label{sec:wf}

This section describes the variational wavefunctions, whose parameters are optimized by variational Monte Carlo (VMC, Sec.~\ref{sec:vmc}); the parameter-free analytical wavefunctions follow in Sec.~\ref{sec:analytic-wf}.

\subsection{Sphere and statistics}\label{sec:sphere}
We place particle $i$ at $\mathbf r_i=R\,\mathbf n_i$, with polar angles $(\theta_i,\varphi_i)$ and flavor $b_i$, and use units $\hbar=m=1$, density $n=1/\pi$ and mean spacing $a_0=n^{-1/2}=\sqrt\pi$. The corresponding spinor and pair functions are
\begin{equation}
\xi_i=\begin{pmatrix}u_i\\ v_i\end{pmatrix}=\begin{pmatrix}\cos\frac{\theta_i}{2}\,\ee^{\ii\varphi_i}\\[2pt]
\sin\frac{\theta_i}{2}\end{pmatrix},\qquad w_{ij}=u_iv_j-v_iu_j,\qquad |w_{ij}|=\frac{d_{ij}}{2R},
\end{equation}
where $d_{ij}=|\mathbf r_i-\mathbf r_j|$ is the straight-line chord length. The shortest arc on the sphere has length $2R\arcsin[d_{ij}/(2R)]$, which agrees with $d_{ij}$ to leading order at small separations. We use the chord length for pair separations. The phase of $w_{ij}$, the spherical counterpart of $z_i-z_j$ \cite{Haldane1983}, enters the wavefunction through
\begin{equation}
\log\Psi=\ii\alpha_s\sum_{i<j}\arg w_{ij}+\log\Phi.
\label{eq:logpsi}
\end{equation}
Here $\alpha_s=\alpha$ in the CB gauge, where $\Phi$ is symmetric within a flavor, or $\alpha_s=\alpha-1$ in the CF gauge, where $\Phi$ is antisymmetric. The other $N-1$ particles contribute $-\alpha_s(N-1)$ statistical flux quanta, giving a total effective flux of $2Q_\Phi=\qbg-\alpha_s(N-1)$. The CF flux is therefore $2Q^*=\qbg+(1-\alpha)(N-1)$, whereas the CB flux is $-N_v$, with $N_v=\alpha(N-1)-\qbg$ antivortices per particle. We construct $\Phi$ from the orbitals of a monopole of strength $Q$, namely the monopole harmonics $Y_{Q,n,m}$ defined in Sec.~\ref{sec:nets}. Following Ref.~\cite{Haldane1983}, we label a monopole by $2Q$, the number of flux quanta through the sphere, which is an integer for single-valued orbitals; the lowest level has angular momentum $|Q|$, which can be half-integer. For $2Q\ge0$, the lowest level is spanned by the $2Q+1$ monomials $u_i^{Q+m}v_i^{Q-m}$, which are holomorphic in the spinor, while level $n$ consists of polynomials of degree $2Q+n$ in $(u_i,v_i)$ and degree $n$ in $(\bar u_i,\bar v_i)$. For $2Q<0$, the roles of $\xi_i$ and $\bar\xi_i$ are exchanged, as required for the antivortices in the CB gauge.

Variational Monte Carlo samples $|\Psi|^2$ and averages the local energy
\begin{equation}
E_L=\frac{H\Psi}{\Psi}=-\frac12\sum_i\Big[\nabla_i^2\log\Psi+(\nabla_i\log\Psi)^2\Big]
\end{equation}
where the derivatives include the monopole vector potential. Although $\Psi$ is multivalued because the branch of $\arg w_{ij}$ in Eq.~\eqref{eq:logpsi} is arbitrary,  $|\Psi|^2$ and $E_L$ are  single valued, and either gauge yields a rigorous variational upper bound.

\subsection{Building blocks}\label{sec:nets}
In each variational family, one network describes the environment of every particle. Its output determines the
Jastrow factor that changes the amplitude and phase, the backflow that shifts the orbital coordinates, and,
where needed, the configuration-dependent orbital coefficients. We optimize these parts together using real
network weights $W$ and biases $\mathbf b$, while the coefficients coupling the network to the wavefunction can
be complex. We start the biases at zero and draw the weights uniformly from an interval centered on zero.
This interval is narrower for layers with more inputs or outputs (Glorot initialization).

\paragraph{Features.} The network represents the environment of particle $i$ by a vector $h_i\in\mathbb R^H$.
To construct its input, we count nearby particles on several length scales. For each pair we evaluate six Gaussians
$g_s(d)=\ee^{-d^2/\varsigma_s^2}$ and the six functions $x_s(d)=(d^2/\varsigma_s^2)g_s(d)$, with widths
$\varsigma_s$ spaced geometrically from $0.3a_0$ to $1.5a_0$. For spinless anyons the input is
$\mathbf h^{\rm in}_i=\sum_{j\ne i}[\mathbf g(d_{ij}),\mathbf x(d_{ij})]\in\mathbb R^{12}$, where the brackets join
the two vectors. For anyons with flavor, we sum separately over neighbors of the same and of different flavor.
Appending $2S_z=\pm1$ gives 25 inputs for SU(2), while appending a three-component vector with a 1 at the occupied
flavor gives 27 inputs for SU(3).

The three networks differ in how particle $i$ weights the information it receives from particle $j$. Each starts
from $h_i^0=\tanh(W_1\mathbf h^{\rm in}_i+\mathbf b_1)$. The simplest network uses only the pair distance and flavors
to weight the contribution of $h_j^0$. The deeper networks update these weights as information passes between
particles. All three depend on positions through chord distances and sum over neighbors, so a rotation leaves their
outputs unchanged and exchanging particles of the same flavor exchanges their output vectors. The smooth distance functions and activations allow us to
differentiate these features when computing the kinetic energy.

The \textbf{MP1} network performs one such sum, called a message-passing step, with $H=32$
\cite{Schutt2018,Pescia2024}. Particle $i$ receives the vector
$m_i=N^{-1/2}\sum_{j\ne i}\tanh[W_g\mathbf g(d_{ij})]\odot h_j^0$, where $\odot$ multiplies corresponding
components. With flavors we use separate matrices $W_g$ for same-flavor and different-flavor pairs. The network
combines this sum with the initial features to give $h_i=h_i^0+\tanh(W_2[h_i^0,m_i]+\mathbf b_2)$. We use this
single step for the theory-guided states.

The \textbf{MP3} network repeats the update three times and keeps a vector $e_{ij}$ for each pair, as in the
two-electron stream of FermiNet \cite{Pfau2020}. The particle vectors have $H=32$ components ($H=64$ for the SU(3)
CF networks) and the pair vectors have $H_e=16$. We initialize $h_i=h_i^0$ and
$e_{ij}=\tanh(W_e[\mathbf g(d_{ij}),\mathbf x(d_{ij}),\delta_{b_ib_j}]+\mathbf b_e)$, omitting the flavor indicator
for spinless particles. In layer $l$ the current pair vector sets the weight of $h_j$ in the sum
$m_i=N^{-1/2}\sum_{j\ne i}\tanh(W_g^l e_{ij})\odot h_j$. We then update both vectors using their values at the
start of the layer,
\[
\begin{aligned}
h_i&\leftarrow h_i+\tanh(W_2^l[h_i,m_i]+\mathbf b_2^l),\\
e_{ij}&\leftarrow e_{ij}+\tanh(W_{ee}^l e_{ij}+W_{he}^l(h_i+h_j)+\mathbf b_{ee}^l).
\end{aligned}
\]
The updated pair vector thus carries information about the environments of both particles into the next layer.

The \textbf{attention} network also uses three layers and the same initial pair vectors. It compares the features of
$i$ and $j$ to assign normalized weights to the neighbors of $i$. Four independent comparisons, called heads,
produce four weighted sums in each layer \cite{vonGlehn2023}. We take $H=64$ or 128 and $H_e=16$. Before each
comparison, layer normalization subtracts the mean of the components of $h_i$ and divides by their standard
deviation with a small numerical regularizer. It then multiplies each normalized component by a learned scale
and adds a learned offset, giving $x_i=\mathrm{LN}(h_i)$. These scales start at one and the offsets at zero,
so the network initially passes on the normalized components directly. In head $\mu$, the matrices $W_Q^\mu$ and $W_K^\mu$ project $x_i$ and $x_j$ onto
$H/4$ components. Their dot product and the pair features give the score $a_{ij}^\mu$, which determines the
weight $A_{ij}^\mu$ and the vector received by $i$,
\[
\begin{aligned}
a_{ij}^\mu&=(W_Q^\mu x_i)\cdot(W_K^\mu x_j)/\sqrt{H/4}+(W_b e_{ij})_\mu,\qquad
A_{ij}^\mu=\frac{\exp a_{ij}^\mu}{\sum_{k\ne i}\exp a_{ik}^\mu},\quad j\ne i,\\
m_i^\mu&=\sum_{j\ne i}A_{ij}^\mu\big[W_V^\mu x_j+(W_{ev}e_{ij})^\mu\big].
\end{aligned}
\]
Once the four attention heads have computed $m_i^\mu$, we concatenate their outputs into $m_i$.
Adding the projected output to the incoming particle features gives $h_i^{\rm att}=h_i+W_Om_i$.
The feedforward step acts on this vector, giving
$h_i\leftarrow h_i^{\rm att}+W_{m2}\tanh[W_{m1}\mathrm{LN}(h_i^{\rm att})+\mathbf b_{m1}]+\mathbf b_{m2}$.
The layer then updates $e_{ij}$ by the MP3 pair rule using the resulting $h_i$ and $h_j$.
Before training, we draw the entries of $W_O$ and $W_{m2}$ using the initialization described above and
multiply each entry by $1/2$. For a given input this halves the contribution that each output matrix adds to
the particle features at initialization. In both deeper networks the final
$h_i$ supplies the particle features, while the final $e_{ij}$ also enters the backflow coefficient below.

\paragraph{Jastrow factor.} The simplest use of $h_i$ is to adjust the amplitude and phase at a given configuration.
We multiply the wavefunction by $\ee^J$, where $J=J_2+J_h$. The pair contribution is
\begin{equation}
J_2=\sum_{i<j}\sum_{r=1}^{10}\bigl(c_r+[b_i\neq b_j]\,c'_r\bigr)\,G_r(d_{ij}),\qquad G_r(d)=\ee^{-d^2/\sigma_r^2},
\end{equation}
where seven widths $\sigma_r$ from $0.15a_0$ to $1.5a_0$ resolve the correlation hole and three widths from $0.5R$
to $2R$ describe correlations across the sphere \cite{Jastrow1955}. Within each range the widths are geometrically
spaced. The indicator $[b_i\ne b_j]$ allows different-flavor pairs to have their own coefficients. The remaining
term, $J_h=\sum_i(\boldsymbol\lambda+\ii\boldsymbol\lambda')\cdot h_i$, adds correlations that depend on the other
particles through $h_i$. The real parts of $J_2$ and $J_h$ change the wavefunction amplitude, and their
imaginary parts change its phase. We initialize $c_r$, $c'_r$, $\boldsymbol\lambda$ and $\boldsymbol\lambda'$
to zero, so the factor $\ee^{J_2+J_h}$ equals one before training.
For the $N=12$, $\qbg=-1/3$ skyrmion Hall crystal used in the extrapolation in Table~\ref{tab:main}, we also include
$J_{\rm lr}=\sum_{i<j}(\gamma_{ij}/2)\log(1+d_{ij}^2/r_s^2)$, with $r_s=0.8a_0$ and separate real exponents for
same-flavor and different-flavor pairs. The crystal energies are given in Fig.~\ref{fig:su3} and Table~\ref{tab:su3}.

\paragraph{Backflow.} The same features allow the other particles to move the position at which we evaluate an
orbital. For each pair we form the complex coefficient
\begin{equation}
c_{ij}=\sum_r\eta_rG_r(d_{ij})+\ee^{-d_{ij}^2/a_0^2}\,\boldsymbol\eta_h\cdot h_i+\boldsymbol\eta_e\cdot e_{ij},
\end{equation}
where the learned coefficients start at zero and the last term is present for networks with pair features.
The backflow transformation on the sphere is
\begin{equation}
\xi'_i=\xi_i+\sum_{j\neq i}c_{ij}\,(1-P_i)P_j\,\xi_i=\xi_i+\sum_{j\neq i}c_{ij}\,S_{ji}\,(\xi_j-S_{ij}\,\xi_i),
\label{eq:bf}
\end{equation}
where $S_{ij}=\xi_i^\dagger\xi_j$ and $P_i=\xi_i\xi_i^\dagger$. The projector $1-P_i$ picks out the change in the
direction of the spinor. A positive real $c_{ij}$ shifts particle $i$ toward $j$, while an imaginary coefficient moves
it transverse to that direction. The spinors $\xi_i'$ rotate together with the original spinors. For nearby particles
the transformation has the planar form $z'_i=z_i+\sum_jc_{ij}(z_j-z_i)$ \cite{FeynmanCohen1956,Kwon1993,Kwon1998}.
All orbitals of particle $i$ are evaluated at this shifted position. Real tangent shifts have also been used for CF
states on the sphere \cite{Gattu2025}.

For CF states we want the environment to move an isolated close pair together. This preserves the relative
coordinate that fixes the anyon wavefunction near contact. Write the backflow of Eq.~\eqref{eq:bf} as
$\xi'_l=M_l\xi_l$, with $M_l=1+\sum_kc_{lk}(1-P_l)P_k$. We average the maps of nearby particles before applying
them to $\xi_i$, and use the same weights to average their features,
\begin{equation}
\xi'_i=\sum_l\pi_{il}M_l\,\xi_i,\qquad \bar h_i=\sum_l\pi_{il}h_l,\qquad
\pi_{il}=\frac{\delta_{il}+s_{il}}{1+\sum_ms_{im}},\qquad
s_{ij}=\Big(1-\frac{d_{ij}^2}{\rho^2}\Big)_+^3 ,
\label{eq:avg}
\end{equation}
where $(x)_+=\max(x,0)$, $s_{ii}=0$, and $\rho$ is either the contact-window radius $r_w$ or the fixed radius
$r_c=0.4a_0$ defined below. The cubic power makes the weight and its first two derivatives continuous at the
boundary of the window. For an isolated particle, $\pi_{il}=\delta_{il}$, so $\xi'_i=M_i\xi_i$ and $\bar h_i=h_i$.
For an isolated pair at contact, both particles give weight $1/2$ to each map and experience the common
$\tilde M=(M_i+M_j)/2$.

To see why the common map matters, write $w_{ij}=\xi_i^T\epsilon\xi_j$ with
$\epsilon=\bigl(\begin{smallmatrix}0&1\\-1&0\end{smallmatrix}\bigr)$. With separate maps the transformed pair
coordinate is $\xi_i^TM_i^T\epsilon M_j\xi_j$. A difference between $M_i$ and $M_j$ that is linear in the
separation can mix $\bar z_{ij}$ into $z_{ij}$. For an isolated close pair the weights in Eq.~\eqref{eq:avg} agree
up to $O(d_{ij}^2)$. To leading order the two particles therefore share a map, and
$\tilde M^T\epsilon\tilde M=\det(\tilde M)\epsilon$ gives $w'_{ij}=\det(\tilde M)w_{ij}$ at that order.
The environment can change the pair amplitude while preserving its winding. Averaging $h_i$ in the same way keeps
the orbital coefficients common to the pair to leading order.

\paragraph{Modification of short-distance behavior.} Consider two anyons with relative coordinate
$z_{ij}=r\ee^{\ii\phi}$. Exchanging particles of the same flavor rotates $z_{ij}$ by $\pi$ and multiplies $\Psi$ by
$\ee^{\ii\pi\alpha}$. A term with angular dependence $\ee^{\ii\ell\phi}$ must therefore have
$\ell\in\alpha+2\mathbb Z$. For different flavors, taking one particle around the other fixes the phase after a
$2\pi$ rotation, giving $\ell\in\alpha+\mathbb Z$. In each channel the kinetic energy fixes the regular radial
solution near contact to $J_{|\ell|}(kr)\propto r^{|\ell|}$. For the statistics studied here, the leading channel
has $\ell=\alpha$. For different flavors at $\alpha=1/2$ the channel $\ell=-1/2$ has the same radial power.

The contact power follows directly from the two-dimensional Laplacian. In polar coordinates the radial part is
$r^{-1}\partial_r(r\partial_r)$ and the angular part is $r^{-2}\partial_\phi^2$. For a trial term
$\psi_\ell=r^s\ee^{\ii\ell\phi}$, the radial derivative gives $r\partial_r\psi_\ell=s\psi_\ell$,
while two angular derivatives give $\partial_\phi^2\psi_\ell=-\ell^2\psi_\ell$. Adding the two contributions gives
\[
\nabla^2\psi_\ell=\frac{1}{r}\partial_r(s\psi_\ell)-\frac{\ell^2}{r^2}\psi_\ell
=(s^2-\ell^2)\,r^{s-2}\ee^{\ii\ell\phi}.
\]
The two particles have reduced mass $m/2$, so their relative kinetic operator is $H_{\rm rel}=-\nabla^2$
in our units $\hbar=m=1$. The corresponding local energy for this power law is
\[
E_{L,\rm rel}=\frac{H_{\rm rel}\psi_\ell}{\psi_\ell}
=\frac{\psi_\ell^* H_{\rm rel}\psi_\ell}{|\psi_\ell|^2}
=\frac{(\ell^2-s^2)r^{2s-2}}{r^{2s}}=\frac{\ell^2-s^2}{r^2}.
\]
The $1/r^2$ term cancels for the regular power $s=|\ell|$. When this channel dominates at contact, the
Monte Carlo sampling probability is proportional to $|\psi_\ell|^2\,r\,dr\,d\phi=r^{2s}\,r\,dr\,d\phi$.
The near-contact contribution to $\langle |E_L|^2\rangle$ is therefore proportional to
$(\ell^2-s^2)^2\int_0^\epsilon r^{2s-3}\,dr$, which diverges for $s\le1$ when $s^2\ne\ell^2$.

For a pair of the same flavor, the two channels with the smallest $|\ell|$ are $\ell=\alpha$ and $\ell=\alpha-2$
for $0<\alpha<1$. Their leading regular terms in the physical wavefunction are
\[
\Psi_{\rm pair}\simeq a\,r^\alpha\ee^{\ii\alpha\phi}
                    + b\,r^{2-\alpha}\ee^{\ii(\alpha-2)\phi},
\]
where $a$ and $b$ can depend on the pair center and the other particles. To obtain the auxiliary wavefunction
in either gauge, we remove the statistical phase $\ee^{\ii\alpha_s\phi}$ of this pair. This leaves the radial
power $r^{|\ell|}$ intact and changes the angular dependence to $\ee^{\ii(\ell-\alpha_s)\phi}$.
In the CB gauge ($\alpha_s=\alpha$) the two angular indices are $0$ and $-2$. In the CF gauge
($\alpha_s=\alpha-1$) they are $1$ and $-1$.

For a pair of different flavors, a full braid fixes the statistical phase, but exchange symmetry imposes no
even or odd parity on the relative orbital at fixed flavor labels. The two lowest channels are now
$\ell=\alpha$ and $\ell=\alpha-1$. Writing $z_{ij}=r\ee^{\ii\phi}$, their leading regular terms are
$\Psi_{\rm pair}\simeq a\,z_{ij}^\alpha+b\,\bar z_{ij}^{\,1-\alpha}$. For $\alpha=1/3$ their magnitudes scale
as $r^{1/3}$ and $r^{2/3}$, respectively. For semions both scale as $r^{1/2}$, with opposite angular windings.

The table compares a generic smooth auxiliary amplitude with the regular forms obtained by removing the
statistical phase from these two physical channels in each flavor sector.
\begin{center}
\begin{tabular}{llccc}
\toprule
pair flavors & gauge & smooth $\Phi$ near contact & $|\Psi|$ for smooth $\Phi$ & regular terms in $\Phi$ \\
\midrule
same & CB & $A$ & const. & $a\,r^\alpha+b\,\bar z_{ij}^{\,2}r^{-\alpha}$ \\
     & CF & $a'\,z_{ij}+b'\,\bar z_{ij}$ & $\sim r$ & $a\,z_{ij}r^{\alpha-1}+b\,\bar z_{ij}r^{1-\alpha}$ \\
\midrule
different & CB & $A$ & const. & $a\,r^\alpha+b\,\bar z_{ij}r^{-\alpha}$ \\
          & CF & $A'$ & const. & $a\,z_{ij}r^{\alpha-1}+b\,r^{1-\alpha}$ \\
\bottomrule
\end{tabular}
\end{center}
For either choice of flavors, multiplying the regular CB amplitude by $\ee^{\ii\phi}=z_{ij}/r$ gives the CF
amplitude with the same $a$ and $b$, up to phase factors from the other pairs. Restoring the respective
statistical phase gives the same physical wavefunction in both gauges.

For a same-flavor pair, a smooth CB amplitude generically approaches a constant, giving
$E_L\simeq\alpha^2/r^2$ and a logarithmically divergent mean energy. A smooth CF amplitude vanishes linearly,
so its mean energy is finite but its variance diverges logarithmically. For a different-flavor pair, a smooth
amplitude can approach a nonzero constant in either gauge. Its physical wavefunction then has only the statistical
phase at leading order, giving $E_L\simeq\alpha^2/r^2$ in the CB gauge and $E_L\simeq(1-\alpha)^2/r^2$ in the
CF gauge. In both cases the uncorrected constant term gives a logarithmically divergent mean energy.

\paragraph{CB gauge.} The CB networks and color states carry a pair factor $\prod_{i<j}(d_{ij}/a_0)^\alpha$.
Since $d_{ij}=2R|w_{ij}|$, combining it with the statistical phase gives
\begin{equation}
\prod_{i<j}\Big(\frac{w_{ij}}{|w_{ij}|}\Big)^{\alpha}\Big(\frac{d_{ij}}{a_0}\Big)^{\alpha}
=\Big(\frac{2R}{a_0}\Big)^{\alpha N(N-1)/2}\prod_{i<j}w_{ij}^{\alpha},
\end{equation}
For a close pair $w_{ij}\propto z_{ij}$, so the leading behavior is $z_{ij}^{\alpha}$ times a smooth amplitude.
The Laplacian of $z_{ij}^\alpha$ vanishes away from the origin, eliminating the singular contribution to the
local energy in this channel.

For a same-flavor pair, the smooth amplitude multiplying $z_{ij}^\alpha$ is even under exchange. Its first term
in the $\ell=\alpha-2$ channel is therefore proportional to $\bar z_{ij}^{\,2}$. The product
$z_{ij}^\alpha\bar z_{ij}^{\,2}$ vanishes as $r^{2+\alpha}$, while the regular power in this channel is
$r^{2-\alpha}$. If the leading $z_{ij}^\alpha$ term has nonzero amplitude, the contribution to $E_L$ stays bounded.

For a different-flavor pair, the smooth amplitude can already contain a term linear in $\bar z_{ij}$.
The wavefunction $z_{ij}^\alpha(A+B\bar z_{ij})$ then has an $\ell=\alpha-1$ component that vanishes as
$r^{1+\alpha}$, faster than the regular power $r^{1-\alpha}$. For $A\ne0$ the relative kinetic energy
$-\nabla^2_{z_{ij}}$ gives $E_L\simeq-4\alpha(B/A)/z_{ij}$. The weight $|\Psi|^2\sim r^{2\alpha}$ makes
both the mean and the second moment of this $1/r$ contribution finite.

We tested whether correcting this different-flavor channel lowers the variational energy. At $\alpha=1/3$,
the basic CB ansatz gives the $\ell=-2/3$ component a radial power $r^{4/3}$. We added a pair factor that
allows the regular $r^{2/3}$ term, using the same construction as in Eq.~\eqref{eq:ifx} adapted to the CB gauge.
For the SU(3) skyrmion Hall crystal and color superconductor at $N=12$, the change in $E/N$ was at most 0.002
in the $\qbg=-1/3$ and $5/3$ sectors, comparable to the statistical errors. We therefore retain the simple
pair factor for the CB networks. 

\paragraph{CF gauge.} For a same-flavor pair, the linear zero of a smooth determinant contains two terms, $a z_{ij}+b\bar z_{ij}$.
After including the statistical phase they have angular momenta $\alpha$ and $\alpha-2$, so their regular radial
powers are $\alpha$ and $2-\alpha$, respectively. Multiplying by $r^{\alpha-1}$ fixes the first term but leaves
the second proportional to $r^\alpha$. Multiplying by $r^{1-\alpha}$ fixes the second and suppresses the first
to $r^{2-\alpha}$. Both terms occur when the CFs occupy higher Landau levels. We therefore correct the
holomorphic and antiholomorphic coordinates separately, using
\begin{equation}
F_{ij}=\Big(\frac{d_{ij}}{a_0}\Big)^{\alpha-1}\ee^{\sum_rf_rG_r(d_{ij})},\qquad
\tilde F_{ij}=\Big(\frac{d_{ij}}{a_0}\Big)^{1-\alpha}\ee^{\sum_r\tilde f_rG_r(d_{ij})},
\end{equation}
The powers of $d_{ij}$ fix the contact behavior, while the learned coefficients $f_r,\tilde f_r$ adjust the shape
at finite separation and start at zero. A window $\kappa_{ij}$ of radius $r_w$ limits the deformation to nearby
particles of the same flavor. The two coordinates entering the orbitals are
\begin{equation}
\zeta_i=\xi'_i+\frac12\sum_{j\neq i}\kappa_{ij}(F_{ij}-1)(1-P'_j)\,\xi'_i,\qquad
\omega_i=\bar\xi'_i+\frac12\sum_{j\neq i}\kappa_{ij}(\tilde F_{ij}-1)(1-\bar P'_j)\,\bar\xi'_i,
\label{eq:contactcoords}
\end{equation}
where $P'_j=\xi'_j\xi_j'^\dagger/|\xi'_j|^2$. For an isolated close pair, $\kappa_{ij}\to1$ and the leading
relative coordinates are proportional to $z_{ij}r^{\alpha-1}$ and $\bar z_{ij}r^{1-\alpha}$.
Expanding the determinant to first order in those coordinates gives the two regular terms in the same-flavor CF row of the table,
with independent amplitudes $a$ and $b$. For different flavors, $\kappa_{ij}=0$ and the opposite-flavor factor
in Sec.~\ref{sec:ans} supplies the contact behavior.

We use either a Gaussian window $\kappa_{ij}=\ee^{-d_{ij}^2/r_w^2}$ or the compact window of Eq.~\eqref{eq:avg}
with $\rho=r_w$, starting the optimization at $r_w=0.34a_0$. A third particle in the window can spoil the
isolated-pair cancellation and give $E_L\sim r^{-2}$. For an isolated pair, higher-order terms in the determinant can still give a weaker
residual $E_L\sim r^{2\alpha-2}$. For semions we obtain stable estimates by
capping $r_w$ at $0.45a_0$. For spinless $\theta=\pi/3$ liquids, allowing $r_w$ to exceed the mean spacing
lowers $E/N$ by 0.01--0.03. We leave $r_w$ free in those runs and in the color-flavor-locked states, and
evaluate the energy with 200 rounds of 1024 walkers (Sec.~\ref{sec:vmc}). A further check separates the regular
deformation, which can extend over $r_w$, from the singular part, which acts only inside $0.4a_0$.
This preserves the isolated-pair exponents for any $r_w$ and reduces the variance per iteration by a factor of
five to ten. The energies agree within 0.006, and the tables give the lowest energy among the variants.

\paragraph{Landau-level orbitals.} With the coordinates specified, we construct the orbitals from the monopole
harmonics $Y_{Q,n,m}$ for $2Q$ flux quanta \cite{WuYang1976,Haldane1983}. For CF states the flux is $2Q^*$.
For $Q\ge0$, the harmonic in level $n$ with $L_z=m$ has the following polynomial form in the contact coordinates
of particle $i$ \cite{JainKamilla1997},
\begin{equation}
Y_{Q,n,m}(i)\propto(\zeta_{i,1}\omega_{i,1}+\zeta_{i,2}\omega_{i,2})^{n_{\max}-n}\sum_{s=0}^{n}(-1)^s\binom ns
\binom{2Q+n}{Q+n-m-s}\,\zeta_{i,1}^{Q+m+s}\zeta_{i,2}^{Q-m+n-s}\,\omega_{i,1}^{s}\,\omega_{i,2}^{n-s},
\label{eq:ylm}
\end{equation}
where $\zeta_{i,1},\zeta_{i,2}$ and $\omega_{i,1},\omega_{i,2}$ are the components of $\zeta_i$ and $\omega_i$.
At the undeformed coordinates $(\zeta_i,\omega_i)=(\xi_i,\bar\xi_i)$ the first factor in
Eq.~\eqref{eq:ylm} equals one.
It gives every polynomial the same total degree, so the contact deformation acts consistently across levels.
For $Q<0$ we exchange $\zeta$ and $\omega$, using $Y_{-|Q|,n,m}(\zeta,\omega)=Y_{|Q|,n,-m}(\omega,\zeta)$.
We normalize the harmonics on the sphere. An orbital expanded through level $n_{\max}$ takes the form
\begin{equation}
\phi(i)=\sum_{n=0}^{n_{\max}}\sum_{m=-(Q+n)}^{Q+n}c_{nm}\,Y_{Q,n,m}(i)
\label{eq:llexp}
\end{equation}
and contains $N_b=(2|Q|+n_{\max}+1)(n_{\max}+1)$ complex coefficients, since a level has angular momentum
$l=|Q|+n$ and hence $2l+1$ states.

The lowest level illustrates what the coefficients control. For $Q>0$ an orbital factorizes as
$\sum_m c_{0m}Y_{Q,0,m}=c\prod_{k=1}^{2Q}w(\xi,\Omega_k)$, where $w(\xi,\Omega)=\xi^T\epsilon\Omega$.
Choosing the coefficients fixes the positions $\Omega_k$ of its $2Q$ zeros. In the stereographic coordinate
$z=u/v$, the higher levels are the spherical counterparts of the planar orbitals
$(\bar z-2\ell^2\partial_z)^nz^{m+n}\ee^{-|z|^2/4\ell^2}$. Each additional conjugate spinor factor is accompanied
by a direct spinor factor, preserving the net winding $2Q$. Mixing levels gives additional freedom in the radial
profile. We store the expansion in the equivalent monomial basis
$\zeta_{i,1}^p\zeta_{i,2}^{2Q+n_{\max}-p}\omega_{i,1}^q\omega_{i,2}^{n_{\max}-q}$ for $Q\ge0$.
On the sphere we verified orthonormality and the $L^2$ and $L_z$ eigenvalue equations to $10^{-15}$.

\paragraph{Generalized orbitals.} To let the zeros move with the surrounding particles, we allow the orbital
coefficients to depend on the same features that enter the Jastrow factor and backflow. Orbital $a$ in determinant
$k$ is then
\begin{equation}
v^k_a(i)=\sum_{n,m}c^k_{a,nm}(i)\,Y_{Q,n,m}(i),\qquad c^k_{a,nm}(i)=C^k_{a,nm}+\tilde h_i\cdot D^k_{a,nm},\qquad
\tilde h_i=\tanh(W_r\bar h_i).
\label{eq:cfnetD}
\end{equation}
where the learned matrix $W_r$ reduces the $H$ components of $\bar h_i$ in Eq.~\eqref{eq:avg} to $H_r=8$ or 16.
The complex number $C^k_{a,nm}$ is the configuration-independent part of the coefficient, and its environmental correction is the
contraction of $\tilde h_i$ with the learned complex vector $D^k_{a,nm}$. The index $a$ runs over the orbitals
in the determinant, or in one flavor block when flavors are present. Exchanging two particles of the same flavor
exchanges their features and hence two rows of the determinant, preserving its antisymmetry.

This construction follows the generalized orbitals of FermiNet \cite{Pfau2020}. Here the Landau-level basis carries
the winding required by the flux. The network changes the mixture of basis states for each particle, allowing its
zeros to follow the other particles [Fig.~\ref{fig:ansatzcf}(c)]. On a lattice the site index takes the role of the
orbital label $(n,m)$ in the related construction $\Phi_{i\alpha}=\sum_\beta y_{i\beta}W_{i\alpha\beta}$ of
Ref.~\cite{Rende2026}. Ref.~\cite{Qian2025} uses configuration-dependent lowest-level coefficients for electrons
on the sphere.

For the smaller networks we use a factorized version of Eq.~\eqref{eq:cfnetD},
\begin{equation}
v_a(i)=\Big[\sum_{n,m}C_{a,nm}Y_{Q,n,m}(i)\Big]\bigl(1+\tilde h_i\cdot U_a\bigr)
+\Big[\sum_{n,m}C'_{a,nm}Y_{Q,n,m}(i)\Big]\,\tilde h_i\cdot W_a,
\label{eq:cfnet}
\end{equation}
The environment now rescales a single reference orbital, as in the multiplicative backflow of PauliNet
\cite{Hermann2020}, and sets the weight of a second orbital. In Eq.~\eqref{eq:cfnetD} this amounts to taking
$D_{qa,nm}=C_{a,nm}U_{a,q}+C'_{a,nm}W_{a,q}$. Each orbital requires two sets of $N_b$ complex coefficients and
two vectors of length $H_r$, a total of $2N_b+2H_r$ parameters per determinant and orbital.
In the full form the environmental correction alone contains $N_bH_r$ coefficients.

We use the full form in Eq.~\eqref{eq:cfnetD} for the attention and SU(3) MP3 networks, the CB networks, the
general holon orbitals of the color states and the color-flavor-locked states. The CF networks with MP1 features
and the CF liquids initialized in an exact second-level state use the factorized form in Eq.~\eqref{eq:cfnet}.
For the semion CF Landau-level states, the occupied orbitals have fixed coefficients shared by every particle.

\subsection{The four families}\label{sec:ans}
\begin{figure}[h]
\centering
\includegraphics[width=\textwidth]{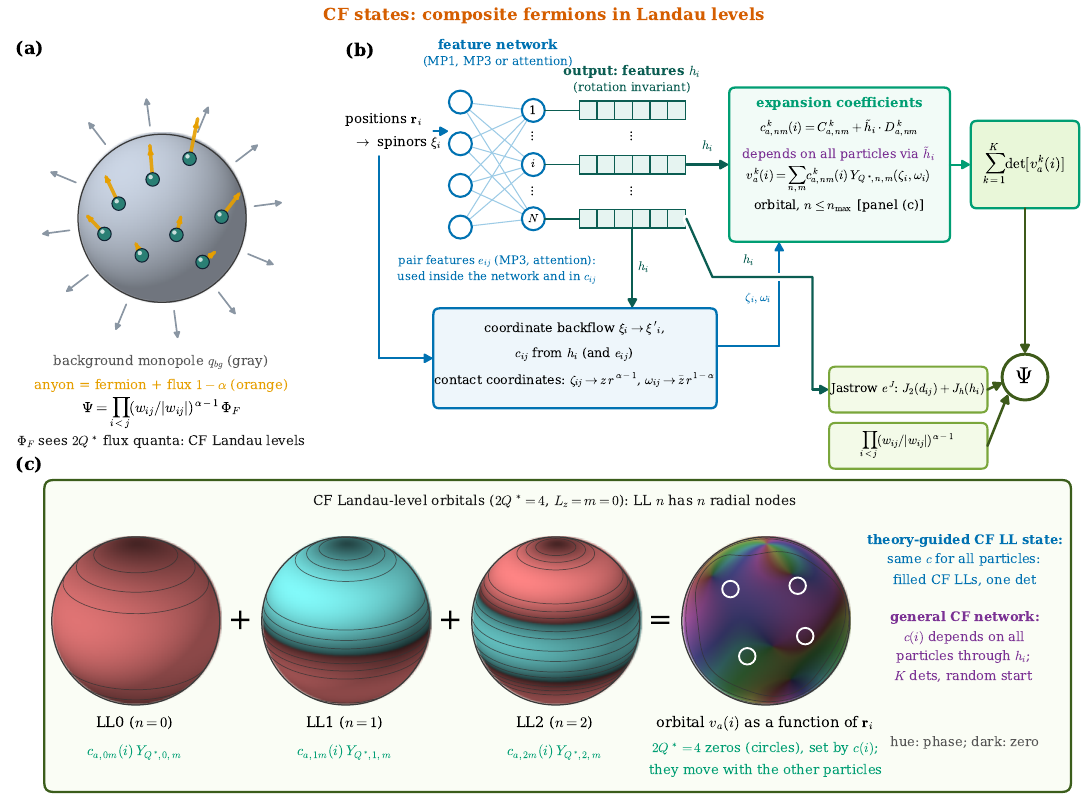}
\caption{The diagrams illustrate the construction of the CF states. (a) In the CF gauge, each anyon is represented by a fermion carrying flux $1-\alpha$ (orange) in a background monopole field $\qbg$ (gray), so that $\Phi_F$ experiences $2Q^*$ flux quanta and the composite fermions occupy Landau levels. (b) The general CF neural network converts particle positions to spinors and uses message passing or attention to generate rotation-invariant features $h_i$. These features determine the LL expansion coefficients $c^k_{a,nm}(i)=C^k_{a,nm}+\tilde h_i\cdot D^k_{a,nm}$ of the generalized orbitals $v^k_a(i)$, the backflow coefficients $c_{ij}$ together with the pair features $e_{ij}$ of the deeper networks, and the Jastrow term $J_h$. Each orbital therefore depends on the full configuration rather than only on $\mathbf r_i$. We evaluate the orbitals at the backflowed contact coordinates, sum the $K$ determinants and multiply by the Jastrow and statistical factors. The CF LL states instead use a single determinant with the filled-LL coefficients shared by all particles. (c) CF LL orbitals are shown for $2Q^*=4$ and $m=0$, with $n$ radial nodes in level $n$. A linear combination defines an orbital $v_a(i)$, plotted as a function of $\mathbf r_i$ with the remaining particles fixed. Its phase winds $2Q^*$ times around the sphere, giving a net $2Q^*$ zeros, or vortices (circles), whose positions are set by the coefficients and move with the other particles in the general network. Hue represents phase, while brightness and thin contours indicate amplitude.}
\label{fig:ansatzcf}
\end{figure}

\begin{figure}[h]
\centering
\includegraphics[width=\textwidth]{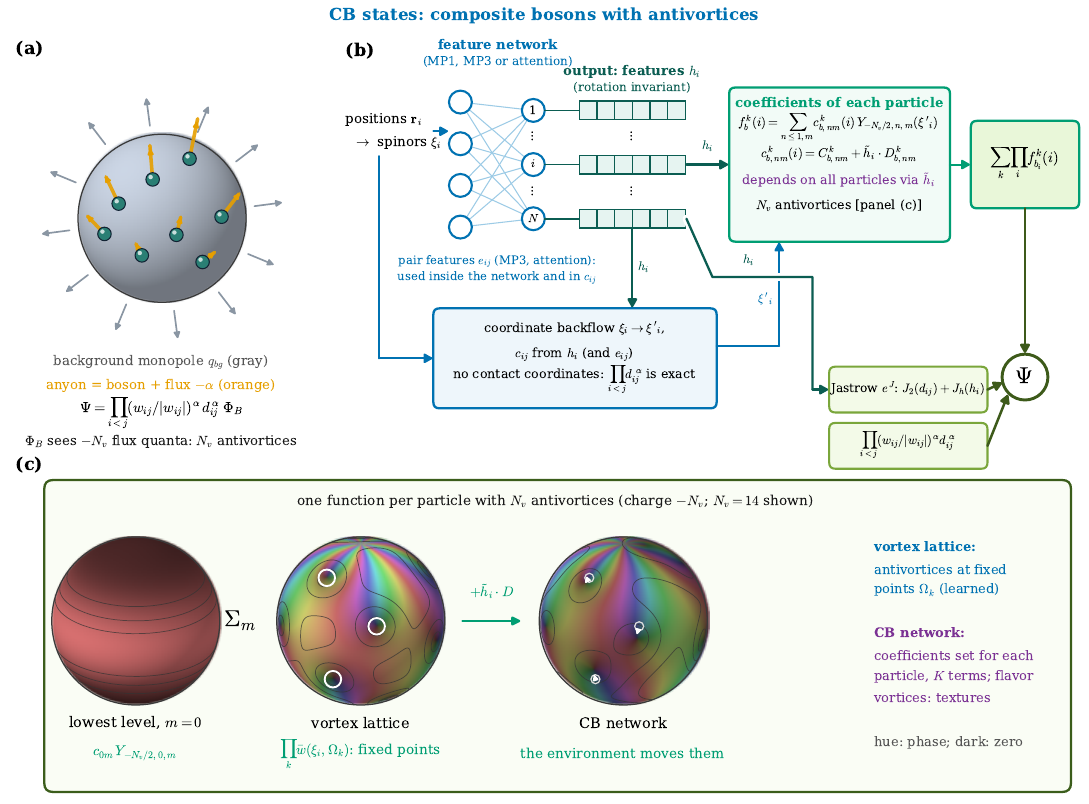}
\caption{The diagrams illustrate the construction of the CB states. (a) In the CB gauge, each anyon is represented by a boson carrying flux $-\alpha$, while $\prod_{i<j}d_{ij}^\alpha$ supplies the exact contact behavior. The wavefunction $\Phi_B$ experiences $-N_v$ flux quanta, requiring $N_v$ antivortices in the function associated with each particle. (b) In the CB network of Eq.~\eqref{eq:cbnet}, the features $h_i$ determine the coefficients of the functions $f^k(i)$ in the lowest two LLs at negative flux, as well as the backflow and Jastrow term. These functions are evaluated at the backflowed spinors and combined as a sum of $K$ products. (c) The functions shown contain $N_v=14$ antivortices (circles), which occupy fixed sites in the vortex lattice but move with each particle's environment in the CB network. Assigning a separate function to each flavor allows the state to develop a flavor texture.}
\label{fig:ansatzcb}
\end{figure}

\begin{figure}[h]
\centering
\includegraphics[width=\textwidth]{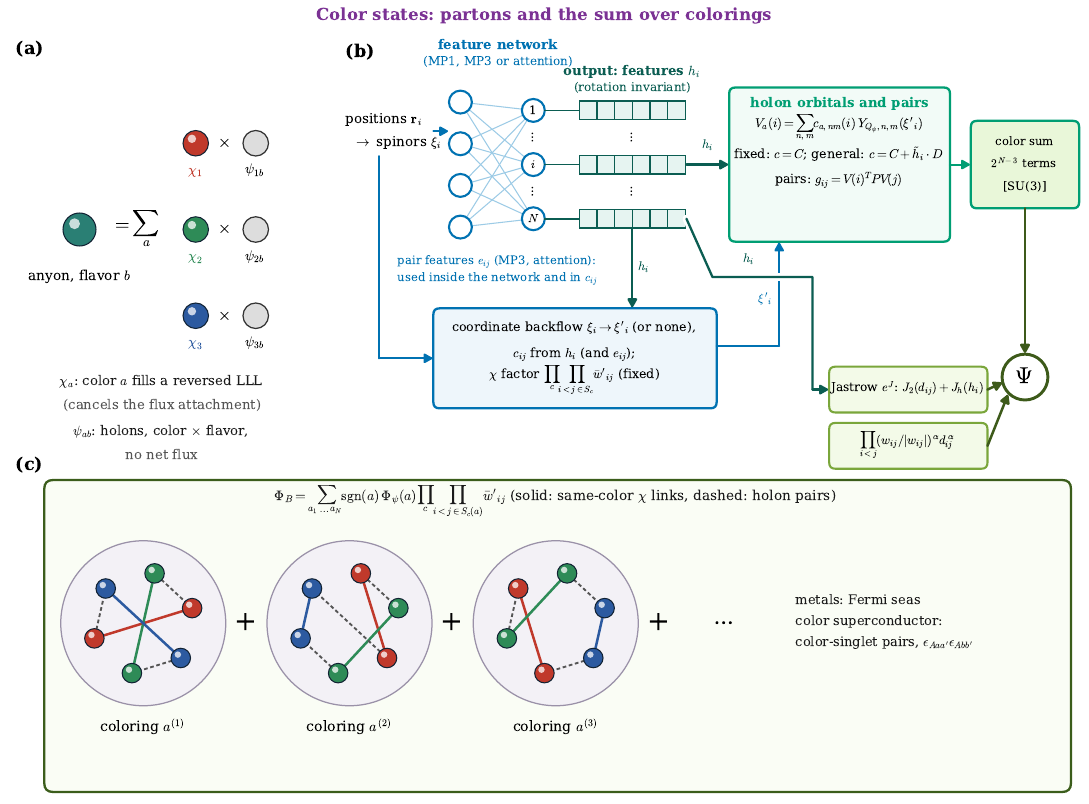}
\caption{The diagrams illustrate the construction of the color states for $N_c=3$. (a) An anyon of flavor $b$ is decomposed as $\sum_a\chi_a\psi_{ab}$, with each color of the $\chi$ partons filling a reversed lowest LL that cancels the flux attachment. The holons $\psi_{ab}$ carry both color and flavor and experience zero net flux per particle. (b) Holon orbitals are expanded in LL functions at the backflowed spinors, with either fixed coefficients or general coefficients determined by the features $h_i$, which also enter the backflow and Jastrow term. For a holon pair at particles $i$ and $j$, the orbital amplitude is $g_{ij}=\sum_{p,q}P_{pq}V_p(i)V_q(j)$. Here $p,q$ label holon orbitals, and $P_{pq}=-P_{qp}$ is their pairing amplitude. The $\chi$ factor is fixed, and we sum over the color assignments exactly. (c) Each coloring contributes one term in which the $\chi$ partons connect anyons of the same color (solid lines), while the holons form Fermi seas or pairs (dashed lines).}
\label{fig:ansatzcolour}
\end{figure}

Figures~\ref{fig:ansatzcf}--\ref{fig:ansatzcolour} illustrate the CF, CB and color families.

\paragraph{CF LL states (semions).} The theory-guided semion state is the integer quantum Hall state of composite fermions
at $\nu^*=2$ : in the CF flux $2Q^*$ the CFs fill the two lowest LLs (spinless) or the lowest LL of each
spin (SU(2)),
\begin{equation}
\Psi=\prod_{i<j}\Big(\frac{w_{ij}}{|w_{ij}|}\Big)^{\alpha-1}\det[Y_{Q^*\!,n_a,m_a}(i)]\;\ee^{J},
\label{eq:cfll}
\end{equation}
where $(n_a,m_a)$ labels the occupied LL orbitals, evaluated at the contact coordinates of the backflowed spinors. For SU(2), we use one determinant per spin and multiply their product by the opposite-spin factor in Eq.~\eqref{eq:ifx}. The orbital functions are shared by all particles, leaving the variational freedom in the Jastrow factor, backflow and contact functions. Allowing admixture of the next unoccupied level has negligible effect, since its weight in the occupied space remains below $2\times10^{-4}$ after training.

\paragraph{CF liquids with an exact second-level start ($\theta=\pi/3$).} At $\nu^*=3/2$ the CFs fill LL0, and the
remaining $N_1=N-2Q^*-1$ CFs half fill the second CF level, where they form a paired or Fermi-liquid state; each shift
sector selects one candidate (Sec.~\ref{sec:res-pi3}). Such a state is not one determinant but a superposition of the ways
to place the $N_1$ CFs in the $2Q^*+3$ orbitals of the second level, and we write it as
\begin{equation}
\Psi=\prod_{i<j}\Big(\frac{w_{ij}}{|w_{ij}|}\Big)^{\alpha-1}\sum_\kappa c_\kappa\det[v_{\rm LL0}(i)\,|\,v_\kappa(i)]\;
\ee^{J}.
\end{equation}
Each configuration $\kappa$ selects $N_1$ second-level orbitals with total $L_z=0$. The matrix $[v_{\rm LL0}(i)\,|\,v_\kappa(i)]$ has dimension $N\times N$ and contains the $2Q^*+1$ LL0 orbitals together with the $N_1$ orbitals in $\kappa$. All are generalized orbitals of the factorized form in Eq.~\eqref{eq:cfnet}, initialized at the LL orbitals with random $C'$, $U=W=0$ and $n_{\max}=2$. The number of configurations ranges from 3 at $N=8$ to 1514 at $N=24$, and we retain those with nonzero coefficients in the target state. The initial amplitudes $c_\kappa$ are the expansion coefficients of the Moore--Read, anti-Pfaffian, Rezayi--Read (CFL) or $J=0$ paired state of $N_1$ particles in the lowest LL at flux $2Q^*+2$. Mapping $m\to m$ onto the second level, which has the same $l=Q^*+1$, places the target state in the second CF level. Stochastic reconfiguration then optimizes these complex amplitudes together with all other parameters, allowing the second-level state, the orbitals and their correlations to adapt during training.  We train each state with the contact treatments described above and report the lowest energy in Table~\ref{tab:pi3}.

\paragraph{General CF neural networks.} We write the wavefunction as
\begin{equation}
\Psi=\prod_{i<j}\Big(\frac{w_{ij}}{|w_{ij}|}\Big)^{\alpha-1}
F_{\ne}\,\ee^{J}\sum_{k=1}^K \mathcal D_k.
\end{equation}
For spinless anyons, $\mathcal D_k=\det[v^k_a(i)]$ and $F_{\ne}=1$.
For $N_f=2$ or 3 flavors, $\mathcal D_k=\prod_{b=1}^{N_f}\det M_b^k$.
The square matrix $M_b^k$ has one row for each particle $i$ of flavor $b$ and one column for each
occupied orbital $a$ of that flavor. Its entries are $(M_b^k)_{ia}=v^k_{b,a}(i)$.
The sum over $k$ combines these products of determinants. A single factor $F_{\ne}$, defined below,
accounts for contacts between different flavors and multiplies the whole sum.
We evaluate the generalized orbitals at the contact coordinates of the backflowed spinors, using $2Q=2Q^*$ and $n_{\max}=2$.
We initialize random orbitals using complex Gaussian coefficients for $C^1$, with each orbital normalized on the sphere, $C^k=C^1+0.3\times$(random) for $k\ge2$, and $D=0$ (or $U=W=0$ with random $C'$). These unconstrained initial states test whether the phases obtained from the theory-guided states depend on their imposed structure. The sum over $K$ orbital configurations gives a multireference wavefunction \cite{Pfau2020}, while particle-dependent coefficients allow the orbitals to leave the occupied subspace of the CF LL states. The networks with MP1 features use the factorized orbitals of Eq.~\eqref{eq:cfnet} with $K=2$: these are the
message-passing networks of the semions and of spinless anyons at $\theta=\pi/3$, and those of SU(3) anyons at $N=18$
and 24. The attention networks (semions, $K=4$; SU(3), $K=16$) and the SU(3) MP3 networks ($K=64$ at $N=6$, $K=16$ at
$N=9$ and 12) use the full orbitals of Eq.~\eqref{eq:cfnetD}.

\paragraph{Contacts between different flavors in the CF gauge.}
Particles of different flavors label rows of separate determinants in the CF states above.
The product of determinants can therefore remain finite at contact. The factor $F_{\ne}$ supplies the
short-distance behavior of the physical wavefunction for these pairs.
Let $x_{ij}=d_{ij}^2/(d_{ij}^2+r_w^2)$. The simplest choice is the radial factor
$F_{\ne}=\prod_{i<j,\,b_i\ne b_j}x_{ij}^{(1-\alpha)/2}$.
Together with the CF statistical phase, it gives $r^{1-\alpha}\ee^{\ii(\alpha-1)\phi}$ for a pair
at relative position $r\ee^{\ii\phi}$. In the $\ell=\alpha$ channel it gives a radial power $r^{2-\alpha}$,
where the regular two-anyon solution scales as $r^\alpha$.

To allow both regular channels, we use
\begin{equation}
F_{\ne}=\prod_{i<j,\,b_i\ne b_j}\Big[x_{ij}^{(1-\alpha)/2}
+\mathcal K(d_{ij})\,x_{ij}^{(\alpha-1)/2}\,\frac{2R}{a_0}\,w'_{ij}\bar\nu_{ij}\Big],
\qquad
\nu_{ij}=\sum_{l\neq i,j}\tau(d_{il},d_{jl})\,(\xi'^T_i\epsilon\,\xi_l)(\xi_l^\dagger\xi'_j).
\label{eq:ifx}
\end{equation}
The relative coordinate $w'_{ij}$ supplies one extra winding. The environment factor $\bar\nu_{ij}$
cancels its spinor gauge phase and is generically nonzero at contact. For an isolated close pair, the
second term thus gives $r^\alpha\ee^{\ii\alpha\phi}$ in the physical wavefunction.
We expand $\mathcal K(d)=\sum_r k_rG_r(d)$ and
$\tau(d_1,d_2)=\sum_s t_s\ee^{-(d_1^2+d_2^2)/(2\varsigma_s^2)}$, with $k_r=0$ and $t_s=1$ initially.
The SU(2) CF states in Table~\ref{tab:semions} and the SU(3) CF message-passing networks use this two-channel form.
The SU(3) attention networks use it at $\qbg=2/3$ and keep $\mathcal K=0$ at $\qbg=-1/3$.

\paragraph{CB networks (crystals).} We first place $N_v$ antivortices at prescribed points on the sphere.
Let $\xi=(u,v)^T$ denote the position of a particle and $\Omega_t=(U_t,V_t)^T$ the position of antivortex $t$.
The function $w(\xi,\Omega_t)=uV_t-vU_t$ vanishes when the particle reaches that point.
All particles occupy the orbital $f_0(\xi)=\prod_{t=1}^{N_v}\overline{w(\xi,\Omega_t)}$, giving
\begin{equation}
\Psi=\prod_{i<j}\Big(\frac{w_{ij}}{|w_{ij}|}\Big)^{\alpha}\Big(\frac{d_{ij}}{a_0}\Big)^{\alpha}
\prod_i f_0(\xi_i)\;\ee^{J}.
\end{equation}
For fixed $\Omega_t$, $f_0$ is a homogeneous polynomial of degree $N_v$ in $\bar u,\bar v$.
It therefore lies in the lowest LL at flux $-N_v$, with one antivortex at each $\Omega_t$.
To start with nearly uniform vortex density, we distribute the points over the sphere using a Fibonacci lattice.
We space their heights $\cos\theta_t=1-(2t-1)/N_v$ uniformly and advance the azimuth by the golden angle
$\pi(3-\sqrt5)$ between successive points. We then optimize their positions as variational parameters shared by all
particles and flavors.

The crystal energies quoted in the main text use a CB network that lowers the energy by allowing the vortex
positions to depend on the surrounding particles. We write its wavefunction as
\begin{equation}
\Psi=\prod_{i<j}\Big(\frac{w_{ij}}{|w_{ij}|}\Big)^{\alpha}\Big(\frac{d_{ij}}{a_0}\Big)^{\alpha}\sum_{k=1}^{K}\prod_i
f^k_{b_i}(i)\;\ee^{J},\qquad f^k_b(i)=\sum_{n=0}^{1}\sum_m\bigl(C^k_{b,nm}+\tilde h_i\cdot D^k_{b,nm}\bigr)\,
Y_{-N_v/2,n,m}(\xi'_i),
\label{eq:cbnet}
\end{equation}
Here the monopole harmonics $Y_{-N_v/2,n,m}$ span the lowest two LLs at flux $-N_v$. We evaluate the orbitals at the backflowed
spinor $\xi'_i$ without contact deformation. Each product $k$ contains a separate orbital $f^k_b$ for each
flavor $b$, so the flavors can develop distinct vortex configurations. We learn the complex coefficients
$C^k_{b,nm}$ and vectors $D^k_{b,nm}\in\mathbb C^{H_r}$. Through the feature vector $\tilde h_i$ in
Eq.~\eqref{eq:cfnetD}, the coefficients of the orbital sampled by particle $i$ depend on its neighbors.

We initialize the first product with the lowest-LL coefficients of $f_0$ and set $D=0$.
The other products start from rotated copies of the lattice. When flavors have independent vortex positions,
we also rotate their initial lattices independently. We use $K=2$ to 16 products, initializing the $K=8$
and 16 networks from trained networks with fewer products. During training the network can shift the zeros
and mix in the next LL.

\paragraph{Color states.} The color ansatz in Equation~\eqref{eq:colour} on the sphere, in the CB gauge with $N_c=1/\alpha$ colors and
$k=N/N_c$ anyons per color, is
\begin{equation}
\Psi=\prod_{i<j}\Big(\frac{w_{ij}}{|w_{ij}|}\Big)^{\alpha}\Big(\frac{d_{ij}}{a_0}\Big)^{\alpha}\sum_{a_1\ldots a_N}
\sgn(a)\,\Phi_\psi(a)\prod_{c=1}^{N_c}\prod_{i<j\in S_c(a)}\bar w'_{ij}\;\ee^{J},
\label{eq:colstate}
\end{equation}
Here $a=(a_1,\dots,a_N)$ is one color configuration. $S_c(a)=\{i:a_i=c\}$ is the set of anyons of color $c$, and $\sgn(a)$ is the sign of the permutation that sorts the anyons by color. The holon wavefunction for each color configuration is $\Phi_\psi(a)$, while $\bar w'_{ij}$ is the conjugate of $w_{ij}$ evaluated at the backflowed spinors. The factor $\prod_{i<j\in S_c}\bar w'_{ij}$ is the spherical counterpart of $\bar V_{S_c}$ in Eq.~\eqref{eq:colour} and represents a reversed filled lowest LL of color $c$. Up to a sign, it represents the integer quantum Hall state of the $\chi_a$ fermions.

The holons experience a flux $2Q_\psi=\qbg-1+1/N_c$ [Eq.~\eqref{eq:Qpsi}], which is the spherical counterpart of zero field and equals $\qbg-2/3$ at $\theta=\pi/3$. Their orbitals $V_p(i)=\sum_{n,m}c_{p,nm}(i)\,Y_{Q_\psi,n,m}(\xi'_i)$ are evaluated at the backflowed spinors, using either fixed coefficients $c=C$ or general coefficients $c=C+\tilde h_i\cdot D$. In the SU(3) holon metal, each of the nine species $\psi_{ab}$ occupies the same closed shell of holon levels, and $\Phi_\psi(a)$ is the product of the corresponding nine determinants. In the color superconductor, $\Phi_\psi(a)=\Pf[\sigma_{ij}g_{ij}]$. The factor $\sigma_{ij}=\sum_A\epsilon_{Aa_ia_j}\epsilon_{Ab_ib_j}$ pairs the color and flavor indices. The orbital factor $g_{ij}=\sum_{p,q}V_p(i)P_{pq}V_q(j)$ is antisymmetric under $i\leftrightarrow j$. Its indices $p,q$ run over the orbitals of the retained holon levels. We initialize the antisymmetric matrix $P$ with total-angular-momentum-zero pairs within each level, $P_{(n,m),(n,-m)}=(-1)^{l-m}$ with $l=|Q_\psi|+n$. For both states, fixed and general orbitals give energies that agree within 0.006 per particle. The other color states in Secs.~\ref{sec:eft-colour}, \ref{sec:res-pi3} and \ref{sec:res-su3} differ only in their holon wavefunctions. We evaluate the color sum exactly using the polarization form in Sec.~\ref{sec:eft-colour}, or by summing individual color assignments at the smallest sizes.

\paragraph{Color-flavor locked states.} In the U(1)$^2$ metal (and its holon quantum Hall form), the U(1)$^2$
superconductor and the singlet Z$_3$ metal, a Higgs field locks the color index of the holons to the flavor index, so
only the coloring $a_i=b_i$ survives in Eq.~\eqref{eq:colstate}. Its $\chi$ factor contains one reversed
LL determinant per flavor, $\prod_b\prod_{i<j\in b}\bar w'_{ij}$.

Let $\omega_{ij}=u_iv_j-v_iu_j$ denote the spinor product written as $w_{ij}$ in Sec.~\ref{sec:sphere}.
Changing from CB to CF gauge multiplies the auxiliary wavefunction by one factor
$\omega_{ij}/|\omega_{ij}|$ for every pair [Eq.~\eqref{eq:logpsi}].
Before backflow, this phase combines with $\bar\omega_{ij}$ from the $\chi$ determinant to give
$|\omega_{ij}|$ for a same-flavor pair. For a different-flavor pair the phase remains. Thus
\begin{equation}
\prod_{i<j}\left(\frac{\omega_{ij}}{|\omega_{ij}|}\right)^{\alpha}
\prod_{\substack{i<j\\ b_i=b_j}}\bar\omega_{ij}
=
\prod_{i<j}\left(\frac{\omega_{ij}}{|\omega_{ij}|}\right)^{\alpha-1}
\prod_{\substack{i<j\\ b_i\ne b_j}}\frac{\omega_{ij}}{|\omega_{ij}|}
\prod_{\substack{i<j\\ b_i=b_j}}|\omega_{ij}|.
\label{eq:cflid}
\end{equation}

\emph{U(1)$^2$ metal.} Each holon flavor fills its own Fermi sea. Inserting its Slater determinant
and the locked $\chi$ factor into Eq.~\eqref{eq:colstate} gives the CB wavefunction
\begin{equation}
\Psi=\prod_{i<j}\Big(\frac{\omega_{ij}}{|\omega_{ij}|}\Big)^{\alpha}\Big(\frac{d_{ij}}{a_0}\Big)^{\alpha}
\prod_b\Big[\prod_{i<j\in b}\bar \omega'_{ij}\,\det_{i\in b}[V_a(i)]\Big]\,\ee^{J}.
\label{eq:cflcb}
\end{equation}
To see which factors change in the variational ansatz, first strip the CF statistical phase from
Eq.~\eqref{eq:cflcb}. With $I_b=\{i:b_i=b\}$, the auxiliary CF wavefunction is exactly
\[
\Phi_F=\ee^J\prod_{i<j}\left(\frac{d_{ij}}{a_0}\right)^\alpha
\prod_{\substack{i<j\\ b_i\ne b_j}}\frac{\omega_{ij}}{|\omega_{ij}|}
\prod_b\left[
 \prod_{i<j\in I_b}\left(\frac{\omega_{ij}}{|\omega_{ij}|}\bar\omega'_{ij}\right)
 \det_{i\in I_b}[V_a(i)]\right].
\]
Here $\omega'_{ij}$ is the spinor product at the backflowed positions. In the absence of backflow,
the same-flavor pair factor in brackets reduces to $|\omega_{ij}|$, as in Eq.~\eqref{eq:cflid}.
On the sphere, a CF sees $2Q^*=\qbg+(1-\alpha)(N-1)$ flux quanta before inter-flavor flux attachment
(Sec.~\ref{sec:sphere}). Each CF carries one flux quantum as seen by a CF of a different flavor. The pair phase
$\omega_{ij}/|\omega_{ij}|$ winds once when particle $i$ circles such a particle $j$.
For three equally populated flavors there are $2N/3$ such partners per particle. At $\alpha=1/3$ the
pair factors carry $2N/3$ units of the total flux $2Q^*=\qbg+2(N-1)/3$. The determinant of flavor $b$
therefore sees an effective flux $2Q_b^*=2Q^*-2N/3=\qbg-2/3$.

The locked CB state has two zeros at a same-flavor contact, one from $\chi$ and one from the holon
determinant. With the radial prefactor, its amplitude scales as $r^{2+\alpha}$. To allow the
leading $r^\alpha$ channel, we replace those short-distance factors by a contact deformation
of the determinant coordinates for same-flavor pairs and a pair function $g_{ij}$ for different flavors.
The full variational wavefunction in CF gauge is
\begin{equation}
\begin{aligned}
\Psi &=
\prod_{i<j}\left(\frac{\omega_{ij}}{|\omega_{ij}|}\right)^{\alpha-1}
\ee^J\left(\prod_{\substack{i<j\\ b_i\ne b_j}}g_{ij}\right)
\sum_{k=1}^{K}\prod_b\det M_b^{(k)},\\[2pt]
g_{ij} &=
\frac{\omega_{ij}}{|\omega_{ij}|}\left[
\left(\frac{d_{ij}}{a_0}\right)^\alpha
\ee^{\sum_r f^o_r G_r(d_{ij})}\frac{\omega'_{ij}}{\omega_{ij}}
+\left(\frac{d_{ij}}{a_0}\right)^{1-\alpha}
\frac{\bar\omega_{ij}}{|\omega_{ij}|}\,\nu_{ij}\right].
\end{aligned}
\label{eq:cflcf}
\end{equation}
The phase in front of the brackets is the inter-flavor flux attachment of Eq.~\eqref{eq:cflid}.
With backflow and the learned corrections turned off, $\omega'_{ij}=\omega_{ij}$ and $\nu_{ij}=0$,
so $g_{ij}=(d_{ij}/a_0)^\alpha\,\omega_{ij}/|\omega_{ij}|$. The ratio $\omega'_{ij}/\omega_{ij}$
describes the backflow deformation of the relative coordinate, and the exponential adjusts its radial profile.
This form follows from $|\omega_{ij}|=d_{ij}/(2R)$.
The second term allows the other regular contact channel. The environment factor $\nu_{ij}$ of Eq.~\eqref{eq:ifx}
transforms under spinor rephasing just like $\omega_{ij}$, so both terms in brackets carry zero net flux.
Near an isolated contact, $\nu_{ij}$ is generically finite. Its multiplier $\bar\omega_{ij}/|\omega_{ij}|$
then cancels the explicit pair winding in this term. Including the CF statistical phase, the first and
second terms give the physical channels $\ell=\alpha$ and $\ell=\alpha-1$, with radial powers
$r^\alpha$ and $r^{1-\alpha}$, respectively. We initialize $f^o_r$ and the coefficients $t_s$ in $\nu_{ij}$ to zero.

For each $k$, the matrix $(M_b^{(k)})_{ia}=v^{(k)}_{b,a}(i)$ has one row for each particle $i\in I_b$
and one column for each of the $|I_b|$ occupied orbitals. We use the generalized orbitals of
Eq.~\eqref{eq:cfnetD} at flux $2Q_b^*$ and initialize them at the filled LLs with $D=0$.
The same-flavor contact coordinates [Eq.~\eqref{eq:contactcoords}] allow radial powers
$r^\alpha$ and $r^{2-\alpha}$ within each determinant. The U(1)$^2$ superconductor uses the same construction with paired holons, replacing the determinant sum by $\Pf[v(i)^TP^{b_ib_j}v(j)]$ over complete LLs; $P^{bb'}$ is initialized with $J=0$ pairs between different flavors and vanishes within a flavor.

\emph{SU(3) singlet Z$_3$ metal.} The flavor-singlet holon $\mathrm{tr}\,\psi/\sqrt{3}$ forms one
Fermi sea shared by all $N$ particles. The locked CB construction therefore has a single holon determinant
and one filled $\chi$ level per flavor,
\begin{equation}
\Psi=\prod_{i<j}\Big(\frac{\omega_{ij}}{|\omega_{ij}|}\Big)^{\alpha}\Big(\frac{d_{ij}}{a_0}\Big)^{\alpha}
\prod_{\substack{i<j\\ b_i=b_j}}\bar\omega'_{ij}\;\det_{i=1\dots N}[V_a(i)]\;\ee^{J}.
\label{eq:z3singlet-cb}
\end{equation}
We also use a second variational wavefunction for the same one-pocket metal. In the CF gauge we write
\begin{equation}
\Psi=\prod_{i<j}\Big(\frac{\omega_{ij}}{|\omega_{ij}|}\Big)^{\alpha-1}
\prod_{\substack{i<j\\ b_i\ne b_j}}\left[\frac{2R}{a_0}\,
\ee^{\sum_r f^o_rG_r(d_{ij})}\,\omega'_{ij}\right]\;
\det_{i=1\dots N}[v_a(i)]\;\ee^{J}.
\label{eq:z3singlet-cf}
\end{equation}
Equations~\eqref{eq:z3singlet-cb} and \eqref{eq:z3singlet-cf} implement the same inter-flavor flux
attachment but have different pair amplitudes. To see the difference, turn off backflow and the contact
deformation, set $v_a=V_a$ and $f^o_r=0$, and use the same $J$ in both states. An exact change of gauge
in Eq.~\eqref{eq:z3singlet-cb} leaves a radial factor $(d_{ij}/a_0)^\alpha$ for every pair.
Equation~\eqref{eq:cflid} also leaves $|\omega_{ij}|$ for a same-flavor pair and
$\omega_{ij}/|\omega_{ij}|$ for a different-flavor pair. In Eq.~\eqref{eq:z3singlet-cf}, the corresponding
factors outside the determinant are $1$ for a same-flavor pair and
$(2R/a_0)\omega_{ij}=(d_{ij}/a_0)\,\omega_{ij}/|\omega_{ij}|$ for a different-flavor pair.
In the optimized CF ansatz, the contact deformation acts on all pairs in each holon determinant.
For the singlet Z$_3$ metal in Fig.~\ref{fig:su3}, we use Eq.~\eqref{eq:z3singlet-cf} with a sum of two
$N\times N$ holon determinants. 

%=======================================================================================================================
\section{Variational Monte Carlo}\label{sec:vmc}
We regard each wavefunction as a map $X\mapsto\log\Psi_\theta(X)\in\mathbb C$ from a configuration $X=(\mathbf r_1,\dots,\mathbf r_N)$, with $|\mathbf r_i|=R$. We minimize $E(\theta)=\langle\Psi_\theta|H|\Psi_\theta\rangle/\langle\Psi_\theta|\Psi_\theta\rangle$ using samples from $|\Psi_\theta|^2$, and evaluate all derivatives exactly by automatic differentiation in JAX rather than by finite differences.

\paragraph{Local energy on the sphere.} With the monopole vector potential, $H=\frac12\sum_i(-\ii\nabla_i-\mathbf A_i)^2$
and
\begin{equation}
E_L(X)=\frac{H\Psi}{\Psi}=-\frac12\sum_i\Big[\nabla_i^2\log\Psi+(\nabla_i\log\Psi)^2\Big]
+\ii\sum_i\mathbf A_i\cdot\nabla_i\log\Psi+\frac12\sum_i\mathbf A_i^2 ,
\label{eq:eloc}
\end{equation}
where $(\nabla_i\log\Psi)^2$ denotes the complex square and we have used $\nabla\cdot\mathbf A=0$. In the gauge chosen for the spinor sections, the Dirac string lies at the north pole and $\mathbf A_i=(\qbg/2R)\cot(\theta_i/2)\,\hat{\boldsymbol\varphi}_i$. Every factor of $\Psi$ depends on the particle positions through the unit vectors $\mathbf n_i=\mathbf r_i/|\mathbf r_i|$ and their spinors. Extending the wavefunction off the sphere in this way makes $\log\Psi$ constant in each radial direction, so its Cartesian gradient in $\mathbb R^3$ is tangent to the sphere. At $|\mathbf r_i|=R$, its Cartesian Laplacian therefore reduces to the Laplace--Beltrami operator,
\begin{equation}
\nabla_i^2f=\Big(\partial_r^2+\frac2r\,\partial_r+\frac1{r^2}\Delta_{S^2}\Big)f=\frac1{R^2}\,\Delta_{S^2}f .
\end{equation}
This allows us to evaluate Eq.~\eqref{eq:eloc} using Cartesian derivatives in $\mathbb R^{3N}$, avoiding angular coordinates and their coordinate singularities. Although $E_L$ is complex, its expectation value is real because $H$ is Hermitian. We therefore report energies obtained from $\mathrm{Re}\,E_L$ and use $\langle\mathrm{Im}\,E_L\rangle$, which vanishes within statistical errors, as a consistency check.

\paragraph{Coordinate derivatives and the forward Laplacian.}
For each sampled configuration, we differentiate $f_\theta(X)=\log\Psi_\theta(X)$ in two ways.
To update the parameters, we hold the particle coordinates $X$ fixed and backpropagate from the output to
each weight $\theta_k$. This gives $O_k(X)=\partial f_\theta(X)/\partial\theta_k$ for stochastic
reconfiguration. To evaluate the kinetic energy in Eq.~\eqref{eq:eloc}, we hold the weights fixed and
differentiate with respect to the $3N$ Cartesian coordinates $x_d$ of $X$. Backpropagation to the inputs
can give all components of $\nabla_X f_\theta$. The kinetic energy also requires the sum of the second
spatial derivatives, $\Delta_X f_\theta=\sum_d\partial_{x_d}^2 f_\theta$.
Differentiating $O_k$ with respect to $x_d$ would give a mixed weight--coordinate derivative.
One could obtain the Laplacian by differentiating each component $\partial_{x_d}f_\theta$ once more
with respect to the same $x_d$. The forward calculation below accumulates this sum layer by layer.

For example, consider a single real neuron $h=\tanh u$ with $u=\sum_d w_d x_d+b$.
Varying $w_d$ at fixed $X$ gives $\partial h/\partial w_d=(1-h^2)x_d$. Varying $x_d$ at fixed weights
gives $\partial h/\partial x_d=(1-h^2)w_d$. A second coordinate derivative and a sum over $d$ then give
$\Delta_X h=-2h(1-h^2)\sum_d w_d^2$. In a deeper layer, the input $u$ to the activation can itself
have a nonzero Laplacian. For $y=\sigma(u)$ the same chain rule gives
\begin{equation}
\nabla_X y=\sigma'(u)\nabla_X u,\qquad
\Delta_X y=\sigma''(u)\sum_d(\partial_{x_d}u)^2+\sigma'(u)\Delta_X u.
\label{eq:fwdlap}
\end{equation}
We use this rule in the forward Laplacian method~\cite{Li2024}. At the input, each coordinate
$x_d$ has gradient $\mathbf e_d$ and Laplacian zero. As we evaluate each layer, we carry the value,
coordinate gradient and Laplacian of each intermediate quantity into the next layer. For an affine layer
$y=Wu+b$ the two derivatives are simply $W\nabla_X u$ and $W\Delta_X u$. Nonlinear operations use
the second-order chain rule, as in Eq.~\eqref{eq:fwdlap}. For a Slater determinant, for example, let
$A_d=M^{-1}\partial_{x_d}M$. Then $\partial_{x_d}\log\det M=\mathrm{tr}\,A_d$ and
$\Delta_X\log\det M=\mathrm{tr}(M^{-1}\Delta_X M)-\sum_d\mathrm{tr}(A_d^2)$.
The output gives both $\nabla_X f_\theta$
and the trace of its coordinate Hessian in one forward evaluation, without storing the full
$3N\times3N$ Hessian. We evaluate the real and imaginary parts as two real outputs and combine their
derivatives in Eq.~\eqref{eq:eloc}. For $N\ge25$, we store gradients of intermediate quantities
sparsely when they depend on only a few coordinates.

\paragraph{Color sums.} In the color states of Eq.~\eqref{eq:colstate} the sum over colorings,
$\Phi_B=\sum_a\sgn(a)\,\Phi_\psi(a)\prod_c\prod_{i<j\in S_c(a)}\bar w'_{ij}$, is evaluated in the polarization form of
Sec.~\ref{sec:eft-colour}, with $2^{N-3}$ terms, and widening each term by $3N$ would be too costly. Instead, we exploit the fact that anyon $i$ enters the sum only through its row $Z_i=\phi(i)\otimes V(i)$, whose entries $Z_{i\mu}=\phi_q(i)V_a(i)$,
$\mu=(q,a)$, are the products of its $\chi$ orbitals $\phi_q(i)=\bar u'^{\,q}_i\bar v'^{\,k-1-q}_i$ and its holon
orbitals $V_a(i)$. The term of each coloring is linear in $\phi(i)$ (one row of a $\chi$ determinant) and in $V(i)$ (one
row of a holon determinant, or one pair of the Pfaffian), so $\Phi_B$ is linear in each $Z_i$. One reverse pass through
the sum gives all derivatives $G_{i\mu}=\partial\log\Phi_B/\partial Z_{i\mu}$, and
$\langle G_i,A\rangle=\sum_\mu G_{i\mu}A_\mu$ denotes the contraction with an array $A$ of the shape of $Z_i$, without
complex conjugation (a vector for $A=\nabla_iZ_i$). When each row depends only on its own anyon (no backflow and fixed
orbitals), the chain rule gives
\begin{equation}
\frac{\nabla_i\Phi_B}{\Phi_B}=\langle G_i,\nabla_iZ_i\rangle,\qquad
\frac{\nabla_i^2\Phi_B}{\Phi_B}=\langle G_i,\nabla_i^2Z_i\rangle,\qquad
\nabla_i^2\log\Phi_B=\frac{\nabla_i^2\Phi_B}{\Phi_B}-(\nabla_i\log\Phi_B)^2 .
\label{eq:colchain}
\end{equation}
The second-derivative chain rule also contains $\sum_{\mu\nu}(\partial^2\Phi_B/\partial Z_{i\mu}\partial Z_{i\nu})\,\nabla_iZ_{i\mu}\cdot\nabla_iZ_{i\nu}$, which vanishes because $\Phi_B$ is linear in $Z_i$. Both relations are thus exact and require only the first derivatives $G_i$ of the color sum, together with $\nabla_iZ_i$ and $\nabla_i^2Z_i$, which are derivatives of one-particle functions with respect to the three coordinates of $\mathbf r_i$. Linearity further gives $\langle G_i,Z_i\rangle=1$, providing a consistency check. With backflowed spinors or environment-dependent orbitals, however, each row depends on all anyons, so $Z=(Z_1,\dots,Z_N)$ is a function of the full configuration $X$. The second-order chain rule then becomes
\begin{equation}
\nabla^2\log\Phi_B=\sum_{d=1}^{3N}D^2F\,[\partial_dZ,\partial_dZ]+DF\,[\Delta Z],\qquad F(Z)=\log\Phi_B(Z),
\end{equation}
with the derivatives of $F$ along $A=(A_1,\dots,A_N)$, $DF[A]=\sum_i\langle G_i,A_i\rangle$ and
$D^2F[A,A]=\partial_t^2F(Z+tA)|_{t=0}$. If each row depends only on its own anyon, $\partial_dZ$ has only the row of the
anyon that coordinate $d$ belongs to, linearity gives $D^2F[\partial_dZ,\partial_dZ]=-(DF[\partial_dZ])^2$, and this
reduces to Eq.~\eqref{eq:colchain}. Here $\partial_dZ$ and $\Delta Z$ come from the forward Laplacian of the cheap map $X\mapsto Z$. The $3N$ second derivatives $D^2F$ are obtained from nested forward passes through the streamed sum. $F$ is
holomorphic in $Z$, so the directions can be complex. The rest of $\log\Psi$ (statistics factor, Jastrow and
message-passing factors) goes through the forward Laplacian. The gradients and Laplacians of the two parts add.

\paragraph{Sampling.} We estimate expectation values from configurations sampled from $|\Psi|^2$ using parallel Metropolis walkers. Each walker is a configuration $X=(\mathbf r_1,\dots,\mathbf r_N)$ of all $N$ anyons, with fixed flavor labels $b_i$, and follows a Markov chain whose stationary distribution is $|\Psi(X)|^2/\langle\Psi|\Psi\rangle$. We use $N_w=1024$ independent walkers, reduced to 256 for the color superconductor at $N\ge12$. Each training update uses one configuration per walker. For the final energy we freeze the trained wavefunction and retain one configuration per walker in each of $N_r$ measurement rounds, giving $N_wN_r$ samples. At each Metropolis step, we propose simultaneous moves of all anyons in a walker,
\begin{equation}
\mathbf r_i'=R\,\frac{\mathbf r_i+\sigma\boldsymbol\eta_i}{|\mathbf r_i+\sigma\boldsymbol\eta_i|},\qquad
\boldsymbol\eta_i\sim\mathcal N(0,I_3),
\end{equation}
and accept the move with probability $\min\{1,|\Psi(X')/\Psi(X)|^2\}$, retaining $X$ otherwise. Because the density of the proposed position $\mathbf r_i'$ depends only on the angle between $\mathbf r_i$ and $\mathbf r_i'$, the proposal is symmetric and satisfies detailed balance with respect to $|\Psi|^2$. We initialize the walkers uniformly on the sphere and thermalize them for 100 blocks of 10 steps. Each training iteration uses one such block, giving one sample per walker separated from the previous sample by 10 steps. The proposal scale starts at $\sigma=0.3a_0$ and is updated after each training block according to $\sigma\leftarrow\sigma\cdot\min\{1.25,\max[0.8,\ee^{1.5(a-1/2)}]\}$, where $a$ is the block acceptance rate; this keeps $a$ near $1/2$. We hold $\sigma$ fixed during evaluation. For some disk CF liquids (Sec.~\ref{sec:res-semion}), we instead sample $|\Psi|^2g$ with $g=1+\sum_{i<j}a^2/r_{ij}^2$ and $a=0.14a_0$, then reweight by $1/g$ to bound the weighted local energy at contact.

\paragraph{Parameter derivatives.} We compute the complex logarithmic derivatives $O_k(X)=\partial\log\Psi/\partial\theta_k$ by reverse-mode differentiation, using separate passes for the real and imaginary parts. Walkers are processed in batches of 32, or 16 for $N\ge25$, and the resulting $N_w\times P$ matrix is stored for the parameter update.

\paragraph{Stochastic reconfiguration.} We train all neural-network states by stochastic reconfiguration \cite{Sorella1998}. At each iteration, we assign weights $p_s=1/N_w$ to the walkers $X_s$, or use the guiding weights when appropriate, and define
\begin{equation}
\bar E=\sum_sp_sE_L(X_s),\qquad\bar O_{sk}=O_k(X_s)-\sum_{s'}p_{s'}O_k(X_{s'}),\qquad
\varepsilon_s=\tilde E_L(X_s)-\sum_{s'}p_{s'}\tilde E_L(X_{s'}).
\end{equation}
Here $\tilde E_L$ is $E_L$ with its real part clipped to $[m-5D,m+5D]$, where $m$ is the median and $D$ the
mean absolute deviation from it. The clipping enters only the update; $\bar E$ and every reported energy
are unclipped. Stacking real and imaginary parts,
\begin{equation}
Y=\begin{pmatrix}\sqrt p\,\mathrm{Re}\,\bar O\\ \sqrt p\,\mathrm{Im}\,\bar O\end{pmatrix}\in\mathbb R^{2N_w\times P},
\qquad
e=\begin{pmatrix}\sqrt p\,\mathrm{Re}\,\varepsilon\\ \sqrt p\,\mathrm{Im}\,\varepsilon\end{pmatrix},
\end{equation}
gives $S=Y^TY$ with $S_{kl}=\mathrm{Re}\sum_sp_s\bar O^*_{sk}\bar O_{sl}$, the quantum metric, and
$(Y^Te)_k=\mathrm{Re}\sum_sp_s\bar O^*_{sk}\varepsilon_s=\frac12\partial_{\theta_k}\bar E$. The update direction
solves
\begin{equation}
(S+\lambda I)\,\delta=Y^Te,\qquad\lambda=10^{-2}\,\mathrm{tr}\,S/P .
\label{eq:sr}
\end{equation}
Equivalently, $\delta$ minimizes $|Y\delta-e|^2+\lambda|\delta|^2$. Changing the parameters by $-\delta$ changes $\log\Psi$ by $-\bar O\delta\approx-\varepsilon$, giving the best tangent-space approximation to a short step of imaginary-time evolution $\ee^{-\tau H}$. For all reported optimizations, we solve Eq.~\eqref{eq:sr} by preconditioned conjugate gradients. We apply $S$ as $Sv=Y^T(Yv)$ and use the diagonal preconditioner $(\mathrm{diag}\,S+\lambda)^{-1}$. The solver uses at most 60 iterations and a relative tolerance of $10^{-6}$, after which we update the parameters according to
\begin{equation}
\theta\leftarrow\theta-\eta_ts_t\,\delta,\qquad\eta_t=\frac{0.02}{1+t/2000},\qquad
s_t=\min\Big(1,\frac{0.05}{\eta_t|Y\delta|}\Big).
\end{equation}
Since $|Y\delta|^2=\delta^TS\delta$, the quantity $\eta_ts_t|Y\delta|$ is the Fubini--Study length of the update to leading order, and the rescaling limits each change in the normalized state to 0.05. We skip updates with non-finite entries.

\paragraph{Cost and runs.} Each training iteration requires, per walker, 10 evaluations of $\mathrm{Re}\log\Psi$, one forward-Laplacian pass and one reverse pass for $O$, with the forward Laplacian accounting for most of the cost. Runs typically last 1000--3000 iterations, and larger systems are initialized from trained states at smaller sizes.

\paragraph{Evaluation.} After training, we hold the variational parameters and proposal scale $\sigma$ fixed. A standard evaluation uses $N_r=20$ rounds separated by 20 Metropolis steps. We first average the real local energy over rounds within each walker, then use the spread of these independent walker means to estimate the standard error,
\begin{equation}
E=\frac1{N_rN_w}\sum_{r,w}\mathrm{Re}\,E_L(X_w^{(r)}),\qquad
\sigma_E=\frac{\mathrm{std}_w(\bar E_w)}{\sqrt{N_w}},\qquad\bar E_w=\frac1{N_r}\sum_r\mathrm{Re}\,E_L(X_w^{(r)}).
\end{equation}
If successive measurements within each walker are independent, the standard error is $\sigma_L/\sqrt{N_wN_r}$,
where $\sigma_L^2=\langle(\mathrm{Re}\,E_L)^2\rangle-E^2$ is the variance of one measurement.
For correlated measurements and runs long compared with the correlation time, the effective sample count is
$N_{\rm eff}\simeq N_wN_r/[1+2\sum_{t\ge1}\rho_E(t)]$, where $\rho_E(t)$ is the normalized autocorrelation
of the real local energy at a lag of $t$ measurement rounds. The reported errors use
$\mathrm{std}_w(\bar E_w)$ directly, so the spread of walker means includes these correlations.
The angular momentum
$\mathbf L=\sum_i[\mathbf r_i\times(-\ii\nabla_i-\mathbf A_i)+\frac{\qbg}2\mathbf n_i]$ is Hermitian, so
\begin{equation}
\langle L^2\rangle=\sum_{a=x,y,z}\Big\langle\Big|\frac{L_a\Psi}{\Psi}\Big|^2\Big\rangle,\qquad
\frac{\mathbf L\Psi}{\Psi}=\sum_i\Big[-\ii\,\mathbf r_i\times\nabla_i\log\Psi-\mathbf r_i\times\mathbf A_i
+\frac{\qbg}2\mathbf n_i\Big],
\end{equation}
which needs first derivatives only. We classify a state as rotation invariant (uniform) when $\langle L^2\rangle\le3$. States with $\mathrm{var}/N>1.5$ near the crystal are re-evaluated with 200 rounds
of 1024 walkers after 100 thermalization rounds (``reeval'' in the data). Running means at 20, 50, 100 and 200 rounds
are stored to show convergence.

\paragraph{Extrapolation.} We extrapolate within each shift sector $\qbg$ using linear fits $E/N=e_\infty+b/N$, since the shift enters only at order $1/N$. For the disk, the boundary adds a term $a/\sqrt N$ (Sec.~\ref{sec:res-semion}). We estimate fit uncertainties from the residual scatter, as optimization errors dominate the VMC uncertainties.

%=======================================================================================================================
\section{Parameter-free analytical wavefunctions}\label{sec:analytic-wf}
Here we give the analytical wavefunctions of the color-superconductor ansatz. They arise from pairing of the holon
metal, whose partons carry a color index with SU($N_c$) gauge symmetry. We use the physical particle coordinates
without backflow. With $N_c=1/\alpha$ colors, a color configuration $a=(a_1,\ldots,a_N)$ assigns $N/N_c$ particles
to each color, and $S_c(a)$ denotes the set of particles of color $c$. In terms of the spinor products $w_{ij}$ of
Sec.~\ref{sec:sphere}, the three wavefunctions read
\begin{align}
\Psi_{\rm s} &= \prod_{i<j}w_{ij}^{1/2}J_{ij}\sum_a\sgn(a)\,
  \Pf\!\left[\epsilon_{a_i a_j}\,\bar w_{ij}^{\,2}|w_{ij}|^{-1}\right]
  \prod_c\prod_{i<j\in S_c(a)}\bar w_{ij}|w_{ij}|^{-1},
  \label{eq:pfree-spinless}\\
\Psi_{\rm SU(2)} &= \prod_{i<j}w_{ij}^{1/2}J_{ij}\sum_a\sgn(a)\,
  \Pf\!\left[\epsilon_{a_i a_j}\epsilon_{s_i s_j}\,\bar w_{ij}|w_{ij}|^{-1}\right]
  \prod_c\prod_{i<j\in S_c(a)}\bar w_{ij}|w_{ij}|^{-1},
  \label{eq:pfree-su2}\\
\Psi_{\rm SU(3)} &= \prod_{i<j}w_{ij}^{1/3}J_{ij}\sum_a\sgn(a)\,
  \Pf\!\left[\sigma_{ij}\,\bar w_{ij}|w_{ij}|^{-2/3}\right]
  \prod_c\prod_{i<j\in S_c(a)}\bar w_{ij}|w_{ij}|^{-2/3},
  \label{eq:pfree}
\end{align}
for spinless semions, SU(2) semions with spins $s_i$, and SU(3) anyons with flavors $b_i$ at $\theta=\pi/3$. Here
$\sigma_{ij}=\sum_{A=1}^3\epsilon_{Aa_ia_j}\epsilon_{Ab_ib_j}$ locks the colors of each Cooper pair to its flavors,
$J_{ij}$ is the long-range Jastrow factor of Eq.~\eqref{eq:Jlr}, and $\sgn(a)$ is the sign of the permutation that
groups the particles by color. The first product carries the anyon statistics, the Pfaffian pairs the holons of each
coloring, and the last product, over same-color pairs, is the contact-corrected integer quantum Hall factor of
$\chi$. Below we abbreviate the pair function as $u_{ij}=\bar w_{ij}|w_{ij}|^{-2\alpha}$ and the $\chi$ factor as
$U_a=\prod_c\prod_{i<j\in S_c(a)}u_{ij}$, as in the main text. 

\paragraph{Holon flux and shift sector.}
In the CB gauge, the auxiliary wavefunction of Eq.~\eqref{eq:colstate} sees $2Q_\Phi=\qbg-\alpha(N-1)$ flux
quanta (Sec.~\ref{sec:sphere}). For each coloring the auxiliary wavefunction is a product of the wavefunction for 
$\chi_a$ and the holon $\psi_{ab}$, so their fluxes add, $2Q_\chi+2Q_\psi=2Q_\Phi$. Each color of $\chi$ fills the
reversed lowest level with $N/N_c$ particles. This requires $|2Q_\chi|+1=N/N_c$ with $2Q_\chi<0$, so
$2Q_\chi=1-N/N_c$. With $N_c=1/\alpha$,
\begin{equation}
2Q_\psi=2Q_\Phi-2Q_\chi=\qbg-1+\alpha .
\label{eq:Qpsi}
\end{equation}
The integer quantum Hall factor of $\chi$  cancels the extensive part $-\alpha N$ of the statistical flux. Because the holon orbitals are
single valued, $2Q_\psi$ must be an integer. The color states therefore exist only for $\qbg\in\mathbb Z+1-\alpha$, that is, for
half-integer $\qbg$ for semions and for $\qbg\in\mathbb Z+\frac23$ at $\theta=\pi/3$.

The pairing channel then fixes the $\qbg$ sector. We use the standard convention of Sec.~\ref{sec:eft-colour}: the
pair orbital $w_{ij}^{\,\ell_\psi}$, like $z_{ij}^{\,\ell_\psi}$ on the plane, has relative angular momentum
$\ell_\psi$, so an antiholomorphic pair orbital has $\ell_\psi<0$. Since the pair orbital must transform like a holon
orbital in each argument, $2Q_\psi=\ell_\psi$ and $\qbg=1-\alpha+\ell_\psi$. The exchange parity required by the color and flavor tensors selects $\ell_\psi$. For spinless semions, the color
singlet pairing is in the $d-id$ channel, which requires $\qbg=-3/2$. For SU(2) semions,
the color-singlet, spin-singlet pairing is in the $p-ip$ channel, which requires $\qbg=-1/2$. Both semion states
require an even $N$. For SU(3) anyons with $\theta=\pi/3$, $b_i$ is the physical flavor, and the symmetric tensor $\sigma_{ij}$ enforces a color-antisymmetric, flavor-antisymmetric pairing in the $p-ip$ channel, $\ell_\psi=-1$, which requires $\qbg=-1/3$. Here $N$ is a multiple of six so that the three colors have equal populations and all holons can
pair. The gauge-invariant condensate combines three holon pairs and carries the charge of six anyons.

\subsection{Fixing the radial power at contact}
In local planar coordinates $z_{ij}=r\ee^{\ii\varphi}$, the regular two-anyon solutions with relative angular
momentum $\ell=\alpha$ and $\ell=\alpha-1$ are $z_{ij}^{\alpha}$ and $\bar z_{ij}^{\,1-\alpha}=r^{1-\alpha}\ee^{\ii(\alpha-1)\varphi}$.
Exchange symmetry restricts two anyons of the same flavor to $\ell\in\alpha+2\mathbb Z$, so the $\ell=\alpha-1$ channel is
allowed only for different flavors. In one coloring of Eqs.~\eqref{eq:pfree-spinless}--\eqref{eq:pfree}, a pair can
receive an antiholomorphic factor from the Pfaffian, if the two anyons are Cooper partners, which requires different
colors, or from $U_a$, if they have the same color. Otherwise it enters only through the common factor
$z_{ij}^{\alpha}$. The radial power $|z_{ij}|^{-2\alpha}$ in $u_{ij}$ makes each antiholomorphic factor a regular
two-anyon solution. Contact regularity fixes this power only as $r\to0$, and our variational choice continues it as a
pure power at all separations.

\emph{Spinless semions}, Eq.~\eqref{eq:pfree-spinless}. All anyons have the same flavor. A Cooper pair carries
$\bar w_{ij}u_{ij}$ from the Pfaffian and enters as $z_{ij}^{\alpha}\bar z_{ij}u_{ij}=\bar z_{ij}^{\,2-\alpha}$, the
regular solution with $\ell=\alpha-2$. The factor $u_{ij}$ of a same-color pair in $U_a$ does not produce the $\ell=\alpha-1$
channel, as shown below, so every pair approaches as a combination of $z_{ij}^{1/2}$ and $\bar z_{ij}^{\,3/2}$.

\emph{SU(2) semions}, Eq.~\eqref{eq:pfree-su2}. The Pfaffian pairs only opposite spins, and a Cooper pair enters as
$z_{ij}^{1/2}u_{ij}=\bar z_{ij}^{\,1/2}$. A same-color pair of opposite spins carries the same factor from $U_a$. Two
opposite spins therefore approach as a combination of $z_{ij}^{1/2}$ and $\bar z_{ij}^{\,1/2}$, whereas two equal spins
approach as $z_{ij}^{1/2}$.

\emph{SU(3) anyons}, Eq.~\eqref{eq:pfree}. Since $\sigma_{ij}=0$ for $b_i=b_j$, the Pfaffian pairs only different
flavors, and a Cooper pair enters as $z_{ij}^{1/3}u_{ij}=\bar z_{ij}^{\,2/3}$. A same-color pair of different flavors
carries the same factor from $U_a$. Two anyons of different flavors therefore approach as a combination of
$z_{ij}^{1/3}$ and $\bar z_{ij}^{\,2/3}$, and two anyons of the same flavor approach as $z_{ij}^{1/3}$.

\emph{Same-flavor contact.} For one coloring, let $M$ be the antisymmetric $N\times N$ matrix inside the Pfaffian,
with rows and columns labeled by the anyons, for example $M_{kl}=\sigma_{kl}u_{kl}$ in Eq.~\eqref{eq:pfree}. Its
Pfaffian $\Pf M=\sum_\pi\sgn(\pi)\prod_{(k,l)\in\pi}M_{kl}$ runs over all pairings $\pi$ of the $N$ anyons into Cooper
pairs, and $(\Pf M)^2=\det M$. Two anyons $i$ and $j$ of the same flavor can carry $\bar z_{ij}^{\,1-\alpha}$ only
through $u_{ij}$ in $U_a$, when they also share a color: in Eqs.~\eqref{eq:pfree-su2} and \eqref{eq:pfree} anyons of
the same flavor are never Cooper partners, and a Cooper pair of spinless semions gives $\bar z_{ij}^{\,2-\alpha}$. With
identical color and flavor labels, $M_{ij}=0$, and the color and flavor tensors in rows $i$ and $j$ coincide, for
example $\sigma_{ik}=\sigma_{jk}$ for every $k$. The two rows then differ only through $u_{ik}$ and $u_{jk}$, which
become equal as $z_i\to z_j$. Exchanging the positions of $i$ and $j$ therefore amounts to exchanging rows $i,j$ and
columns $i,j$ of $M$, which reverses the sign of the Pfaffian. The Pfaffian is thus odd in the relative coordinate and
vanishes linearly, as a combination of $z_{ij}$ and $\bar z_{ij}$. Multiplied by $z_{ij}^{\alpha}u_{ij}=\bar z_{ij}^{\,1-\alpha}$,
it gives $\bar z_{ij}^{\,1-\alpha}z_{ij}$ and $\bar z_{ij}^{\,2-\alpha}$, which lie in the allowed channels $\ell=\alpha$
and $\ell=\alpha-2$ and vanish at least as $r^{2-\alpha}$. The leading same-flavor behavior is therefore
$z_{ij}^{\alpha}$, from the colorings in which the two anyons have different colors.

\subsection{Flavor symmetry and projection}\label{sec:flavor-proj}
The above wavefunction for the SU(3) anyons with $\theta=\pi/3$ actually weakly breaks the SU(3) flavor symmetry. A flavor rotation $g\in{\rm SU(3)}$ acts on Eq.~\eqref{eq:pfree} only through $\sigma_{ij}$, which it replaces by
$\sigma_{ij}(W)=\sum_{A,B}\epsilon_{Aa_ia_j}W_{AB}\epsilon_{Bb_ib_j}$ with $W=g^{-1}$. The same-color factor in $U_a$
depends on the individual color labels, so Eq.~\eqref{eq:pfree} is not exactly invariant. The exact flavor singlet is
the Haar average over the locking matrix,
\begin{equation}
\Psi_{\rm SU(3)}^{(0)}=\int_{\rm SU(3)}dW\;\Psi_{\rm SU(3)}\Big|_{\sigma_{ij}\to\sigma_{ij}(W)}.
\label{eq:pfree-proj}
\end{equation}
If $U_a$ were a color singlet, as the filled lowest levels $\prod_c\prod_{i<j\in S_c(a)}\bar w_{ij}$ without the radial
power are, this average would return Eq.~\eqref{eq:pfree} unchanged. The integrand is a polynomial of degree $N/2$ in
$W$. We use Eq.~\eqref{eq:pfree} without the projection, because the non-singlet admixture it would remove is small:
it vanishes at $N=6$ and is about 0.15\% at $N=12$, according to the Casimir test below.

A simpler construction restores the symmetry exactly by switching color indices instead of averaging over $W$.
Writing the radial part of $U_a$ as $\prod_{i<j}[1+(f_{ij}-1)\delta_{a_ia_j}]$ with $f_{ij}=|w_{ij}|^{-2\alpha}$, the
breaking comes from the color-diagonal $\delta_{a_ia_j}$, which is not invariant under color rotations. We replace it
by the color-symmetric projector $(1+P^c_{ij})/2$, where $P^c_{ij}$ switches the color indices of anyons $i$ and $j$.
This gives the pair factor
\begin{equation}
F_{ij}=\frac{1+f_{ij}}{2}+\frac{f_{ij}-1}{2}\,P^c_{ij},
\label{eq:Fexchange}
\end{equation}
which equals $f_{ij}$ for a same-color pair, where $P^c_{ij}=1$, and commutes with color rotations, so the state is
an exact flavor singlet. Because the $\chi$ partons are fermions, $P^c_{kl}$ acting on the $\chi$ factor equals minus the
exchange of the coordinates of $k$ and $l$ in it. Applying $1+\sum_{k<l}(F_{kl}-1)$ to the plain $\chi$ factor
$\bar V_a=\prod_c\prod_{i<j\in S_c(a)}\bar w_{ij}$ gives the flavor-symmetric wavefunction
\begin{equation}
\Psi_{\rm SU(3)}^{\rm sym}=\prod_{i<j}w_{ij}^{1/3}J_{ij}\sum_a\sgn(a)\,
\Pf\!\left[\sigma_{ij}\,\bar w_{ij}|w_{ij}|^{-2/3}\right]
\Big[\bar V_a+\sum_{k<l}\frac{f_{kl}-1}{2}\big(\bar V_a-\bar V_a\big|_{k\leftrightarrow l}\big)\Big],
\label{eq:pfree-sym}
\end{equation}
where $\bar V_a|_{k\leftrightarrow l}$ is $\bar V_a$ with the positions of anyons $k$ and $l$ exchanged. For a
same-color pair $\bar V_a|_{k\leftrightarrow l}=-\bar V_a$, so the bracket multiplies $\bar V_a$ by $f_{kl}$, as $U_a$
does. We keep the radial factor within the first neighbor shell, $f_{kl}=(d_{kl}/1.5a_0)^{-2\alpha s_{kl}}$ with
$s_{kl}=[1-(d_{kl}/1.5a_0)^2]^2$ for $d_{kl}<1.5a_0$ and $f_{kl}=1$ otherwise. We also use the additive form instead of
the product of the $F_{kl}$, which keeps the number of terms linear in the number of close pairs. Since $\bar V_a$
contributes $|w_{ij}|$ for each distant same-color pair, $J_{ij}$ uses $\gamma=1/3+(N-2)/[3(N-1)]$.
Equation~\eqref{eq:pfree-sym} is the flavor-covariant state whose energies are quoted in
Sec.~\ref{sec:res-su3}. A plain average over permutations of the color labels would not help, since these
permutations act on $\sigma_{ij}$ as a finite subgroup of the $W$ in Eq.~\eqref{eq:pfree-proj} and Eq.~\eqref{eq:pfree}
is already invariant under them, so the average would return it unchanged, non-singlet admixture included. 

This problem does not arise for SU(2) semions. In
Eq.~\eqref{eq:pfree-su2} the spin tensor $\epsilon_{s_is_j}$ shares no index with the color tensor, so every term of the
Pfaffian is a product of spin singlets whatever the color factor, and the state is an exact spin singlet. In the SU(3)
state, by contrast, $\sigma_{ij}$ ties the flavor index of each Cooper pair to a color index, so flavor invariance
relies on a color-singlet color factor.

\paragraph{Casimir test.}
We measure the flavor symmetry with the quadratic Casimir of the total flavor rotation,
$C_2=\sum_iC_2^{(i)}+2\sum_{i<j}\mathbf T_i\cdot\mathbf T_j$, where $\mathbf T_i$ are the SU(3) generators acting on the
flavor of anyon $i$. In the fundamental representation $C_2^{(i)}=4/3$ and $2\mathbf T_i\cdot\mathbf T_j=P_{ij}-1/3$,
where $P_{ij}$ exchanges the flavors of anyons $i$ and $j$. With $N/3$ anyons of each flavor, $P_{ij}=1$ for the
$N(N-3)/6$ pairs of equal flavor, so $C_2=N+\sum_{i<j,\,b_i\neq b_j}P_{ij}$. In the CB gauge $\Phi$ is symmetric under
exchanging both the positions and the flavors of two anyons, so $P_{ij}\Phi(R)=\Phi(R^{(ij)})$, where $R^{(ij)}$ is the
configuration $R$ with the positions of $i$ and $j$ exchanged. The statistical and Jastrow factors do not depend on the
flavors, so
\begin{equation}
D(R)\equiv\frac{[C_2\Psi](R)}{\Psi(R)}=N+\sum_{i<j,\,b_i\neq b_j}\frac{\Phi(R^{(ij)})}{\Phi(R)} .
\label{eq:casimir}
\end{equation}
A flavor singlet has $C_2=0$ and therefore $D(R)=0$ at every configuration $R$. In general, we decompose $\Psi$ into
components $\Psi_\rho$ in the irreducible representations $\rho$ of SU(3), with weights $p_\rho$. Averages over
$|\Psi|^2$ then give $\langle D\rangle=\sum_\rho C_2(\rho)\,p_\rho$ and $\langle|D|^2\rangle=\sum_\rho C_2(\rho)^2p_\rho$.
Equation~\eqref{eq:pfree} is invariant under every permutation of the three flavors, because the same permutation of
the colors in the sum over colorings compensates it, and we confirm this numerically. Its components are zero-weight
states, with $N/3$ anyons of each flavor, and the zero-weight states of the octet and of the decuplets contain no state
invariant under these permutations. These representations therefore cannot appear, and every admixture has
$C_2\ge8$, the value of the 27-dimensional representation. Both $\langle D\rangle/8$ and $\langle|D|^2\rangle/64$ thus bound
the total non-singlet weight from above, and both equal it for a pure 27-plet admixture. Equation~\eqref{eq:pfree}
gives $|D|<10^{-14}$ at $N=6$. At $N=12$, 768 configurations give $\langle D\rangle/8=0.16(7)\%$ and
$\langle|D|^2\rangle/64=0.15(2)\%$, consistent with a 27-plet admixture of about 0.15\%.

\subsection{Fixing the long-distance correlations}
The pair amplitudes also contribute powers of distance to the modulus of the wavefunction. For SU(2) semions
$|u_{ij}|=1$, so the only such power is the common $|w_{ij}|^{1/2}$, and $|\Psi|^2$ contains
\[
\prod_{i<j}|w_{ij}|\propto\exp\Big(\sum_{i<j}\ln d_{ij}\Big).
\]
This is the Boltzmann weight of a classical two-dimensional plasma in which all particles repel logarithmically. For
$|\Psi|^2\propto\exp[-\sum_{i<j}v(d_{ij})]$, the random-phase approximation of this classical fluid, which gives the
exact small-$q$ behavior for a long-range $v$, yields
\begin{equation}
S(q)=\frac{1}{1+n\tilde v(q)},
\label{eq:Srpa}
\end{equation}
where $\tilde v(q)$ is the two-dimensional Fourier transform of $v$. The logarithm $v=-\ln d$ has $\tilde v=2\pi/q^2$,
so $S(q)=q^2/(q^2+2\pi n)\simeq q^2a_0^2/2\pi$ with $a_0=n^{-1/2}$. The plasma screens perfectly, and its density is
incompressible. A superfluid with a sound mode instead has $S(q)\simeq s\,q$ at small $q$, which by
Eq.~\eqref{eq:Srpa} requires $\tilde v\propto1/q$, that is, a $1/d$ tail of $v$ \cite{Reatto1967}. We remove the
mean long-distance power and supply this tail with a flavor-independent Jastrow factor
\begin{equation}
J_{ij}=\left(1+\frac{d_{ij}^2}{a_0^2}\right)^{-\gamma/2}
\exp\!\left[-\frac{\lambda a_0}{\sqrt{d_{ij}^2+a_0^2}}\right].
\label{eq:Jlr}
\end{equation}
Both factors are smooth at contact. The first contributes $d_{ij}^{-\gamma}$ at long distance and cancels the mean
power. The second adds $v=2\lambda a_0/d$, with $\tilde v=4\pi\lambda a_0/q$, so that Eq.~\eqref{eq:Srpa} gives
$S(q)\simeq qa_0/(4\pi\lambda)$ at small $q$.

We fix $\lambda$ by matching this slope to the phonon of the superfluid. The ideal anyon gas has no length scale other
than $a_0$, so its energy per particle in the thermodynamic limit has the form $E/N=e_\infty\pi\hbar^2n/m$. The energy
density $e_\infty\pi\hbar^2n^2/m$ gives the chemical potential $\mu=2\pi e_\infty\hbar^2n/m$, and the compressibility
gives the sound speed through $mc^2=n\,\partial\mu/\partial n=2\pi e_\infty\hbar^2n/m$. At small $q$ the phonon
exhausts the $f$-sum rule $\int d\omega\,\hbar\omega\,S(q,\omega)=\hbar^2q^2/2m$, which gives the Feynman relation
$S(q)=\hbar q/(2mc)$. Hence $s=\hbar/(2mc)=a_0/(2\sqrt{2\pi e_\infty})$, and equating it to $a_0/(4\pi\lambda)$ gives
\[
\lambda=\sqrt{e_\infty/2\pi}.
\]
We take $e_\infty$ from independently optimized states of the same phase, so that $\lambda$ is fixed before the trial
state is evaluated: $e_\infty=0.784$ and $0.541$ from the CF $\nu^*=2$ states of spinless and SU(2) semions
(Table~\ref{tab:main}), and $e_\infty=0.492$ for SU(3) anyons from a preliminary extrapolation of the analytical state
with a trained Jastrow factor and backflow, close to the $0.498(1)$ of the optimized color superconductor.

For SU(2) semions the common power gives $\gamma=1/2$. For spinless semions the extra
$\bar w_{ij}$ in each Cooper pair adds one power per particle. Distributed over its $N-1$ partners
this gives $\gamma=1/2+1/(N-1)$. For SU(3), each particle has $N/3-1$ same-color partners and
one Cooper partner, each contributing $|u_{ij}|=|w_{ij}|^{1/3}$. Including the common $1/3$
gives $\gamma=1/3+N/[9(N-1)]$. The values of $\lambda$ used for spinless semions,
SU(2) semions and SU(3) anyons are $0.353$, $0.293$ and $0.280$, respectively.

The energies without $J_{ij}$, which need no input, lie only 0.5--2\% higher: $e_\infty=0.550(3)$ instead of
$0.5405(16)$ for SU(2) semions (Sec.~\ref{sec:res-semion}), $e_\infty=0.509(1)$ from the four sizes in
Sec.~\ref{sec:res-su3} instead of $0.498(3)$ for SU(3) anyons, and $E/N=0.8007(30)$ instead of $0.7963(15)$ at
$N=24$ for spinless semions. Fixing $\lambda$ from the state without $J_{ij}$ would give $\lambda=0.296$ and $0.285$
for SU(2) semions and SU(3) anyons, within 2\% of the values above. The construction can therefore be made
self-contained, with $e_\infty$ estimated from the ansatz without $J_{ij}$; we have not re-evaluated the energies
with these slightly different values of $\lambda$.

%=======================================================================================================================
\section{Low-energy effective theories}\label{sec:eft}

\subsection{Flux attachment as a Chern--Simons theory}\label{sec:eft-cs}
Writing the anyon wavefunction as $\Psi=\prod_{i<j}(z_{ij}/|z_{ij}|)^{\alpha_s}\Phi$ isolates the statistical factor $\ee^{\ii\alpha_s\sum_{i<j}\theta_{ij}}$, where $\theta_{ij}=\arg z_{ij}$. Acting on this factor with the kinetic operator gives
\begin{equation}
\mathcal H\Psi=\ee^{\ii\alpha_s\sum_{i<j}\theta_{ij}}\sum_i\frac{\hbar^2}{2m}\bigl[-\ii\nabla_i-\mathbf A(\mathbf r_i)
-\mathbf a(\mathbf r_i)\bigr]^2\Phi,\qquad \mathbf a(\mathbf r_i)=-\alpha_s\sum_{j\neq i}\frac{\hat z\times\mathbf r_{ij}}
{r_{ij}^2},
\label{eq:hstat}
\end{equation}
so $\Phi$ describes particles that carry $-\alpha_s$ flux quanta of $\mathbf a$,
$\nabla\times\mathbf a=-2\pi\alpha_s\sum_j\delta^2(\mathbf r-\mathbf r_j)$. In field theory, we introduce $a_\mu$ as a U(1) gauge field with a self Chern-Simons (CS) term,
\begin{equation}
\mathcal L=\mathcal L_\Phi[a]+\frac{1}{4\pi\alpha_s}\,a\,da ,
\label{eq:cs}
\end{equation}
The equation of motion by varying $a_0$ gives the internal flux $b=-2\pi\alpha_s\rho$. Coupling every flavor to the same field $a$ implements the flavor-independent statistics. At mean field, $\bar b=-2\pi\alpha_s n$. The signed filling $\nu^*=2\pi n/\bar b=-1/\alpha_s$ therefore gives $-1/\alpha$ for CBs and $1/(1-\alpha)$ for CFs. Each variational family below is a mean-field state of Eq.~\eqref{eq:cs}, or of its parton representation. We use the notation $X\,dY=\epsilon^{\mu\nu\lambda}X_\mu\partial_\nu Y_\lambda$.

\paragraph{Parent FCI.} If the anyons are the quasiparticles of a parent FCI with topological order
U(1)$_m$, $m=1/\alpha$, the CB form of Eq.~\eqref{eq:cs} is its Chern--Simons theory, and the
electromagnetic field $A_c$ couples to the flux of $a$,
\begin{equation}
\mathcal L=\mathcal L_B[a]+\frac{m}{4\pi}a\,da+\frac{1}{2\pi}A_c\,da .
\label{eq:csAc}
\end{equation}
Integrating out $a$ gives $\sigma_{xy}=-\frac1m\,e^2/h$, and the Gauss law gives the anyon charge $q=-1/m$ (in units
of $e$, as all charges below).

The CF gauge follows from the exact transmutation of a boson into a fermion bound to one flux quantum,
$\mathcal L_B[X]=\mathcal L_F[X+\gamma]-\frac{1}{4\pi}\gamma\,d\gamma$: the CFs couple to $u=a+\gamma$ and see
$\alpha_s=\alpha-1$. For semions ($m=2$) the unimodular change of variables to $u$ and $w=2a+\gamma$ gives
$\frac{2}{4\pi}a\,da-\frac1{4\pi}\gamma\,d\gamma=-\frac{2}{4\pi}u\,du+\frac1{4\pi}w\,dw$, so $w$ decouples from the
matter, and integrating it out leaves the CF form of Eq.~\eqref{eq:cs} with the parent FCI ($u$ is the CF-gauge $a$ of
the main text),
\begin{equation}
\mathcal L=\mathcal L_F[u]-\frac{2}{4\pi}u\,du-\frac{1}{2\pi}A_c\,du-\frac{1}{4\pi}A_c\,dA_c .
\label{eq:csF}
\end{equation}
At $\theta=\pi/3$ the CF level $1/\alpha_s=-3/2$ is not an integer, and we keep $a$ and $\gamma$ separate
(Sec.~\ref{sec:eft-trion}).

\subsection{Semions: superfluid of pairs}\label{sec:eft-sf}
At $\nu^*=2$, the CFs fill two Landau levels under the effective magnetic field $\bar b$. Integrating out these gapped fermions produces their Hall response together with a Maxwell term set by the CF cyclotron gap, which combines with Eq.~\eqref{eq:csF} to give
\begin{equation}
\begin{split}
\mathcal L&=\frac{2}{4\pi}u\,du-\frac{2}{4\pi}u\,du-\frac{1}{2\pi}A_c\,du-\frac{1}{4\pi}A_c\,dA_c
+\frac{\epsilon}{2}\mathbf e^2-\frac{1}{2\mu}b^2\\
&=-\frac{1}{2\pi}A_c\,du-\frac{1}{4\pi}A_c\,dA_c+\frac{\epsilon}{2}\mathbf e^2-\frac{1}{2\mu}b^2 .
\end{split}
\label{eq:sfS}
\end{equation}
The exact cancellation of the Chern--Simons term leaves a pure Maxwell field in 2+1 dimensions, which is dual to a compact scalar $\sigma$ through $\frac{1}{2\pi}\epsilon^{\mu\nu\lambda}\partial_\nu u_\lambda\propto\partial^\mu\sigma$. The coupling $-\frac{1}{2\pi}A_c\,du$ identifies $\sigma$ as the phase of a superfluid \cite{Chen1989,Fetter1989}, whose condensed boson consists of a pair of semions. For SU(2) semions, the spin sector also has a quantized Hall response. A probe $A_s$ coupled to $S_z=\pm\frac12$ gives $\frac1{4\pi}\sum_\pm\big(u\pm\frac12A_s\big)d\big(u\pm\frac12A_s\big)=\frac{2}{4\pi}u\,du+\frac{1}{8\pi}A_s\,dA_s$, which reproduces Eq.~\eqref{eq:sfS} in the charge sector and leaves the spin Hall term $\frac{1}{8\pi}A_s\,dA_s$ for the gapped spin sector.

\paragraph{The pairs in the wavefunction.} On the plane, with $\ell^*=(\pi n)^{-1/2}$ and
$G=\prod_i\ee^{-|z_i|^2/4\ell^{*2}}$, the semion states are $\Phi_F=\chi_2G$ with
$\chi_2=\det[z_i^k\,|\,\bar z_iz_i^k]$, $k=0,\ldots,N/2-1$ (spinless), and $\Phi_F=\Delta(z)\Delta(w)G$ [SU(2), with $z$
($w$) the up (down) coordinates and $\Delta(z)=\prod_{i<j}(z_i-z_j)$]. We follow one pair whose members sit at $R$ and
$R+\varepsilon$, and denote the other $N-2$ coordinates by $Z'$.

\emph{SU(2) semions.} With the up particle $z_1$ and the down particle $w_1$ both at $R$, each Vandermonde factorizes,
$\Delta(z)=\pm\prod_k(z_k-R)\,\Delta(z')$ and $\Delta(w)=\pm\prod_k(w_k-R)\,\Delta(w')$, with the products over the
remaining up and down particles, and the two Gaussians combine into $\ee^{-|R|^2/2\ell^{*2}}$. Writing $z_k$ for the
coordinate of a remaining particle of either spin,
\begin{equation}
\Phi_F(R,R,Z')=\pm\ee^{-|R|^2/2\ell^{*2}}\prod_{k\in Z'}(z_k-R)\,\Phi_F^{(N-2)}(Z').
\label{eq:insert}
\end{equation}
An up--down pair has no Pauli zero, so this amplitude is the $\varepsilon\to0$ limit.

\emph{Spinless semions.} Two members of the same species vanish at contact, so we expand to first order in
$\varepsilon$. Subtracting the row of the member at $R$ from that of the member at $R+\varepsilon$ leaves
$(k\varepsilon R^{k-1}\,|\,\bar\varepsilon R^k+k\varepsilon\bar RR^{k-1})$, so
$\chi_2(R,R+\varepsilon,Z')=\bar\varepsilon A+\varepsilon B+O(|\varepsilon|^2)$. In $A$ the first two rows are
$(R^k\,|\,\bar RR^k)$ and $(0\,|\,R^k)$, and subtracting $\bar R$ times the second from the first turns them into
$(R^k\,|\,0)$ and $(0\,|\,R^k)$. If another particle sits at $z_k=R$, its row $(R^k\,|\,\bar z_kR^k)$ is a combination
of these two, so $A$ vanishes linearly in $z_k-R$ and contains $\prod_{k\in Z'}(z_k-R)$. The remaining factor is
antisymmetric in $Z'$ and lies in the two lowest LLs with one orbital fewer each, which the $N-2$ fermions fill in a
unique way, and a Laplace expansion along the first two rows fixes the prefactor:
$A=\pm\prod_{k\in Z'}(z_k-R)\,\chi_2^{(N-2)}(Z')$ exactly, the analog of Eq.~\eqref{eq:insert}. The pair therefore
condenses in the $\bar\varepsilon$ channel.

\emph{Cancellation of the vortex.} In the anyon wavefunction the two members carry the statistical factor
$[(z_k-R)/|z_k-R|]^{\alpha_s}$ with every other particle, which gives
$\prod_k[(z_k-R)/|z_k-R|]^{2\alpha_s}=\prod_k(z_k-R)^{-1}|z_k-R|$ at $\alpha_s=-1/2$. This removes the vortex
$\prod_k(z_k-R)$ and leaves
\begin{equation}
\Psi(\text{pair at }R;Z')=\phi_{\rm pair}(\varepsilon)\,W_R(Z')\,\Psi^{(N-2)}(Z'),\qquad
W_R=\ee^{-|R|^2/2\ell^{*2}}\prod_k|z_k-R|>0 ,
\end{equation}
with $\phi_{\rm pair}=(\varepsilon/|\varepsilon|)^{\alpha_s}$ for SU(2) and $(\varepsilon/|\varepsilon|)^{\alpha_s}\bar\varepsilon$
for spinless semions. The amplitude for the pair at $R$ is positive and nodeless, as in Feynman's argument for bosons.
For a uniform density, two-dimensional electrostatics gives
$\langle\sum_k\ln|z_k-R|\rangle=n\int d^2r\,\ln|\mathbf r-R|=\frac{\pi n}{2}|R|^2+{\rm const}$, which cancels the Gaussian
$\ee^{-|R|^2/2\ell^{*2}}=\ee^{-\pi n|R|^2/2}$. 

\emph{BCS form.} For SU(2) semions, Cauchy's identity
\[
\det\Big[\frac{1}{z_i-w_j}\Big]=\pm\frac{\Delta(z)\Delta(w)}{\prod_{i,j}(z_i-w_j)}
\]
gives $\Phi_F=\pm\prod_{i,j}(z_i-w_j)\,\det[1/(z_i-w_j)]\,G$. The determinant
$\det[g(z_i-w_j)]=\sum_P\sgn(P)\prod_ig(z_i-w_{P(i)})$ with $g(z)=1/z$ is the component with $N/2$ pairs of the BCS state
$\exp[\int d^2r\,d^2r'\,g(\mathbf r-\mathbf r')\,c^\dagger_\uparrow(\mathbf r)c^\dagger_\downarrow(\mathbf r')]|0\rangle$.
The remaining factor, $\prod_{i,j}(z_i-w_j)\,G$ together with the statistical phases, vanishes only at contact and acts
as a Jastrow factor. The SU(2) state is therefore a number-projected BCS state of up--down semion pairs with the CF pair
function $1/(z-w)$, times a Jastrow factor \cite{Read2000}.

\paragraph{Pair coherence.} The pair density matrix $\rho_2(R,R')\propto\langle W_RW_{R'}\rangle_{N-2}$ can be evaluated without a sign problem. In the bare state, the CF density is incompressible, with $S(q)=q^2\ell^{*2}/2$, so that $\ln W_R=\int\rho\ln|\mathbf r-R|$ has logarithmic correlations and $C_2=\rho_2(R,R')/\rho_2(R,R)\propto(\ell^*/d)^{\nu^*/2}$ exhibits only algebraic pair order. True off-diagonal long-range order requires $S(q)\propto q$, which in a Jastrow description can be obtained by combining $\prod_{i<j}|w_{ij}|^{-1/2}$, which removes the logarithmic kernel, with a Reatto--Chester tail \cite{Reatto1967}. In our wavefunctions, the backflow and Jastrow factor modify these correlations so that the pair correlator approaches a plateau, indicating long-range pair order (Sec.~\ref{sec:res-semion}).

\subsection{CF states at \texorpdfstring{$\nu^*=3/2$}{nu*=3/2}: trion Fermi liquid}\label{sec:eft-trion}
At $\alpha=1/3$ the CFs couple to $u=a+\gamma$ (Sec.~\ref{sec:eft-cs}) and see $\bar b=\frac{4\pi}{3}n$. The lowest
level is filled (density $2n/3$) and the second holds $n/3$, half filling. Integrating out the filled level gives
$\frac{1}{4\pi}u\,du$. The second-level CFs are in the situation of electrons at $\nu=1/2$ \cite{Halperin1993}:
attaching two more flux quanta through a field $\tilde a$, in the integer form
$\frac{1}{2\pi}\tilde a\,dg-\frac{2}{4\pi}g\,dg$ (which gives $\frac{1}{8\pi}\tilde a\,d\tilde a$ when $g$ is integrated
out), makes the final fermions $f$ see zero average field,
\begin{equation}
\mathcal L=\mathcal L_f[u+\tilde a]+\frac{1}{4\pi}u\,du+\frac{3}{4\pi}a\,da-\frac{1}{4\pi}\gamma\,d\gamma
+\frac{1}{2\pi}A_c\,da+\frac{1}{2\pi}\tilde a\,dg-\frac{2}{4\pi}g\,dg .
\label{eq:trion1}
\end{equation}
The gauge fields $x=(a,\gamma,\tilde a,g)$ enter through $\frac{1}{4\pi}x^TK\,dx$, with the integer matrix $K=\bigl(\begin{smallmatrix}4&1\\1&0\end{smallmatrix}\bigr)\oplus\bigl(\begin{smallmatrix}0&1\\1&-2\end{smallmatrix}\bigr)$. Since $\det K=1$, all of these fields are gapped. The fermion $f$ couples through $l=(1,1,1,0)$, for which $l^TK^{-1}l=0$, and therefore retains no statistical interaction. The electromagnetic probe enters through $t_c=(1,0,0,0)$, giving $f$ a charge $-t_c^TK^{-1}l=-1$. Moreover, $t_c^TK^{-1}t_c=0$, so integrating out $x$ leaves no Hall response and yields
\begin{equation}
\mathcal L=\mathcal L_f[-A_c] ,
\label{eq:trion2}
\end{equation}
Therefore the trion Fermi liquid has no Hall conductance, since it cancels the response of the parent FCI in the same way as the Hall crystal (Sec.~\ref{sec:eft-cb}). The fermion $f$ carries the physical charge $-1$ of three anyons and couples to no dynamical gauge field, identifying it as a trion with finite overlap with the physical electron.

\subsection{Color states}\label{sec:eft-colour}
\paragraph{Partons.} In the CB gauge we write the boson of flavor $b$ as $\phi_b=\sum_{a=1}^{N_c}\chi_a\psi_{ab}$. The
decomposition is invariant under $\chi\to U\chi$, $\psi\to U^*\psi$ with $U\in$ U($N_c$), so color is a gauge
label and the parton gauge field $\beta$ is a U($N_c$) field: $\chi$ couples to $a+\beta$ and $\psi$ to $-\beta^T$.
Since U($N_c$)$=[\mathrm{SU}(N_c)\times\mathrm U(1)]/\mathbb Z_{N_c}$, $\beta$ cannot be split into independent
SU($N_c$) and U(1) fields with integer fluxes; its trace $\mathrm{Tr}\,\beta$ is an ordinary U(1) field, and we write its U(1) part as
$\beta_c=\frac{1}{N_c}\mathrm{Tr}\,\beta$.
With the flux attachment of Eq.~\eqref{eq:csAc}, the parton theory reads
\begin{equation}
\mathcal L=\mathcal L_\chi[a+\beta]+\mathcal L_\psi[-\beta^T]+\frac{N_c}{4\pi}a\,da+\frac{1}{2\pi}A_c\,da ,
\label{eq:Lparton}
\end{equation}
where $\mathcal L_\chi$ and $\mathcal L_\psi$ are the nonrelativistic Lagrangians of the partons minimally coupled to
the indicated fields, and $\mathcal L_\psi$ also contains the holon interactions.
Each color of $\chi$ has density $\alpha n$ in the field $\bar b=-2\pi\alpha n$, giving a signed filling $\nu_\chi=-1$.
Each color thus fills one reversed LL with Chern number $-1$. Integrating out the $\chi$ fermions gives the U($N_c$)$_{-1}$ term
\begin{equation}
-\frac{1}{4\pi}\mathrm{Tr}\Big[(a+\beta)d(a+\beta)-\tfrac{2\ii}{3}(a+\beta)^3\Big]
=-\frac{N_c}{4\pi}a\,da-\frac{N_c}{2\pi}a\,d\beta_c-\frac1{4\pi}\mathrm{Tr}\Big[\beta\,d\beta-\tfrac{2\ii}{3}\beta^3\Big].
\end{equation}
Since $m=1/\alpha=N_c$, the $a\,da$ terms cancel exactly, and the full theory becomes
\begin{equation}
\mathcal L=\mathrm{tr}\,\psi^\dagger\Big(\ii D_t+\frac{\mathbf D^2}{2m_\psi}\Big)\psi+\mathcal L_{\rm int}[\psi]
-\frac{1}{4\pi}\mathrm{Tr}\Big(\beta\,d\beta-\frac{2\ii}{3}\beta^3\Big)+\frac{1}{2\pi}a\,d(A_c-N_c\beta_c),
\label{eq:LNc}
\end{equation}
with $D_\mu\psi=(\partial_\mu+\ii\beta^T_\mu)\psi$, which is Eq.~\eqref{eq:Lholon} for $N_c=3$. The field $a$ enters only through the last term, so it acts as a Lagrange multiplier imposing $\beta_c=A_c/N_c$. Locally, we can write $\beta=\hat\beta+\beta_c\openone=\hat\beta+\frac{1}{N_c}A_c\openone$ with traceless $\hat\beta$. The holons are then fermions of charge $-1/N_c$ in zero net field, while the U($N_c$) term separates into the FCI response $-\frac{1}{4\pi N_c}A_c\,dA_c$ and an SU($N_c$)$_{-1}$ gauge theory for $\hat\beta$. Globally, each flux quantum of $A_c$ is accompanied by a $\mathbb Z_{N_c}$ flux of $\hat\beta$, ensuring consistency with the fractional holon charge. The auxiliary fermions $\chi_a$ thus encode the U($N_c$)$_{-1}$ term in a first-quantized anyon wavefunction. For $N_c=3$, Eq.~\eqref{eq:Lholon} reproduces the parton theory of the doped FCI in Ref.~\cite{Zhang2026b} with the opposite chirality.

\paragraph{Projection.} The physical wavefunction is obtained as the boson amplitude in the parton state, $\Phi_B=\langle0|\prod_i\phi_{b_i}(\mathbf r_i)|\Phi_\chi\rangle|\Phi_\psi\rangle$, with the expansion of each $\phi_b$ generating one term per coloring $a=(a_1,\dots,a_N)$. Because each color level contains exactly $k=N/N_c$ fermions, only colorings satisfying $|S_c(a)|=k$ contribute. Their $\chi$ factor is $\sgn(a)\prod_c\Vb_{S_c(a)}$, where $\sgn(a)$ is the sign of the permutation that sorts the particles by color; this sign is essential for Bose symmetry and gives Eq.~\eqref{eq:colour}. In the CB gauge each anyon experiences $-\alpha(N-1)$ statistical flux quanta and has $k-1$ antivortices in $\prod_c\Vb_{S_c}$, one on every other anyon of the same color, leaving zero effective field for the holons in the thermodynamic limit. Since every antivortex is bound to an anyon, these states contain no free vortices and describe liquids.

\paragraph{Descendants of the holon metal.} The phases derived from Eq.~\eqref{eq:Lholon} are discussed in Ref.~\cite{Zhang2026b}, and Table~\ref{tab:descendants} summarizes those considered here. In the parent SU(3) holon metal, each of the nine species $\psi_{ab}$ fills a Fermi sea of density $n/9$. Color-flavor-locked pairing produces the color superconductor, $\Phi_\psi=\Pf[\sigma_{ij}g(\xi_i,\xi_j)]$ with $\sigma_{ij}=\sum_A\epsilon_{Aa_ia_j}\epsilon_{Ab_ib_j}$, by locking color rotations to flavor rotations, SU(3)$_c\times$SU(3)$_f\to$SU(3)$_{\rm diag}$ \cite{Alford1999,Alford2008}. This completely Higgses the gauge field, while the gauge-invariant order parameter, a color singlet of three pairs, binds six anyons with charge $6q=-2$. 

\paragraph{Low-energy theory of the color superconductor.} The color-flavor-locked condensate Higgses the SU(3)
part of $\beta$, so at low energies $\beta=\beta_c\openone$ with integer fluxes of $\beta_c$; a fractional flux of
$\beta_c$ requires a $\mathbb Z_3$ flux of the SU(3) part and binds a vortex of the condensate. A holon pair has charge
$-2$ under $\beta_c$, so the phase $\theta$ of the condensate enters through
$\frac{\rho_s}{2}(\partial_\mu\theta+2\beta_{c,\mu})^2$. We dualize it with a U(1) field $b$ whose flux measures the
pair current, $\frac{1}{2\pi}\epsilon^{\mu\nu\lambda}\partial_\nu b_\lambda=\rho_s(\partial^\mu\theta+2\beta_c^\mu)$,
which replaces the condensate by $-\frac{2}{2\pi}b\,d\beta_c$ and a Maxwell term for $b$. Setting
$\beta=\beta_c\openone$ in Eq.~\eqref{eq:Lholon} and adding this term gives
\begin{equation}
\mathcal L_{\rm SC}=-\frac{3}{4\pi}\beta_c\,d\beta_c-\frac{3}{2\pi}a\,d\beta_c-\frac{2}{2\pi}b\,d\beta_c
+\frac{1}{2\pi}A_c\,da+\cdots ,
\label{eq:Lsc}
\end{equation}
where the ellipsis denotes Maxwell terms. The self Chern--Simons term of $\beta_c$ is the Hall response of the three
filled $\chi$ levels, which couple to $a+\beta$. With $\beta=\beta_c\openone$ they give
$-\frac{3}{4\pi}(a+\beta_c)d(a+\beta_c)$, whose $a\,da$ part cancels the flux attachment and leaves the first two
terms of Eq.~\eqref{eq:Lsc}. The gauge fields $x=(a,\beta_c,b)$ enter through
$\frac{1}{4\pi}x^TK\,dx$ with
$K=\bigl(\begin{smallmatrix}0&-3&0\\-3&-3&-2\\0&-2&0\end{smallmatrix}\bigr)$, which has rank two. Its null vector
$x_0=(2,0,-3)$ carries only a Maxwell term, so this combination is a gapless photon, dual to the phase of a
superfluid. The probe enters through $t_c=(1,0,0)$, and $t_c\cdot x_0=2$ is the charge of the condensed boson, three
holon pairs or six anyons.

To find the remaining quasiparticles, note that an excitation carries an integer charge vector $l\in\mathbb Z^3$, one
integer for each of the three U(1) fields in $x$, since each field has $2\pi$-quantized fluxes. Its charge under the
gapless photon is $l\cdot x_0$. An excitation with $l\cdot x_0\neq0$ is a vortex of the superfluid, whose energy grows
logarithmically with the system size, so finite-energy excitations have $l\cdot x_0=0$. Among them are the charges
that the Chern--Simons terms bind to $2\pi n$ of flux of the internal fields $x$ (not of the probe $A_c$), $l=Kn$ for an
integer vector $n$. For example, a $2\pi$ flux of $b$, which adds one pair, binds a holon pair, $l=(0,-2,0)$, and a
$2\pi$ flux of $a$ binds three holons, $l=(0,-3,0)$, with the charge of an electron of the parent FCI. Since $Kx_0=0$,
these objects have $l\cdot x_0=0$ and are not vortices. They are created by monopole events of the compact fields $x$,
and their Aharonov--Bohm phase with any excitation $l'$ is $2\pi n\cdot l'$, so they are local. The distinct anyon types
are therefore the finite-energy charges modulo $K\mathbb Z^3$. To count them we use the Smith normal form: any integer
matrix can be written as $K=UDV$, with $U$ and $V$ integer matrices of determinant $\pm1$ and
$D=\mathrm{diag}(d_1,d_2,d_3)$, so that $\mathbb Z^3/K\mathbb Z^3\cong\bigoplus_i\mathbb Z/d_i\mathbb Z$. A zero entry
gives a free factor $\mathbb Z$, measured by $l\cdot x_0$, which counts vortices. The anyons are the finite factors: an
entry $d_i>1$ would give $\mathbb Z_{d_i}$ anyons and a factor $d_i$ in the torus degeneracy. Here $D=\mathrm{diag}(1,1,0)$, so the gapped sector has no anyons.

On the other hand, adjoint Higgs fields that lock the color and flavor indices instead reduce the gauge group to U(1)$^2$ or Z$_3$: the former gives the U(1)$^2$ metal and its inter-flavor paired superconductor, while the latter gives the Type I triplet $Z_3$ metal and the singlet Z$_3$ metals.

\paragraph{The U(1)$^2$ metal and its paired descendant.}
A particle--hole Higgs field can select the three diagonal holons
$f_b=\psi_{bb}$ and raise the energies of the six off-diagonal species.
The remaining gauge transformations are diagonal color rotations with
unit determinant. There are two independent phases, giving U(1)$^2$.
Writing the surviving traceless gauge field as
$\hat\beta=\mathrm{diag}(b_1,b_2,b_3)$, Eq.~\eqref{eq:Lholon} reduces locally to
\begin{equation}
\mathcal L_{\mathrm{U}(1)^2}
=\sum_{b=1}^3\mathcal L_{f_b}[-b_b-A_c/3]
-\frac{1}{4\pi}\sum_{b=1}^3 b_b\,db_b
-\frac{1}{12\pi}A_c\,dA_c,
\qquad \sum_b b_b=0 .
\label{eq:u1metal}
\end{equation}
We have three Fermi seas of holons with charge
$-e/3$, each of density $n/3$, couple to two continuous gauge fields
\cite{Zhang2026a,Fan2026,Zhang2026b}.
The diagonal selection chooses a flavor basis and intrinsically breaks
the SU(3) flavor symmetry down to $S_3$.

The mean-field holon wavefunction is a product of three flavor
determinants. Only $a_i=b_i$ contributes to the color projection, giving
Eq.~\eqref{eq:cflcb}, or its CF form in Eq.~\eqref{eq:cflcf}.
The U(1)$^2$ superconductor replaces these determinants by an inter-flavor
$p$-wave Pfaffian. 

\paragraph{The singlet and Type I triplet Z$_3$ metals.}
Higgs fields that lock all color generators to flavor generators leave
only the center of SU(3) as a gauge symmetry. Unlike the U(1)$^2$ metal,
these states preserve a global SU(3), realized by a flavor rotation
combined with a compensating color rotation \cite{Zhang2026b}.
To specify the two locks in our conventions, let $U$ act on color and
$V$ on the physical flavor, so that $\psi\to U^*\psi V^T$.
The ordinary lock $U=V$ selects
$\psi_0=\mathrm{tr}\,\psi/\sqrt3$ and gaps the orthogonal octet.
This flavor singlet fills one Fermi sea of density $n$.
Since its amplitude is proportional to $\delta_{ab}$, the projected
ansatz again has only $a_i=b_i$. It contains three filled lowest-LL
factors and a single holon determinant over all $N$ particles, as in
Sec.~\ref{sec:ans}.

The conjugate lock $U=V^*$ instead selects the antisymmetric subspace,
$\eta_A=\sum_{ab}\epsilon_{Aab}\psi_{ab}/\sqrt2$, and gaps the symmetric
sextet. Filling one Fermi sea of density $n/3$ for each $A$ gives the
Type I triplet Z$_3$ metal.
For a fixed coloring, its holon Slater matrix has entries
$\epsilon_{Aa_i b_i}V_m(i)/\sqrt2$, with $N/3$ occupied orbitals $m$
per pocket $A$.
An anyon of flavor $b$ can therefore carry either color $a\ne b$.
The projection retains a sum over colorings, rather than a single
product. With $k=N/3$ particles of each flavor, the number of allowed
balanced colorings is $\sum_{x=0}^k\binom{k}{x}^3$.
A fixed permutation of color and flavor only relabels the colors and
cannot reproduce this antisymmetric projection.

In both Z$_3$ descendants, the continuous color gauge fluctuations are
massive, while a discrete gauge field remains. The low-energy matter
consists of one or three Fermi pockets of holons with charge $-e/3$ and nontrivial
Z$_3$ gauge charge. The discrete gauge sector inherits its topological
term from SU(3)$_{-1}$ in Eq.~\eqref{eq:Lholon}.
These are fractionalized metals: a gapless holon is not a local
electron, and the Fermi surface need not give an electron quasiparticle
pole.

\paragraph{The Type II triplet Z$_3$ metal.}
The construction of Ref.~\cite{Senthil2026} instead writes
$\phi_b=d\,\Psi_b$, with a flavor-singlet fermion $d$ and a
flavor-triplet fermion $\Psi_b$.
The parton $d$ fills three reversed LLs, with total Chern number $-3$,
while each $\Psi_b$ forms a Fermi sea of density $n/3$ in zero mean field.
The corresponding CB wavefunction is
\begin{equation}
\Phi_B^{\rm II}=\bar\chi_3\prod_{b=1}^3\det_{i\in b}[V_m(i)] .
\label{eq:z3typeii-wf}
\end{equation}
Here $\bar\chi_3$ is one determinant of three filled reversed LLs over
all particles. There is no color sum.
Let $\gamma$ be the parton U(1) gauge field, with $d$ coupled to
$a+\gamma$ and $\Psi_b$ to $-\gamma$.
Integrating out $d$ adds $-3(a+\gamma)d(a+\gamma)/(4\pi)$ to
Eq.~\eqref{eq:csAc}. The $a\,da$ terms cancel, leaving
\begin{equation}
\mathcal L_{\rm II}
=\sum_b\mathcal L_{\Psi_b}[-\gamma]
-\frac{3}{4\pi}\gamma\,d\gamma
+\frac{1}{2\pi}a\,d(A_c-3\gamma) .
\label{eq:z3typeii}
\end{equation}
The compact field $a$ imposes $3\gamma=A_c$ locally, so the fermions
have charge $-e/3$. Globally, the factor of three leaves a Z$_3$ gauge
field; the $\gamma\,d\gamma$ term specifies its twist.
We omit gravitational response terms here.

Both triplet constructions therefore have three fractional-charge
Fermi pockets and discrete gauge structure, but their parton content
and flavor action differ. In Type I, the three filled lowest LLs carry
color, and conjugate locking acts as $\psi\to V\psi V^T$, giving
$\eta\to V^*\eta$.
In Type II, the filled parton is a flavor singlet and the gapless
fermions transform as $\Psi\to V\Psi$.
Taking the physical anyon to be an SU(3) fundamental, the three Fermi pockets therefore form an antifundamental
multiplet in Type I and a fundamental multiplet in Type II.

\paragraph{Sphere flux and closed shells.}\label{par:colour-shells}
On the sphere, a metallic mean-field state is rotation invariant only if every Fermi pocket fills complete
angular-momentum shells. This fixes which sizes $N$ and sectors $\qbg$ we can use for each metal. By
Eq.~\eqref{eq:Qpsi}, the holons of the color construction see the flux $2Q_\psi=\qbg-2/3$, and their shells have
angular momentum $l=|Q_\psi|,|Q_\psi|+1,\ldots$ with $2l+1$ orbitals each. At $\qbg=-1/3$ or $5/3$, where
$|2Q_\psi|=1$, the shells hold $2,4,6,\ldots$ holons, so a pocket is closed with $2,6,12,\ldots$ holons. At
$\qbg=2/3$, where $2Q_\psi=0$, the shells hold $1,3,5,\ldots$ holons, and a pocket is closed with $1,4,9,\ldots$ holons. The
total $N$ is this occupation times the number of pockets: nine for the SU(3) holon metal, three for the U(1)$^2$ and
Type I triplet Z$_3$ metals, and one for the singlet Z$_3$ metal. At $\qbg=-1/3$, for example, the holon metal is
closed at $N=18$ and the three-pocket metals at $N=6$ and 18.

The Type II triplet Z$_3$ metal sits in a different sector because its filled parton differs. The parton $d$ fills three
reversed LLs, which requires the flux $2Q_d=3-N/3$. Subtracting this from the CB flux $2Q_B=\qbg-(N-1)/3$ of
Sec.~\ref{sec:sphere} leaves $2Q_\Psi=\qbg-8/3$ for the fermions $\Psi_b$. Zero-flux Fermi seas of $\Psi_b$ therefore
require $\qbg=8/3$, where the three pockets are closed at $N=12,27,\ldots$ (three filled LLs of $d$ need $N\ge9$). This
shift comes from the different filled-LL factor and does not by itself distinguish the metallic phases. The paired
color states need no closed shells. A $J=0$ pair in the $p\mp ip$ channel, however, requires $2Q_\psi=\mp1$, that is,
$\qbg=-1/3$ or $5/3$.

\begin{table}[h]
\caption{The table summarizes the phases of Eq.~\eqref{eq:Lholon} that descend from the SU(3) holon metal. The Type II triplet Z$_3$ metal uses the separate construction in Eqs.~\eqref{eq:z3typeii-wf} and \eqref{eq:z3typeii}.}
\label{tab:descendants}
\centering\footnotesize
\setlength{\tabcolsep}{3pt}
\begin{tabular}{lllll}
\toprule
descendant & condensate & gauge group & gapless holons & phase \\
\midrule
parent & none & SU(3) & nine Fermi surfaces & SU(3) holon metal (fractionalized) \\
color-flavor-locked pairing & $\langle\psi_{ab}\psi_{a'b'}\rangle\propto\epsilon_{Aaa'}\epsilon_{Abb'}$ & none & none &
color superconductor (chiral, charge $2e$) \\
color $=$ flavor & adjoint Higgs & U(1)$^2$ & $\psi_{bb}$, three Fermi seas & U(1)$^2$ metal \\
\quad + pairing & inter-flavor $p$-wave & none & none & U(1)$^2$ superconductor (chiral) \\
conjugate locking & adjoint Higgs & Z$_3$ & $\epsilon_{Aab}\psi_{ab}$, three pockets & Type I triplet Z$_3$ metal \\
ordinary locking & adjoint Higgs & Z$_3$ & $\mathrm{tr}\,\psi$, one pocket & singlet Z$_3$ metal \\
\bottomrule
\end{tabular}
\end{table}

\paragraph{Chirality and chiral central charge.} We adopt the orientation appropriate to tMoTe$_2$ (Sec.~\ref{sec:eft-cb}), for which the parent FCI in Eq.~\eqref{eq:csAc} has $\sigma_{xy}=-\frac13\,e^2/h$, as does the doped FCI component of the $\nu=2/3$ state. Equation~\eqref{eq:Lholon} therefore has the opposite chirality to the theory in Ref.~\cite{Zhang2026b}, whose FCI has $\sigma_{xy}=+\frac13\,e^2/h$.

We label a paired state by the winding $\ell$ of
$\langle\psi_{\mathbf k}\psi_{-\mathbf k}\rangle\propto\ee^{\ii\ell\vartheta_{\mathbf k}}$, the eigenvalue of
$L_z=-\ii\partial_\vartheta$. A weakly paired pocket has BdG Chern number $-\ell$ and contributes $c_-=-\ell/2$
\cite{Zhang2026b}, in the orientation in which a filled LL in a positive field contributes $+1$. A $J=0$ pair in holon
flux $2Q_\psi$ has $\ell=2Q_\psi$, so the color superconductor is $p-ip$ ($\ell=-1$) at $\qbg=-1/3$ and $p+ip$ at
$\qbg=5/3$. The Chern--Simons term selects $\ell=-1$ in our orientation \cite{Zhang2026b}, the sector in which the color
superconductor lies furthest below the crystal (Sec.~\ref{sec:res-su3}). With SU(3) gauge field Higgsed completely, the three
$\chi$ levels contribute $3\times(-1)$ and the nine holon pockets $9\times\frac12$: the doped FCI contributes
$c_-=\frac32$. Together with the filled $C=1$ band of the $\nu=2/3$ state, we have  $c_-=\frac52$ in total. Three anyons make one carrier of the
parent FCI, $c\sim\psi^\dagger\psi^\dagger\psi^\dagger$ up to the filled $\chi$ levels, so the charge-$2e$ pairs have
angular momentum $L=-3\ell=3$: they are in $f+if$ channel\cite{Zhang2026b}. A BCS superconductor of the same carriers with the
same $L=3$ has one pocket with $c_-=-\frac32$, and together with the $\nu=1$ background, we have $c_-=1-\frac32=-\frac12$. The two superconductors have the same pairing
symmetry, but their $c_-$ differ by 3.

\paragraph{Semions with $N_c=2$: the color superconductor.} For $N_c=2$ the same construction, with $\beta_c=A_c/2$, leaves color
doublets of charge $-1/2$ coupled to SU(2)$_{-1}$, at holon flux $2Q_\psi=\qbg-\frac12$. A $J=0$ pair in flux $2Q_\psi$
has relative angular momentum $2Q_\psi$ in the planar coordinate $z$ (holomorphic for $2Q_\psi>0$), so the closed-shell
sectors $\qbg=-3/2$ (spinless) and $-1/2$ [SU(2)] give $d-id$ and $p-ip$ pairs ($\ell=-2$ and $-1$ in the notation above). Color-singlet pairing,
$\Phi_\psi=\Pf[\epsilon_{a_ia_j}g_{ij}]$ (spinless, $d-id$) or $\Pf[\epsilon_{a_ia_j}\epsilon_{f_if_j}g_{ij}]$
[SU(2), spin-singlet $p-ip$], gives a condensate of charge $-1$, a pair of semions, which leads to the same paired
superfluid phase as in the CF $\nu^*=2$ description. The final phase has chiral central charge $c_-=0$: the two $\chi$
levels contribute $2\times(-1)$, the paired holon pockets $2\times1$ (spinless, $\ell=-2$) or $4\times\frac12$
[SU(2), $\ell=-1$], and the color SU(2) gauge field, whose level $-1$ CS term is canceled by the pairing of  the holons, confines and adds nothing. In the context of doping a bosonic Laughlin state (or a chiral spin liquid), the final phase is an ordinary superfluid of the physical bosons, without topological order. For SU(2) semions the spin sector stays chiral, with the
spin Hall response of Sec.~\ref{sec:eft-sf}, and the charge sector compensates to reach $c_-=0$.

\subsection{Hyperdeterminant structure and color channels}\label{sec:eft-hyperdet}
The color projection gives a direct connection between the SU(3) holon
metal and hyperdeterminant wavefunctions \cite{Hu2024,Lin2026,Hu2026b}.
We first consider the unpaired mean-field state and omit the common
statistical phase, contact factor and Jastrow factor.
The parton decomposition is
\begin{equation}
\phi_b(\mathbf r)=\sum_{a=1}^3
\chi_a(\mathbf r)\psi_{ab}(\mathbf r),
\label{eq:hyperdet-partons}
\end{equation}
where $\chi$ and $\psi$ are fermions, $a$ is the gauge color and $b$
is the physical flavor.
For $N$ composite bosons, each fermionic parton sector contains $N$
particles. Let $\chi_{\mu a}(\mathbf r)$ and
$u_{\nu,ab}(\mathbf r)$ be its occupied one-particle orbitals, with
$\mu,\nu=1,\ldots,N$.
These labels run over the chosen occupied subspaces, including their
color and flavor indices.
For the SU(3) holon metal, a basis with definite color and flavor has
$\mu=(c,m)$ and $\nu=(d,s,\ell)$.
Here $c,d$ label color, $s$ labels flavor, and $m,\ell$ label the spatial
orbitals of $\chi$ and $\psi$, respectively.
The local field components $a_i,b_i$ are matched to these occupied-state
labels through
\begin{equation}
\chi_{(c,m),a_i}(\mathbf r_i)
=\delta_{c a_i}\varphi_m(\mathbf r_i),
\qquad
u_{(d,s,\ell),a_i b_i}(\mathbf r_i)
=\delta_{d a_i}\delta_{s b_i}v_\ell(\mathbf r_i) .
\label{eq:hyperdet-component-indices}
\end{equation}
Thus the occupied $\chi$ and holon orbitals must have the same color
$c=d=a_i$, and the holon flavor must equal the physical flavor $s=b_i$.
The color $a_i$ is summed over, while $b_i$ is fixed by the sampled
configuration $x_i=(\mathbf r_i,b_i)$.
Projection gives, up to an overall normalization and a
configuration-independent sign,
\begin{equation}
\Phi_B^{(0)}(x_1,\ldots,x_N)
=\sum_{a_1,\ldots,a_N}
\det_{i\mu}[\chi_{\mu a_i}(\mathbf r_i)]\,
\det_{i\nu}[u_{\nu,a_i b_i}(\mathbf r_i)] .
\label{eq:hyperdet-projection}
\end{equation}
The color-permutation sign in Eq.~\eqref{eq:colour} is already present
in the full $\chi$ determinant here.
Only colorings with $N/3$ particles of each color contribute.
For these colorings, order the columns by color and sort the rows
into the sets $S_c(a)=\{i:a_i=c\}$. This gives
\begin{equation}
\det_{i\mu}[\chi_{\mu a_i}(\mathbf r_i)]
=\operatorname{sgn}(a)\prod_{c=1}^3
\det_{i\in S_c(a),m}[\varphi_m(\mathbf r_i)] .
\label{eq:hyperdet-color-sign}
\end{equation}
Rows within each set retain their original particle order.
This is the explicit sign used when writing the $\chi$ wavefunction
as a product of color determinants in Eq.~\eqref{eq:colour}.
In Eq.~\eqref{eq:hyperdet-projection}, it remains inside the unsplit
determinant, so inserting another $\operatorname{sgn}(a)$ would count
it twice. If both parton determinants are split into color blocks,
their two row-sorting signs cancel.

Expand the two determinants as sums over permutations $\sigma,\tau$
of their occupied columns. For fixed $\sigma,\tau$, the color $a_i$
appears only in the local product
$\chi_{\sigma(i),a_i}(\mathbf r_i)u_{\tau(i),a_i b_i}(\mathbf r_i)$.
The sum over all $a_1,\ldots,a_N$ therefore factorizes into a product
of $N$ local color sums, one for each particle.
Defining the fusion tensor
\begin{equation}
T_{i;\mu\nu}
=\sum_{a=1}^3\chi_{\mu a}(\mathbf r_i)
u_{\nu,a b_i}(\mathbf r_i),
\label{eq:hyperdet-tensor}
\end{equation}
we obtain
\begin{equation}
\Phi_B^{(0)}=\operatorname{Hdet}_B(T)
\equiv\sum_{\sigma,\tau\in S_N}
\operatorname{sgn}(\sigma)\operatorname{sgn}(\tau)
\prod_{i=1}^N T_{i;\sigma(i)\tau(i)} .
\label{eq:hyperdet-boson}
\end{equation}
This is the bosonic combinatorial hyperdeterminant: the two occupied-orbital
indices are antisymmetrized, while exchanging two physical particles
produces two canceling signs. The tensor has size $N\times N\times N$ for
each sampled configuration. Its entries are built from parton orbitals.

\paragraph{One channel versus three channels.}
For a fixed physical coordinate and flavor $x_i$, regard the fusion
tensor as an $N\times N$ matrix. Its rows label occupied $\chi$ orbitals
and its columns label occupied holon orbitals:
\begin{equation}
T_i\equiv
\begin{pmatrix}
T_{i;11}&T_{i;12}&\cdots&T_{i;1N}\\
T_{i;21}&T_{i;22}&\cdots&T_{i;2N}\\
\vdots&\vdots&\ddots&\vdots\\
T_{i;N1}&T_{i;N2}&\cdots&T_{i;NN}
\end{pmatrix}
=\sum_{a=1}^{3}\boldsymbol\chi_a(\mathbf r_i)
\bigl[\mathbf u_{a b_i}(\mathbf r_i)\bigr]^T .
\label{eq:hyperdet-channels}
\end{equation}
The $N$-component column vectors use the same orbital amplitudes as
Eq.~\eqref{eq:hyperdet-tensor}:
$[\boldsymbol\chi_a(\mathbf r_i)]_\mu=\chi_{\mu a}(\mathbf r_i)$ and
$[\mathbf u_{a b_i}(\mathbf r_i)]_\nu=u_{\nu,a b_i}(\mathbf r_i)$.
A \emph{channel} is one outer product of these vectors, labeled here
by color $a$. The minimum number of outer products required is the
matrix rank of $T_i$; we use $R$ for its generic value.
Equivalently, in orthonormal bases of the two occupied subspaces,
$T_i$ defines a normalized two-parton amplitude
\begin{equation}
|T_i\rangle
=\frac{1}{\|T_i\|_F}\sum_{\mu,\nu=1}^N
T_{i;\mu\nu}\,|\mu\rangle_\chi\otimes|\nu\rangle_\psi,
\qquad
\|T_i\|_F^2=\sum_{\mu,\nu}|T_{i;\mu\nu}|^2 .
\label{eq:hyperdet-entanglement}
\end{equation}
Its Schmidt rank is $R$. Thus $R=1$ gives independent, separable
$\chi$ and holon amplitudes, whereas $R>1$ entangles their orbital
indices. This entanglement resides in the local fusion map used to
project the two mean-field Slater determinants.
A conventional product construction has $R=1$, namely
$T_{i;\mu\nu}=T^\chi_{i;\mu}T^\psi_{i;\nu}$.
Equation~\eqref{eq:hyperdet-boson} then factorizes exactly into
$\det[T^\chi_{i;\mu}]\det[T^\psi_{i;\nu}]$.
This includes both the singlet Z$_3$ metal, with effective decomposition
$\phi_b=\chi_b f$, and the Type II triplet Z$_3$ metal,
$\phi_b=d\,\Psi_b$ of Eq.~\eqref{eq:z3typeii-wf}.

In Eq.~\eqref{eq:hyperdet-tensor}, the channel is instead the contracted
color $a$. The SU(3) holon metal has $R=3$: its three color contributions
are independent, as the example below makes explicit.
Color is summed over and is not assigned to a physical particle in
advance. The occupied orbitals include complete color triplets, and the
determinants enforce the occupation of every selected orbital once.
Thus the projection sums all compatible color assignments.

\begin{figure}[ht]
\centering
\includegraphics[width=\textwidth]{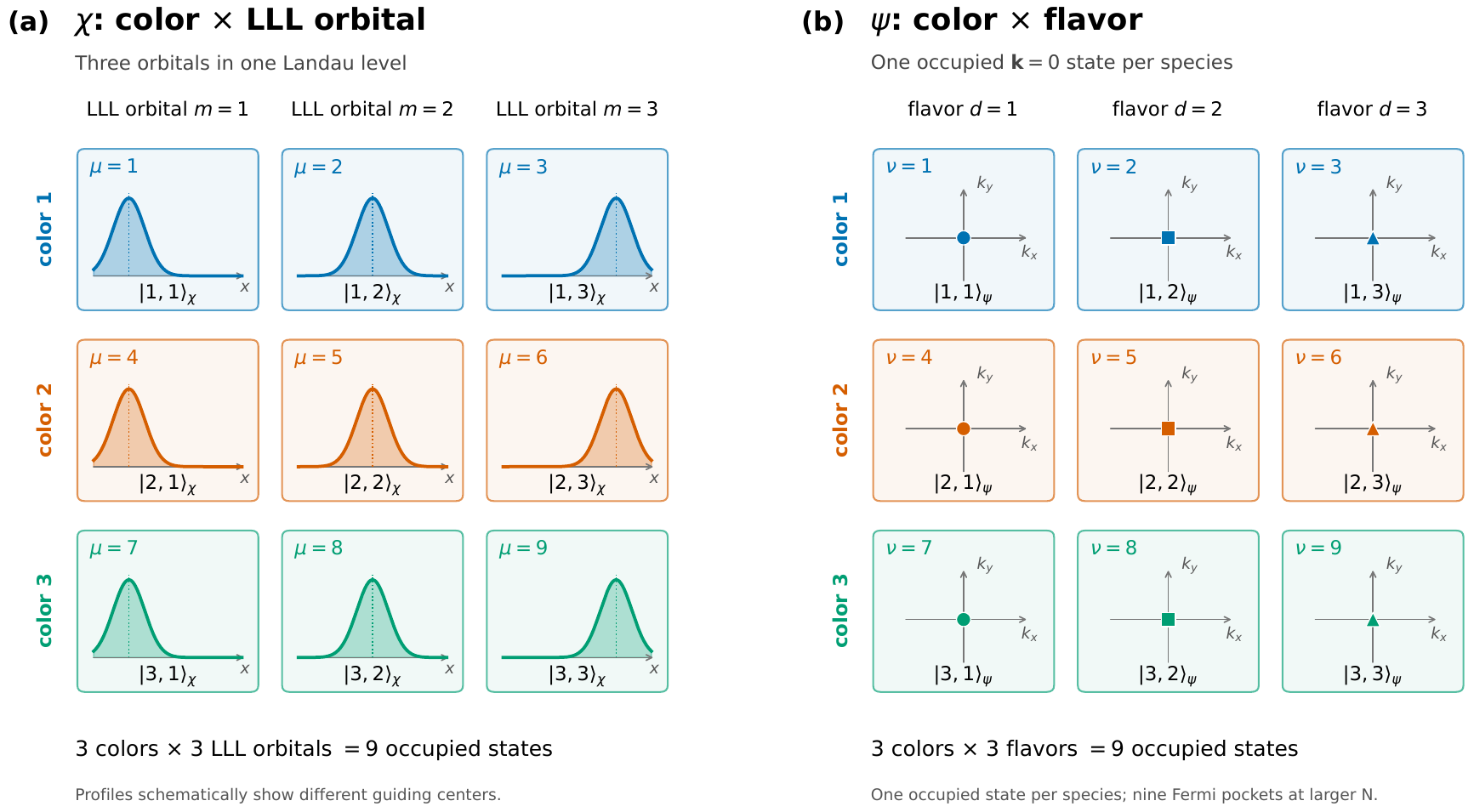}
\caption{The nine occupied orbitals in each parton sector at
$N=9$ and $N_\phi=3$ are organized into $3\times3$ arrays.
Rows and hues label color in both panels.
(a) The LLL with three Landau orbitals can host 9 particles for the fermion $\chi_a$ due to the color index.
(b) Just the $\mathbf k=0$ state can host 9 particles for the fermion $\psi_{ab}$ due to the color and flavor index.
The labels $\mu,\nu=1,\ldots,9$ specify the row and column ordering of
the fusion matrix in Eq.~\eqref{eq:hyperdet-nine-matrix}.
}
\label{fig:hyperdet-basis}
\end{figure}

\paragraph{A simple illustration with nine particles.}
Consider $N=9$ anyons, with three particles of each physical flavor. Let the reversed mean field contain $N_\phi=3$ flux quanta for $\chi_a$,
so its lowest LL has three orbitals $\varphi_m$, $m=1,2,3$.
In the zero-field holon $\psi_{ab}$ sector, we only consider the $\mathbf k=0$ orbital
$v_0(\mathbf r)=\mathcal S^{-1/2}$, where $\mathcal S$ is the system area.
The SU(3) color state occupies all nine $\chi$ orbitals $(a,m)$ and all nine
holon orbitals $(c,d,\mathbf k=0)$, where $a,c$ are colors and $d$ is
flavor. To display the bases, write
$|a,m\rangle_\chi=|a\rangle_{\rm color}\otimes|\varphi_m\rangle$ and
$|c,d\rangle_\psi=|c\rangle_{\rm color}\otimes|d\rangle_{\rm flavor}\otimes|\mathbf k=0\rangle$.
The two occupied bases are
\begin{equation}
\mathcal B_\chi=
\begin{pmatrix}
|1,1\rangle_\chi&|1,2\rangle_\chi&|1,3\rangle_\chi\\
|2,1\rangle_\chi&|2,2\rangle_\chi&|2,3\rangle_\chi\\
|3,1\rangle_\chi&|3,2\rangle_\chi&|3,3\rangle_\chi
\end{pmatrix},
\qquad
\mathcal B_\psi=
\begin{pmatrix}
|1,1\rangle_\psi&|1,2\rangle_\psi&|1,3\rangle_\psi\\
|2,1\rangle_\psi&|2,2\rangle_\psi&|2,3\rangle_\psi\\
|3,1\rangle_\psi&|3,2\rangle_\psi&|3,3\rangle_\psi
\end{pmatrix} .
\label{eq:hyperdet-nine-bases}
\end{equation}
Each row has a fixed color. The columns of $\mathcal B_\chi$ distinguish
LLL orbitals, while those of $\mathcal B_\psi$ distinguish flavors.
Every entry in both arrays is occupied once, as illustrated in
Fig.~\ref{fig:hyperdet-basis}.
We fix the ordering by reading each array row by row:
$\mu=3(a-1)+m$ and $\nu=3(c-1)+d$.
The corresponding orbital amplitudes are
$\chi_{(a,m),a'}(\mathbf r)=\delta_{aa'}\varphi_m(\mathbf r)$ and
$u_{(c,d),a'b}(\mathbf r)=\delta_{ca'}\delta_{db}v_0(\mathbf r)$.
Thus the $9\times9\times9$ fusion tensor is
\begin{equation}
T_{i;(a,m)(c,d)}
=\delta_{ac}\,\varphi_m(\mathbf r_i)
\delta_{d b_i}\,v_0(\mathbf r_i),
\qquad a,c,m,d=1,2,3 .
\label{eq:hyperdet-nine}
\end{equation}

For a fixed particle $i$ with physical flavor $b_i=1$, the matrix in
these bases is explicitly
\begin{equation}
T_i=v_0(\mathbf r_i)
\left(
\begin{array}{ccc|ccc|ccc}
\varphi_1&0&0&0&0&0&0&0&0\\
\varphi_2&0&0&0&0&0&0&0&0\\
\varphi_3&0&0&0&0&0&0&0&0\\ \hline
0&0&0&\varphi_1&0&0&0&0&0\\
0&0&0&\varphi_2&0&0&0&0&0\\
0&0&0&\varphi_3&0&0&0&0&0\\ \hline
0&0&0&0&0&0&\varphi_1&0&0\\
0&0&0&0&0&0&\varphi_2&0&0\\
0&0&0&0&0&0&\varphi_3&0&0
\end{array}
\right),
\qquad \varphi_m\equiv\varphi_m(\mathbf r_i) .
\label{eq:hyperdet-nine-matrix}
\end{equation}
The separators mark the three color sectors.
Only columns $\nu=1,4,7$ are nonzero, since these are the flavor-1
holon orbitals of colors 1, 2 and 3.
All particles use the same ordered lists of nine occupied $\chi$
orbitals and nine occupied $\psi$ orbitals. The matrix $T_i$ evaluates
their amplitudes at particle $i$'s position and physical flavor;
changing $i$ does not change the occupied orbital basis.
The full hyperdeterminant uses every
occupied column once across the nine particles.
Each diagonal $3\times3$ color block has rank one.
The three blocks have disjoint row and column supports, giving
$\operatorname{rank}T_i=3$.
The normalized amplitude in Eq.~\eqref{eq:hyperdet-entanglement}
therefore has three equal Schmidt coefficients $1/\sqrt3$.
The $\chi$ and holon colors are entangled: their colors must match,
and all three matches enter coherently. Neither parton has a fixed
color of its own in this fusion amplitude.
All nine $\chi$ fermions fit in the LLL, with three per color, while
all nine holons fit at $\mathbf k=0$, with one per color and flavor.
Projection combines one
fermion from each parton sector into each composite boson.

\paragraph{The two Z$_3$ fusion matrices.}
For comparison, we write the two product constructions in their own
nine-dimensional occupied bases, again at fixed $b_i=1$.
To obtain nonzero nine-particle states, these bases must include
additional momenta or LLs.
The holons occupy normalized plane waves
$\mathcal S^{-1/2}\ee^{\ii\mathbf k_j\cdot\mathbf r}$,
where the momenta $\mathbf k_j$ are distinct and $\mathbf k_1=0$.
All orbital amplitudes in the following matrices are evaluated at
$\mathbf r_i$.

For the singlet Z$_3$ metal, $\phi_b=\chi_b f$, the $\chi$ basis consists
of $|b,m\rangle_\chi$ with $b,m=1,2,3$, ordered by
$\mu=3(b-1)+m$. Here $b$ is the physical flavor selected by the
color-flavor lock. The flavor-singlet $f$ instead occupies nine
distinct momenta, with column basis
$(|\mathbf k_1\rangle_f,\ldots,|\mathbf k_9\rangle_f)$.
The matrix elements are
$T^{\rm s}_{i;(b,m)j}=\mathcal S^{-1/2}\delta_{b b_i}\varphi_m\ee^{\ii\mathbf k_j\cdot\mathbf r_i}$.
Thus
\begingroup
\setlength{\arraycolsep}{2pt}
\begin{equation}
T_i^{\rm s}=\frac{1}{\sqrt{\mathcal S}}
\left(
\begin{array}{ccc|ccc|ccc}
\varphi_1\ee^{\ii\mathbf k_1\cdot\mathbf r_i}&\varphi_1\ee^{\ii\mathbf k_2\cdot\mathbf r_i}&\varphi_1\ee^{\ii\mathbf k_3\cdot\mathbf r_i}&\varphi_1\ee^{\ii\mathbf k_4\cdot\mathbf r_i}&\varphi_1\ee^{\ii\mathbf k_5\cdot\mathbf r_i}&\varphi_1\ee^{\ii\mathbf k_6\cdot\mathbf r_i}&\varphi_1\ee^{\ii\mathbf k_7\cdot\mathbf r_i}&\varphi_1\ee^{\ii\mathbf k_8\cdot\mathbf r_i}&\varphi_1\ee^{\ii\mathbf k_9\cdot\mathbf r_i}\\
\varphi_2\ee^{\ii\mathbf k_1\cdot\mathbf r_i}&\varphi_2\ee^{\ii\mathbf k_2\cdot\mathbf r_i}&\varphi_2\ee^{\ii\mathbf k_3\cdot\mathbf r_i}&\varphi_2\ee^{\ii\mathbf k_4\cdot\mathbf r_i}&\varphi_2\ee^{\ii\mathbf k_5\cdot\mathbf r_i}&\varphi_2\ee^{\ii\mathbf k_6\cdot\mathbf r_i}&\varphi_2\ee^{\ii\mathbf k_7\cdot\mathbf r_i}&\varphi_2\ee^{\ii\mathbf k_8\cdot\mathbf r_i}&\varphi_2\ee^{\ii\mathbf k_9\cdot\mathbf r_i}\\
\varphi_3\ee^{\ii\mathbf k_1\cdot\mathbf r_i}&\varphi_3\ee^{\ii\mathbf k_2\cdot\mathbf r_i}&\varphi_3\ee^{\ii\mathbf k_3\cdot\mathbf r_i}&\varphi_3\ee^{\ii\mathbf k_4\cdot\mathbf r_i}&\varphi_3\ee^{\ii\mathbf k_5\cdot\mathbf r_i}&\varphi_3\ee^{\ii\mathbf k_6\cdot\mathbf r_i}&\varphi_3\ee^{\ii\mathbf k_7\cdot\mathbf r_i}&\varphi_3\ee^{\ii\mathbf k_8\cdot\mathbf r_i}&\varphi_3\ee^{\ii\mathbf k_9\cdot\mathbf r_i}\\ \hline
0&0&0&0&0&0&0&0&0\\
0&0&0&0&0&0&0&0&0\\
0&0&0&0&0&0&0&0&0\\ \hline
0&0&0&0&0&0&0&0&0\\
0&0&0&0&0&0&0&0&0\\
0&0&0&0&0&0&0&0&0
\end{array}
\right) .
\label{eq:hyperdet-z3-singlet-matrix}
\end{equation}
\endgroup
The row blocks label flavor and the above tensor is for $b_i=1$.
For $b_i=2$ or 3, the nonzero rows move to $4,5,6$ or $7,8,9$.
Although nine holon columns participate at each $i$, every nonzero
row is proportional to the same momentum vector, so $R=1$.

For the Type II triplet Z$_3$ metal, $\phi_b=d\,\Psi_b$, the
flavor-singlet parton $d$ occupies all three orbitals of each of the
three reversed LLs $n_{\rm LL}=0,1,2$.
Order these nine states by $\mu=3n_{\rm LL}+m$ and denote their
amplitudes by $D_\mu=\varphi_{n_{\rm LL},m}(\mathbf r_i)$.
Each holon flavor occupies three distinct momenta, with basis
$|b,\mathbf k_j\rangle_\Psi$, $b,j=1,2,3$, ordered by
$\nu=3(b-1)+j$.
Now $T^{\rm II}_{i;\mu(b,j)}=\mathcal S^{-1/2}D_\mu\delta_{b b_i}\ee^{\ii\mathbf k_j\cdot\mathbf r_i}$, giving
\begingroup
\setlength{\arraycolsep}{5pt}
\begin{equation}
T_i^{\rm II}=\frac{1}{\sqrt{\mathcal S}}
\left(
\begin{array}{ccc|ccc|ccc}
D_1\ee^{\ii\mathbf k_1\cdot\mathbf r_i}&D_1\ee^{\ii\mathbf k_2\cdot\mathbf r_i}&D_1\ee^{\ii\mathbf k_3\cdot\mathbf r_i}&0&0&0&0&0&0\\
D_2\ee^{\ii\mathbf k_1\cdot\mathbf r_i}&D_2\ee^{\ii\mathbf k_2\cdot\mathbf r_i}&D_2\ee^{\ii\mathbf k_3\cdot\mathbf r_i}&0&0&0&0&0&0\\
D_3\ee^{\ii\mathbf k_1\cdot\mathbf r_i}&D_3\ee^{\ii\mathbf k_2\cdot\mathbf r_i}&D_3\ee^{\ii\mathbf k_3\cdot\mathbf r_i}&0&0&0&0&0&0\\ \hline
D_4\ee^{\ii\mathbf k_1\cdot\mathbf r_i}&D_4\ee^{\ii\mathbf k_2\cdot\mathbf r_i}&D_4\ee^{\ii\mathbf k_3\cdot\mathbf r_i}&0&0&0&0&0&0\\
D_5\ee^{\ii\mathbf k_1\cdot\mathbf r_i}&D_5\ee^{\ii\mathbf k_2\cdot\mathbf r_i}&D_5\ee^{\ii\mathbf k_3\cdot\mathbf r_i}&0&0&0&0&0&0\\
D_6\ee^{\ii\mathbf k_1\cdot\mathbf r_i}&D_6\ee^{\ii\mathbf k_2\cdot\mathbf r_i}&D_6\ee^{\ii\mathbf k_3\cdot\mathbf r_i}&0&0&0&0&0&0\\ \hline
D_7\ee^{\ii\mathbf k_1\cdot\mathbf r_i}&D_7\ee^{\ii\mathbf k_2\cdot\mathbf r_i}&D_7\ee^{\ii\mathbf k_3\cdot\mathbf r_i}&0&0&0&0&0&0\\
D_8\ee^{\ii\mathbf k_1\cdot\mathbf r_i}&D_8\ee^{\ii\mathbf k_2\cdot\mathbf r_i}&D_8\ee^{\ii\mathbf k_3\cdot\mathbf r_i}&0&0&0&0&0&0\\
D_9\ee^{\ii\mathbf k_1\cdot\mathbf r_i}&D_9\ee^{\ii\mathbf k_2\cdot\mathbf r_i}&D_9\ee^{\ii\mathbf k_3\cdot\mathbf r_i}&0&0&0&0&0&0
\end{array}
\right) .
\label{eq:hyperdet-z3-typeii-matrix}
\end{equation}
\endgroup
Here the row blocks label LLs and the column blocks label flavor.
For $b_i=2$ or 3, the nonzero columns move to $4,5,6$ or $7,8,9$.
All nonzero columns are proportional to the same vector of LL
amplitudes, again giving $R=1$.
Thus the singlet and Type II matrices have one nonzero block row and
one nonzero block column, respectively. Each factorizes into a single
outer product. The SU(3) matrix in Eq.~\eqref{eq:hyperdet-nine-matrix}
instead has three independent diagonal color blocks and Schmidt rank
three. Restricting the Z$_3$ constructions to just LLL and
$\mathbf k=0$ subspaces would leave too few one-particle states to host 9 particles.

The table compares the available one-particle states in the three
constructions.
\begin{center}
\begin{tabular}{lccc}
\toprule
construction & $R$ & LL-sector states in the LLL & holon states at $\mathbf k=0$ \\
\midrule
singlet Z$_3$ metal: $\chi_b f$ & 1 & $3\times3=9$ & 1 \\
Type II triplet Z$_3$ metal: $d\,\Psi_b$ & 1 & 3 & 3 \\
SU(3) holon metal: $\sum_a\chi_a\psi_{ab}$ & 3 & $3\times3=9$ & $3\times3=9$ \\
\bottomrule
\end{tabular}
\end{center}
The singlet Z$_3$ metal can keep its $\chi_b$ fermions in the LLL,
but its flavor-singlet holon $f$ must occupy nine distinct momenta.
Type II triplet Z$_3$ metal requires the flavor-singlet parton $d$ to fill three LLs,
each containing three orbitals. Each of its three holon flavors must
also occupy three distinct momenta.
Neither product construction can therefore accommodate nine particles
using only the LLL and $\mathbf k=0$.
The color construction supplies the missing internal states in both
sectors without violating Pauli exclusion.

This counting gives a simple kinetic-energy motivation for the color
ansatz. For the same parabolic parton dispersions, it avoids both the
higher-LL cost of the Type II triplet Z$_3$ metal and the nonzero-momentum cost of the two
product states in this example.
In the thermodynamic limit the holons form Fermi seas, but their nine
pockets have density $n/9$ each, compared with $n$ in the singlet Z$_3$ metal
and $n/3$ in the Type II triplet Z$_3$ metal. The resulting smaller Fermi momenta reduce the
mean-field Fermi energy and $k_F$. Since our Hamiltonian contains only the kinetic energy, this color construction naturally saves energy compared with the two Z$_3$ metals, whose holon wavefunctions necessarily involve a larger $k_F$.

\subsection{CB crystals: Hall crystals and skyrmions}\label{sec:eft-cb}
\paragraph{A composite-boson condensate in statistical flux.}
For the SU(3) anyon gas at $\alpha=1/3$, we introduce a composite-boson
field $\boldsymbol\phi=(\phi_1,\phi_2,\phi_3)^T$.
Its components describe three internal states, as in a spinor Bose condensate
\cite{OhmiMachida1998,Ho1998}. All three flavors couple to the same statistical gauge field $a$.
Including the electromagnetic probe $A_c$, the theory is
Eq.~\eqref{eq:csAc} with $m=3$:
\begin{equation}
\mathcal L=\mathcal L_\phi[a]
+\frac{3}{4\pi}a\,da+\frac{1}{2\pi}A_c\,da .
\label{eq:cb-crystal}
\end{equation}
With the probe turned off, flux attachment gives
$b_a=-2\pi\rho/3$, where $\rho=\sum_b|\phi_b|^2$.
At mean density $n$, the three CB flavors together have signed filling
$\nu^*=2\pi n/\bar b_a=-3$, or $-1$ per flavor for equal populations.
The crystal ansatz accommodates this flux through a vortex lattice,
with one vortex per statistical flux quantum in each flavor.
Each component therefore has vortex density $n/3$, with circulation
of the same sign as $b_a$.
The three vortex sublattices can interleave, allowing the other
flavors to fill each vortex core \cite{Mueller2002}.
The total density can then remain nearly uniform while the flavor
orientation varies periodically. This smooth orientation is the
skyrmion texture of the crystal.

\paragraph{The CP$^2$ texture and its skyrmions.}
Where the total CB density is nonzero, the local flavor orientation is
specified by
\begin{equation}
\mathbf Z=\frac{\boldsymbol\phi}
{\sqrt{\boldsymbol\phi^\dagger\boldsymbol\phi}},
\qquad
\mathbf Z^\dagger\mathbf Z=1,
\qquad
\mathbf Z\sim e^{\ii\chi}\mathbf Z .
\label{eq:cp2-definition}
\end{equation}
Removing the common phase leaves two independent flavor weights and
two relative phases. The resulting space is the complex projective
plane, CP$^2$, which has four real dimensions.
Equivalently, its points are nonzero complex triples
$[\phi_1:\phi_2:\phi_3]$ identified under multiplication by any common
nonzero complex number.
A vortex in one component does not make this orientation singular if
another component remains nonzero.

A texture on the sphere defines a map $S^2\to\mathrm{CP}^2$.
The same description applies on the plane when the orientation is fixed
at infinity. Such maps have an integer skyrmion number because
$\pi_2(\mathrm{CP}^2)=\mathbb Z$.
It is obtained from the Berry connection of the normalized spinor:
\begin{equation}
\mathcal A=-\ii\,\mathbf Z^\dagger d\mathbf Z,
\qquad
Q_{\rm sk}=\frac{1}{2\pi}\int d\mathcal A\in\mathbb Z .
\label{eq:cp2-charge}
\end{equation}
Here $d\mathcal A$ is the Berry curvature through the real-space surface.

\paragraph{One skyrmion from three vortices.}
The two-component case provides useful intuition.
A normalized spinor $(Z_1,Z_2)^T$ defines a unit pseudospin
$\mathbf n=\mathbf Z^\dagger\boldsymbol\sigma\mathbf Z$, so that
CP$^1$ is the ordinary spin sphere.
At a vortex of component 1, component 2 fills the core and $n_z=-1$.
If the components have equal weights away from the core, the spin
approaches the equator, and the relative-phase winding rotates it
around that equator. The texture covers one hemisphere and is a
half-skyrmion, or meron.
A vortex of component 2 has $n_z=+1$ at its core and covers the other
hemisphere. When the two vortices have the same circulation, their
relative-phase windings are opposite and their skyrmion charges have
the same sign. Together they form a unit skyrmion
\cite{Kasamatsu2005}.

For three components, write $Z_b=\sqrt{p_b}\,e^{\ii\theta_b}$, so that
$\mathcal A=\sum_b p_b\,d\theta_b$.
On a loop with constant flavor weights, a $2\pi$ winding in just one
component gives Berry flux $2\pi p_b$.
For equal weights this is half a flux quantum in SU(2), and one third
in SU(3). Three separated component vortices, one of each flavor,
can therefore carry one unit of skyrmion number together
\cite{Garaud2013}. In our reference crystal at $\qbg=-1/3$, each flavor has $N_v=N/3$
vortices. The three lowest-LL orbitals have the same antiholomorphic
degree $N_v$ and no common zero, giving $Q_{\rm sk}=-N_v=-N/3$ in
our convention.
There are consequently $3N_v$ component vortices but only $N_v$ units
of skyrmion number.

\paragraph{Classical energy of the interlaced crystal.}
To estimate the energy of this texture, consider a periodic patch of the plane with mean density $n$,
CB mass $M$ and an SU(3)-invariant contact coupling $g\geq0$. The ideal anyon gas has no contact interaction, but its
classical average-field description acquires one: replacing the point fluxes by a smooth statistical field drops their
short-distance self-interaction, and the resulting Chern--Simons--Schr\"odinger functional for bosons with attached flux
contains an effective contact term whose strength depends on the flux profile and the intrinsic spin-orbit coupling
of the anyons \cite{AtaeiCSS}. We therefore keep $g$ as a parameter of the classical estimate. We use the classical kinetic energy
$\hbar^2\int\sum_b|\boldsymbol\Pi\phi_b|^2/(2M)$ and contact energy $(g/2)\int\rho^2$,
where $\boldsymbol\Pi=-\ii\nabla-\mathbf a$ and $b_a=-2\pi\rho/3$ locally.
Since $[\Pi_x,\Pi_y]=\ii b_a$, integration by parts gives
\[
E_{\rm cl}=
\frac{\hbar^2}{2M}\int d^2r\sum_b|(\Pi_x-\ii\Pi_y)\phi_b|^2+
\left(\frac{\pi\hbar^2}{3M}+\frac{g}{2}\right)\int d^2r\,\rho^2 .
\]
The first term is nonnegative and $\int\rho^2\geq Nn$. Every classical condensate therefore satisfies
$E_{\rm cl}/N\geq e_{\rm lb}$, with $e_{\rm lb}=\pi\hbar^2n/(3M)+gn/2$.
Approaching this bound requires orbitals close to the lowest Landau level and a nearly uniform total density.
Interlacing the vortices allows each component to vanish at its own cores while the other two fill them.

\begin{figure}[h]
\centering
\includegraphics[width=0.48\textwidth]{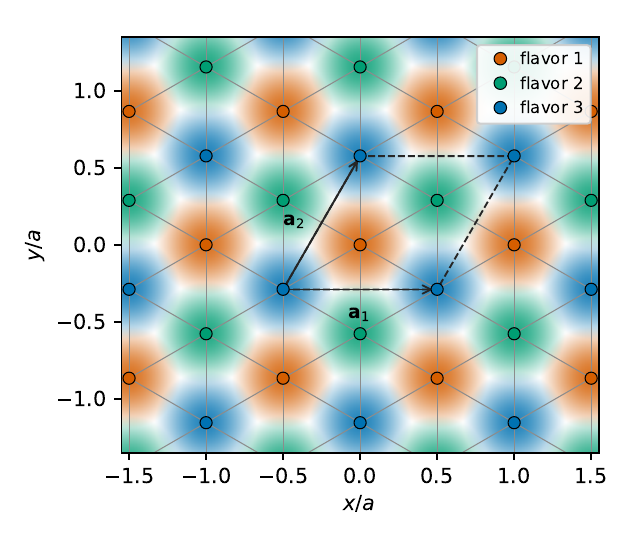}
\caption{Interlaced vortex lattices of the trial field in Sec.~\ref{sec:eft-cb}. Dots mark the vortices of the three
flavors, displaced by $0$, $\mathbf d$ and $2\mathbf d$ with $\mathbf d=(\mathbf a_1+\mathbf a_2)/3$, and the dashed
rhombus is the unit cell of one flavor. The shading around each vortex marks where the weight of that flavor in the
CP$^2$ texture is depleted; the other two flavors fill its core. Together the vortices form a triangular lattice with
spacing $a/\sqrt3$ (gray lines), where $a=|\mathbf a_1|$.}
\label{fig:vortex-texture}
\end{figure}

For an explicit trial field, take three equal-weight triangular LLL vortex lattices in the mean field
$\bar b_a=-B$, with $B=2\pi n/3$. Each component has one vortex per cell of area $2\pi/B$.
With primitive vectors $\mathbf a_1,\mathbf a_2$ at $60^\circ$, displace the three copies by
$0,\mathbf d,2\mathbf d$, where $\mathbf d=(\mathbf a_1+\mathbf a_2)/3$. Together the $3N_v$ vortices then form a
triangular lattice with spacing $|\mathbf a_1|/\sqrt3$, whose three sublattices carry the three flavors
(Fig.~\ref{fig:vortex-texture}). Each vortex has six nearest neighbors, alternating between the other two flavors.
This is not a kagome arrangement, which would require displacements by half lattice vectors.
For a reciprocal vector $\mathbf G$ with integer indices $(h,k)$, the LLL density Fourier coefficient
has magnitude $\exp[-G^2/(4B)]$ relative to its mean \cite{Mueller2002}. Averaging the displaced copies
cancels every coefficient with $h+k\not\equiv0\pmod3$, including the six shortest reciprocal vectors.
For $f=\rho/n$ this gives $\langle f^2\rangle=1.00011267$, where the brackets denote a spatial average.
The sum of the magnitudes of the nonzero harmonics is less than 0.027,
so $f>0$ everywhere for the interlaced state and its flavor orientation
defines a smooth CP$^2$ texture. On the common three-flux magnetic cell it carries $Q_{\rm sk}=-3$.
For comparison we also use a coincident-core trial state, built from the same three lattices without the displacement
($\mathbf d=0$). All three components then vanish at the same points, so the total density vanishes there as well and
$\langle f^2\rangle=1.15959527$. Its flavor orientation is uniform, so it is an ordinary vortex lattice without a
skyrmion texture.

To impose the local Gauss law on this trial field, write $\mathbf a=\bar{\mathbf a}+\delta\mathbf a$
and choose a periodic transverse correction of zero mean with
$\nabla\times\delta\mathbf a=-B(f-1)$. The orbitals obey the LLL condition in $\bar{\mathbf a}$, so the first
term in the completed square reduces to $\hbar^2\int\rho|\delta\mathbf a|^2/(2M)$. Thus
\[
\frac{E_{\rm trial}}{N}=e_{\rm lb}\langle f^2\rangle+
\frac{\hbar^2}{2M}\langle f|\delta\mathbf a|^2\rangle .
\]
For the interlaced lattice the Fourier sum gives $\langle f|\delta\mathbf a|^2\rangle/B=0.00000515$.
At $g=0$ the resulting energy per particle is $0.333373$ in units of $\pi\hbar^2 n/M$, within $0.012\%$
of the general lower bound $1/3$. The coincident-core state gives $0.392635$, $18\%$ above the bound.

\paragraph{The filled-LLL interpretation.}
To understand the Hall response, first place the anyons in a uniform
imposed field. Choose the orientation for which LLL wavefunctions
are holomorphic, and let $\ell$ be the magnetic length.
For a fixed flavor orientation, the hard-core anyon wavefunctions take
the form
\begin{equation}
\Psi_{\rm LLL}
=\prod_{i<j}(z_i-z_j)^{1/3}\,
F(z_1,\ldots,z_N)\,
\ee^{-\sum_i|z_i|^2/(4\ell^2)} .
\label{eq:anyon-filled-lll}
\end{equation}
The fractional power supplies the exchange phase $\ee^{\ii\pi/3}$,
while $F$ is a regular symmetric holomorphic function.
The densest state has $F=1$, with no additional orbital zeros
\cite{Dasnieres1994,Ouvry2021}.
On the sphere, the statistical factor has degree $(N-1)/3$ in each
particle coordinate, so the filled state requires $N_\Phi=(N-1)/3$
flux quanta of the single-particle vector potential.
Here one flux quantum is $2\pi$.
Thus $\nu_a=N/N_\Phi\to3$ in the thermodynamic limit: the LLL
accommodates three anyons per flux quantum because of their statistics. Increasing $N_\Phi$ by one accommodates three more anyons, so the
St\v{r}eda relation gives the Hall coefficient
$C_a=\partial N/\partial N_\Phi=3$.
This is the anyon IQH state, with all anyons fully occupying the lowest LL.

In the skyrmion Hall crystal, the flavor texture supplies the effective field.
Projecting motion onto the local spinor $\mathbf Z$ gives the vector
potential $-\boldsymbol{\mathcal A}$, with $-Q_{\rm sk}$ flux quanta.
The reference value $Q_{\rm sk}=-N/3$ therefore gives the same
density-to-flux ratio $\nu_a=+3$ as the filled anyon LLL. Thus, in this crystal phase, the anyons form a $C_a=3$ integer quantum Hall state due to the spontaneously generated skyrmion texture.
This crystal is analogous to the electronic anomalous Hall crystals proposed
for rhombohedral pentalayer graphene
\cite{DongPatri2024,Zhou2023AHC,Dong2024AHC}.
In a topologically trivial crystal of doped anyons, by contrast, the
anyons are localized and add no Hall response. The total Hall
conductance then remains that of the undoped FCI.

\paragraph{The physical Hall response.}
The physical probe $A_c$ is normalized to unit electron charge.
An anyon of charge $-e_*$, with $e_*=e/3$, couples to $-A_c/3$.
Its $C_a=3$ IQH response therefore contributes
$3(-A_c/3)d(-A_c/3)/(4\pi)=+A_c\,dA_c/(12\pi)$.
The coefficient of $A_c\,dA_c/(4\pi)$ is $1/3$, because the
fractional charge enters twice.
This cancels the parent FCI response $-A_c\,dA_c/(12\pi)$:
\begin{equation}
\sigma_{xy}=\sigma_{\rm FCI}+3\frac{e_*^2}{h}
=-\frac{e^2}{3h}+\frac{e^2}{3h}=0 .
\label{eq:crystal-hall}
\end{equation}

The same conclusion follows directly from the original CB action,
Eq.~\eqref{eq:cb-crystal}.
With the texture pinned, the CB condensate completely Higgses $a_\mu$.
Integrating out its massive fluctuations leaves no $A_c\,dA_c$ term
at long wavelengths. Because $\boldsymbol\phi$ has unit gauge charge,
no residual discrete gauge group remains.
The final phase is therefore a topologically trivial charge
insulator, with broken translation and flavor symmetries.
For the $\nu=2/3$ FCI in tMoTe$_2$, the Hall crystal of doped anyons
adds $e^2/(3h)$ to the parent response $2e^2/(3h)$.
The resulting RIQAH insulator has $\sigma_{xy}=e^2/h$.

\subsubsection{Relation to the commensurate FCI-CDW transition}\label{sec:eft-commensurate}
The transition studied in Ref.~\cite{Song2024} at
fixed filling $\nu=2/3$ has the same three-component condensation
mechanism. In our chirality convention, their three dual-vortex fields play the role of
$\boldsymbol\phi$ in Eq.~\eqref{eq:cb-crystal}, with a common gauge
field and the same level-three Chern--Simons term.
The matter fields are relativistic in that transition theory.
When they are gapped, the fractional Hall response combines with the
filled $C=1$ band to give $\sigma_{xy}=2e^2/(3h)$. Condensation Higgses $a$ and
gives an IQH phase with $\sigma_{xy}=e^2/h$.
Because lattice translations act projectively on the three fields,
their condensation also produces a commensurate CDW.
Its order parameters are gauge-invariant bilinears of the fields,
encoded in the CP$^2$ orientation.

The distinction upon doping is the net anyon density.
In the undoped case at fixed $\nu=2/3$, a relativistic condensate does not require a finite average statistical flux.
Doping populates one sign of the anyons with charge magnitude $e/3$.
Keeping this branch gives the nonrelativistic CB theory used above.
For anyons of charge $-e/3$ at density $n$, the equation of motion gives
\begin{equation}
\frac{\delta\rho_{\rm el}}{e}
=\frac{b_a}{2\pi}
=-\frac{n}{3} ,
\label{eq:doping-flux}
\end{equation}
where $\delta\rho_{\rm el}$ is the charge density
relative to the commensurate state, at zero applied magnetic field.
The resulting flux requires a vortex density $n/3$ in each component
of the condensate. Near the SU(3) limit, the cores can separate to
form the smooth skyrmion texture described above.

The skyrmion Hall crystal is therefore a candidate extension of the
commensurate IQH-CDW phase to finite doping.
The doped density sets the additional texture scale,
$\ell_v\sim(n/3)^{-1/2}$, which can greatly exceed the moir\'e period.

%=======================================================================================================================
\FloatBarrier
\section{Numerical results}\label{sec:results}

Table~\ref{tab:main} compares the lowest-energy liquid and crystal of each system. The subsections below give the detailed results.

\begin{table}[h]
\caption{The table compares the lowest-energy liquid and crystal for each system on the sphere. We give $E/N$ in units of $\Eunit$ at the indicated $(\qbg,N)$, together with the extrapolated energy $e_\infty$ obtained from $E/N=e_\infty+b/N$ within each shift sector using $N\ge8$, or $N=6$--18 for the color superconductor.  Rows labeled analytical wavefunction correspond to the parameter-free states of Eqs.~\eqref{eq:pfree-spinless}--\eqref{eq:pfree}, including the long-range Jastrow factor of Eq.~\eqref{eq:Jlr}, with $e_\infty$ from $N=8$--24 for semions and $N=6$--24 for SU(3) anyons.}
\label{tab:main}
\centering\footnotesize\setlength{\tabcolsep}{3.5pt}
% BEGIN AUTO TABLE MAIN (paper/scripts/make_tables.py)
\begin{tabular}{lllcrrrr}
\toprule
 & & & & \multicolumn{2}{c}{$E/N$} & \multicolumn{2}{c}{$e_\infty$} \\
system & liquid & crystal & $(\qbg,N)$ & liquid & crystal & liquid & crystal \\
\midrule
spinless semions & paired superfluid (CF $\nu^*=2$) & vortex crystal (CB network) & $(-3/2,24)$ & 0.7803 & 0.9033 & 0.7837(4) & 0.909(6) \\
 & analytical wavefunction &  &  & 0.796(1) &  & 0.801(2) &  \\
SU(2) semions & paired superfluid (CF $\nu^*=2$) & vortex crystal (CB network) & $(-1/2,24)$ & 0.5179 & 0.7361 & 0.5407(7) & 0.783(5) \\
 & analytical wavefunction &  &  & 0.514(1) &  & 0.541(2) &  \\
spinless, $\theta=\pi/3$ & trion Fermi liquid (CF) & vortex crystal (CB network) & $(-4/3,24)$ & 0.6582 & 0.6684 & 0.651$^*$ & 0.6571(3) \\
SU(3), $\theta=\pi/3$ & color superconductor (optimized) & skyrmion Hall crystal & $(-1/3,18)$ & 0.4662 & 0.5200 & 0.498(1) & 0.572(2) \\
 & analytical wavefunction &  &  & 0.464(1) &  & 0.498(3) &  \\
\bottomrule
\end{tabular}
% END AUTO TABLE MAIN
\end{table}

\subsection{Semions}\label{sec:res-semion}
\paragraph{Energies.} Table~\ref{tab:semions} compares the energies of all trial states at the background charges of
Fig.~\ref{fig:semions}, $\qbg=-3/2$ for spinless and $\qbg=-1/2$ for SU(2) semions. In these sectors the CF flux
$2Q^*=\qbg+(N-1)/2$ lets the CF $\nu^*=2$ state fill two complete Landau levels, or one per spin, at every even $N$, so
that its shells are closed and it is rotation invariant. Up to $N=24$, the CF $\nu^*=2$ state and the CF networks
initialized with random orbitals converge to the same uniform liquid, with energies within $0.01$ of each other. So
does the optimized $N_c=2$ color superconductor, which we trained only up to $N=12$ because the color sum with backflow
becomes too costly; its parameter-free analytical form needs no training and no backflow, and Table~\ref{tab:semions}
gives it up to $N=24$. For $N\ge32$ the randomly initialized networks lose rotation invariance, and
at $N=40$ they end in non-uniform states with energy  $0.06$ above the CF $\nu^*=2$ state, which indicates that their optimization
is trapped in a local minimum. The parameter-free analytical wavefunctions of Eqs.~\eqref{eq:pfree-spinless} and
\eqref{eq:pfree-su2} come close to these optimized states. For SU(2) semions the analytical state lies below all of
them, by 0.002--0.011 at $N=8$--24, and for spinless semions it lies 0.007--0.016 above them. This suggests that the
color-superconductor wavefunctions are close to the ground state, particularly for SU(2) semions.

\begin{table}[h]
\caption{The table gives the energies $E/N$ of semions on the sphere, in units of $\Eunit$, at the background charges $\qbg$ where the CF $\nu^*=2$ state has closed shells. An asterisk marks a state that is not rotation invariant, with $\langle L^2\rangle>3$. The analytical $N_c=2$ color superconductor is the parameter-free wavefunction of Eq.~\eqref{eq:pfree-spinless} or \eqref{eq:pfree-su2}. The optimized one is the color state of Eq.~\eqref{eq:colstate} with $N_c=2$, whose pair function $g_{ij}$, Jastrow factor and backflow are trained. }
\label{tab:semions}
\centering\footnotesize
\setlength{\tabcolsep}{4pt}
\begin{tabular}{lrrrrrrrr}
\toprule
\multicolumn{9}{l}{\emph{spinless semions, $q_{bg}=-3/2$}} \\
$N$ & 8 & 10 & 12 & 16 & 20 & 24 & 32 & 40 \\
\midrule
CB network (crystal) & 0.8782 &  & 0.8889 & 0.8857 & 0.8964 & 0.9033 &  &  \\
CB vortex lattice & 1.1553 &  & 1.1306 & 1.1085 & 1.1108 & 1.1065 &  &  \\
$N_c=2$ color superconductor (optimized) & 0.7779 & 0.7785 & 0.7795 &  &  &  &  &  \\
$N_c=2$ color superconductor (analytical) & 0.7841 &  & 0.7884 & 0.7913 & 0.7948 & 0.7963 &  &  \\
CF $\nu^*=2$ state & 0.7808 &  & 0.7812 & 0.7817 & 0.7818 & 0.7823 & 0.7835 & 0.7833 \\
CF network (MP), random start & 0.7770 &  & 0.7790 & 0.7815 & 0.7800 & 0.7803 & $0.7868^{\ast}$ & $0.8427^{\ast}$ \\
CF network (attention), random start & 0.7812 &  & 0.7784 & 0.7784 & 0.7798 & 0.7811 & $0.7824^{\ast}$ & $0.8463^{\ast}$ \\
\midrule
\multicolumn{6}{l}{\emph{SU(2) semions, $q_{bg}=-1/2$}} \\
$N$ & 8 & 12 & 16 & 20 & 24 \\
\midrule
CB network (crystal) & 0.6501 & 0.6983 & 0.7227 & 0.7290 & 0.7361 \\
CB vortex lattice & 1.0179 & 1.0295 & 1.0538 & 1.0595 & 1.0605 \\
$N_c=2$ color superconductor (optimized) & 0.4773 & 0.4986 &  &  &  \\
$N_c=2$ color superconductor (analytical) & 0.4641 & 0.4908 & 0.5014 & 0.5113 & 0.5144 \\
CF $\nu^*=2$ state & 0.4792 & 0.5003 & 0.5091 & 0.5161 & 0.5207 \\
CF network (MP), random start & 0.4755 & 0.5003 & 0.5123 & 0.5192 & 0.5261 \\
CF network (attention), random start &  & 0.4970 &  & 0.5135 & 0.5179 \\
\bottomrule
\end{tabular}
\end{table}

\paragraph{Overlaps.} To test whether the liquids obtained from these different constructions describe the same phase,
we compute the fidelity $F=|\langle A|B\rangle|^2/(\langle A|A\rangle\langle B|B\rangle)$ between trained states
(Table~\ref{tab:semionoverlaps}). Here $A$ and $B$ are the full anyonic amplitudes $\Psi$, including the statistical
factor of the gauge in which each state is built, so that states built in the CF and CB gauges can be compared. Since
$\langle A|B\rangle/\langle A|A\rangle=\langle B/A\rangle_{|A|^2}$ and
$\langle B|B\rangle/\langle A|A\rangle=\langle|B/A|^2\rangle_{|A|^2}$, where $\langle\cdots\rangle_{|A|^2}$ averages over
configurations sampled from $|A|^2$, samples of $A$ alone give the one-sided estimate
$F=|\langle B/A\rangle_{|A|^2}|^2/\langle|B/A|^2\rangle_{|A|^2}$. Sampling both states gives the two-sided estimate
$F=\langle B/A\rangle_{|A|^2}\langle A/B\rangle_{|B|^2}$, which we use as a check. The liquids differ in construction:
the CF $\nu^*=2$ state of Eq.~\eqref{eq:cfll} fills CF Landau levels, the CF networks start from random orbitals instead
of filled levels, and the $N_c=2$ color superconductor of Eq.~\eqref{eq:colstate} arises from color-singlet pairing of the holon metal. Nevertheless, they have mutual fidelities of $0.95$--$0.997$ at $N=8$ and 12, whereas their fidelities
with the CB crystal are only $0.02$--$0.5$ and decrease rapidly with $N$. The independent liquid constructions therefore
describe the same state. In particular, the $N_c=2$ color superconductor is in the same phase as Laughlin's original
anyon superfluid, the $\nu^*=2$ IQH state of composite fermions \cite{Laughlin1988}.

\begin{table}[h]
\caption{The table gives the fidelity $F=|\langle A|B\rangle|^2/(\langle A|A\rangle\langle B|B\rangle)$ between trained semion states. We use a one-sided estimate on samples of $A$; the two-sided check agrees within 0.004. The $N_c=2$ color superconductor is the optimized state of Table~\ref{tab:semions}.}
\label{tab:semionoverlaps}
\centering\footnotesize
\begin{tabular}{lcccc}
\toprule
 & \multicolumn{2}{c}{spinless, $\qbg=-3/2$} & \multicolumn{2}{c}{SU(2), $\qbg=-1/2$} \\
$A$ -- $B$ & $N=8$ & $N=12$ & $N=8$ & $N=12$ \\
\midrule
CF $\nu^*=2$ -- CF network (MP) & 0.997(1) & 0.992(1) & 0.995(1) & 0.988(1) \\
CF $\nu^*=2$ -- CF network (attention) & 0.992(1) & 0.986(1) &  & 0.977(1) \\
$N_c=2$ color superconductor -- CF $\nu^*=2$ & 0.988(1) & 0.977(1) & 0.975(1) & 0.948(1) \\
$N_c=2$ color superconductor -- CF network (MP) & 0.991(1) & 0.977(1) & 0.972(1) & 0.948(1) \\
$N_c=2$ color superconductor -- CF network (attention) & 0.993(1) & 0.982(1) &  & 0.951(1) \\
CF $\nu^*=2$ -- CB crystal & 0.479(6) & 0.151(3) & 0.208(3) & 0.025(2) \\
$N_c=2$ color superconductor -- CB crystal & 0.512(4) & 0.167(4) & 0.233(3) & 0.025(1) \\
\bottomrule
\end{tabular}
\end{table}

\begin{figure}[h]
\centering
\includegraphics[width=0.9\textwidth]{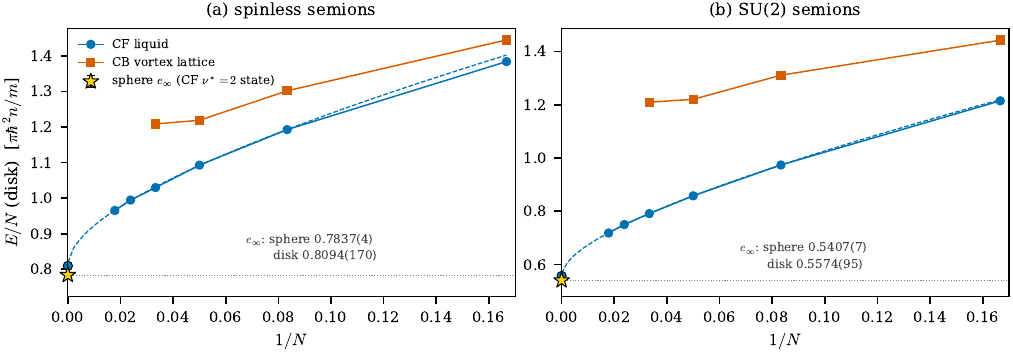}
\caption{We compare spinless (a) and SU(2) (b) semions on a disk with a hard wall, plotting $E/N$ in units of $\Eunit$ against $1/N$. At each $N$, the CF liquid is the lowest of the planar CF $\nu^*=2$ state and the CF networks, and the CB vortex lattice has learned vortex positions (see text). The dashed curve is the fit $E/N=e_\infty+a/\sqrt N+b/N$ to the CF liquid over $N=12$--56, with its intercept marked at $1/N=0$. The star and dotted line give $e_\infty$ of the CF $\nu^*=2$ state on the sphere.}
\label{fig:disk}
\end{figure}

\paragraph{Disk.} To check the sphere results, we study planar versions of the same states on a disk of radius
$R=\sqrt{N/\pi n}$ with a hard wall (Fig.~\ref{fig:disk}). They use the planar coordinates $z_i$ and $z_{ij}$ in place
of the spinors and $w_{ij}$, together with a hard-wall factor $\prod_i(1-r_i^2/R^2)$ and a learned one-body Jastrow
factor. As on the sphere, the CF $\nu^*=2$ state occupies the two lowest Landau levels of the mean CF field, or one per
spin, and the CF networks start from these Landau-level orbitals or from random ones. We call the lowest of these CF
states at each $N$ the CF liquid. The CB vortex lattice is the planar counterpart of the one on the sphere,
$\Phi=\prod_{i<j}r_{ij}^{\alpha}\prod_{i,k}(\bar z_i-\bar v_k)/R$ with learned vortex positions $v_k$. The CF liquid
lies well below the CB vortex lattice at every $N$, with energy differences of $0.18$ for spinless semions and $0.42$ for
SU(2) semions at $N=30$. Accounting for the edge contribution of order $N^{-1/2}$, we fit the CF liquid energies to
$E/N=e_\infty+a/\sqrt N+b/N$ over $N=12$--56. The extrapolated values, $e_\infty=0.809(17)$ for spinless semions and $0.557(10)$ for SU(2) semions, agree within two standard errors with the sphere results $0.7837(4)$ and $0.5407(7)$.

\paragraph{Pair correlator.} The pair density matrix $\rho_2(R,R')=\langle\Delta^\dagger(R')\Delta(R)\rangle$, with
$\Delta(R)$ removing two anyons near $R$, is estimated in the trained state by moving a pair rigidly. For a configuration $Z$ sampled from $|\Psi|^2$, we pick an anyon at random and its nearest partner within $c_2=0.6\,n^{-1/2}$ (for SU(2), of the opposite or of the same spin). We rotate both by an angle $\gamma$ about the axis through the center of the sphere perpendicular to the pair's center. The pair thus moves along a great circle by the arc length $R\gamma$ with its internal configuration unchanged. With $Z_\gamma$ the moved configuration,
\begin{equation}
C_2(\gamma)=\Big\langle\frac{\Phi(Z_\gamma)}{\Phi(Z)}\,\ee^{\ii[S(Z_\gamma)-S(Z)]}\Big\rangle_{|\Psi|^2},\qquad
S=\sum_{j\notin{\rm pair}}\Big[2\alpha_s\arg w_{cj}+\alpha_s\sum_{a\in{\rm pair}}\arg\frac{w_{aj}}{w_{cj}}\Big],
\label{eq:C2}
\end{equation}
where $\Phi$ is the wavefunction without the statistical factor, in the gauge in which the state is built
($\alpha_s=\alpha-1$ for CF states and $\alpha_s=\alpha$ for CB and color states), and $w_{cj}=u_cv_j-v_cu_j$ is
written with the spinor $\xi_c$ of the pair's center. We average over $3\times10^3$--$2.5\times10^4$ pairs per size and
define the plateau as the mean of $|C_2|$ over the outer half of the move, $\gamma\ge\pi/2$.

The phase $S$ specifies the anyonic pair operator. Because $\Psi$ is multivalued, the ratio $\Psi(Z_\gamma)/\Psi(Z)$ is
fixed only once we specify how the statistical factor between the two moved anyons and the others,
$\prod_{a\in{\rm pair}}\prod_{j\notin{\rm pair}}(w_{aj}/|w_{aj}|)^{\alpha_s}$, is continued from $Z$ to $Z_\gamma$.
This choice is the string carried by $\Delta$, the analog of a Jordan--Wigner string. For each anyon $j$ outside the
pair we factor
\[
\prod_{a\in{\rm pair}}\Big(\frac{w_{aj}}{|w_{aj}|}\Big)^{\alpha_s}
=\Big(\frac{w_{cj}}{|w_{cj}|}\Big)^{2\alpha_s}\prod_{a\in{\rm pair}}\Big(\frac{w_{aj}/w_{cj}}{|w_{aj}/w_{cj}|}\Big)^{\alpha_s}.
\]
For semions $2\alpha_s=\pm1$, so the first factor is single valued: seen from outside, the pair carries an integer
statistical charge and braids trivially with every anyon. The second factor approaches 1 when anyon $j$ is far from the
pair. We take it on the principal branch, $x^{\alpha_s}=\ee^{\ii\alpha_s\mathrm{Arg}\,x}$ with
$\mathrm{Arg}\,x\in(-\pi,\pi]$ for a unit complex number $x$. This phase jumps only where $w_{aj}/w_{cj}$ is negative
real, that is, when anyon $j$ lies on the arc between member $a$ and the center $c$, so the branch cut lies inside the
pair. Summed over $j$, the phase of the product is $S$ of Eq.~\eqref{eq:C2}. It is a single-valued function of the
configuration that jumps only when another anyon crosses the cut inside the pair, so $\ee^{\ii S}$ is the string of the
local pair operator $\Delta=\psi(z_1)\psi(z_2)$. The local string is essential. The natural alternative is a path string, which continues each factor
$(w_{aj}/|w_{aj}|)^{\alpha_s}$ continuously along the trajectory of member $a$ during the rotation, that is, it assigns
the statistical phase that the two anyons accumulate when they are transported along their paths. This continued phase
agrees with $\ee^{\ii S}$ until another anyon passes between the two members, crossing the cut. There
the principal-branch factor jumps by $\ee^{\pm2\pi\ii\alpha_s}=-1$, while the continued phase does not, so the two
differ by a factor $-1$ for every anyon that passes between the members during the move. These signs depend on the
positions of the other anyons and average out, so they would suppress $C_2$ even in a genuine pair condensate.

\begin{figure}[h]
\centering
\includegraphics[width=\textwidth]{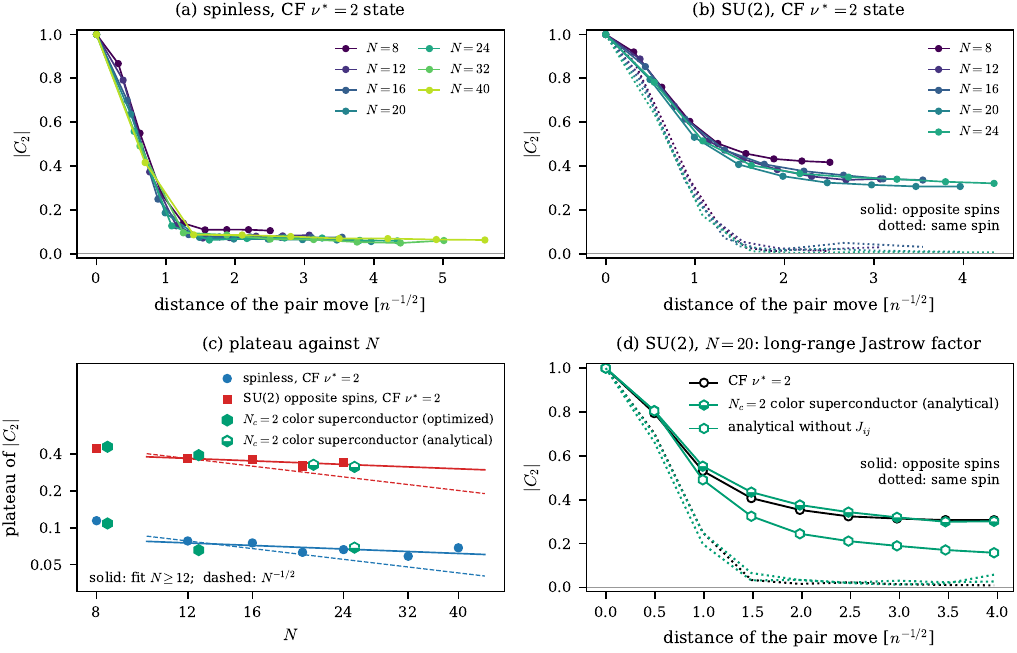}
\caption{The pair correlator $|C_2|$ [Eq.~\eqref{eq:C2}] of semions is plotted against the pair displacement $R\gamma$. Panels (a) and (b) show the trained CF $\nu^*=2$ state of spinless semions at $N=8$--40 and of SU(2) semions at $N=8$--24, with solid and dotted curves for opposite-spin and same-spin pairs, respectively. Panel (c) plots the plateau, defined as the mean over $\gamma\ge\pi/2$, against $N$. For the CF $\nu^*=2$ state, solid lines are power-law fits for $N\ge12$ and dashed lines show the $N^{-1/2}$ decay of the bare CF state. Hexagons show the $N_c=2$ color superconductor, optimized at $N=8$ and 12 (filled) and analytical at $N=24$ (spinless) and $N=20$ and 24 [SU(2)] (half filled), slightly offset in $N$. Panel (d) compares SU(2) semions at $N=20$ in the analytical wavefunction of Eq.~\eqref{eq:pfree-su2} with (half-filled hexagons) and without (open hexagons) its long-range Jastrow factor $J_{ij}$.}
\label{fig:paircorr}
\end{figure}

Figure~\ref{fig:paircorr} shows that $|C_2|$ for the CF $\nu^*=2$ state decreases over one to two interparticle
spacings and then approaches a plateau, both for spinless semions and for opposite-spin pairs of SU(2) semions. The
plateau decreases with $N$ only approximately as $N^{-0.16}$, much more slowly than the $N^{-1/2}$ decay of the bare CF
state (Sec.~\ref{sec:eft-sf}). The $N_c=2$ color superconductor has the same pair correlator, as Fig.~\ref{fig:paircorr}(c) shows. For the analytical wavefunction this agreement
requires the long-range Jastrow factor. Without $J_{ij}$, the color sum keeps its plasma-like long-distance
correlations (Sec.~\ref{sec:analytic-wf}), and the opposite-spin plateau falls well below the CF value
[Fig.~\ref{fig:paircorr}(d)].

\begin{figure}[h]
\centering
\includegraphics[width=\textwidth]{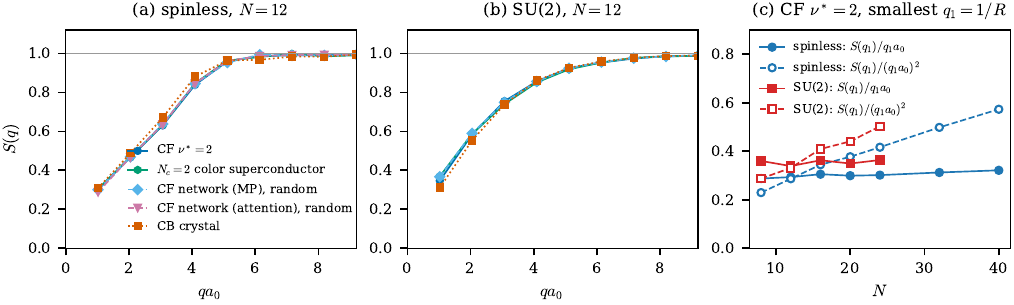}
\caption{The density structure factor $S(q)$ at $q=l/R$ is shown for the trained semion states. Panels (a) and (b) compare spinless and SU(2) semions at $N=12$, showing the CF $\nu^*=2$ state, $N_c=2$ color superconductor, CF networks initialized with random orbitals and CB crystal. Panel (c) shows the CF $\nu^*=2$ state at the smallest wave vector $q_1=1/R$: the filled symbols represent the nearly constant ratio $S(q_1)/q_1a_0$, while the open symbols show the growing ratio $S(q_1)/(q_1a_0)^2$, both plotted against $N$ with $a_0=n^{-1/2}$.}
\label{fig:sq}
\end{figure}

\paragraph{Structure factor.} The density structure factor $S(l)=\frac1N\sum_{ij}P_l(\cos\theta_{ij})$ at $q=l/R$ is
shown in Fig.~\ref{fig:sq}. At $N=12$ the CF $\nu^*=2$ state, the random-start networks and the $N_c=2$ color superconductor
have the same $S(q)$, to within $0.01$ (spinless) and $0.02$ [SU(2)], as their overlaps (Table~\ref{tab:semionoverlaps})
require; at this size the crystal's $S(q)$ is still close to theirs. At the smallest wave vector $q_1=1/R$ the CF state
has $S(q_1)/q_1a_0=0.29$--$0.32$ (spinless, $N=8$--40) and $0.34$--$0.36$ [SU(2), $N=8$--24], with $a_0=n^{-1/2}$,
nearly independent of $N$, while $S(q_1)/(q_1a_0)^2$ grows with $N$. Thus $S(q)$ is linear in $q$, rather than quadratic as for an
incompressible quantum Hall state such as the bare CF state.

A linear $S(q)$ alone does not distinguish the superfluid from the crystal. The statistical vector potentials cancel
in pairs, $\sum_i\mathbf a(\mathbf r_i)=0$, so the total momentum is conserved. At long wavelengths the statistical
fields $b=-2\pi\alpha_s\rho$ and $\mathbf e=-2\pi\alpha_s\hat{\mathbf z}\times\mathbf j$ exert no net force,
$\rho\,\mathbf e+\mathbf j\times b\,\hat{\mathbf z}=0$, so no cyclotron gap appears. The weight of the $f$-sum rule at
small $q$ is then carried by a gapless longitudinal mode, the sound of the superfluid or the longitudinal phonon of the
crystal. If $c$ is its velocity, then in both phases
\[
S(q)\simeq s\,q,\qquad s=\frac{\hbar}{2mc}.
\]
The velocity is set by how strongly the system resists the deformation that a long-wavelength longitudinal wave
produces. Resistance to compression is measured by the bulk modulus $K=n\,\partial P/\partial n$, where $P$ is the
pressure: a small density change $\delta n$ raises the pressure by $\delta P=K\,\delta n/n$. Resistance to a change of
shape at fixed area is measured by the shear modulus $\mu_s$, which vanishes in a liquid and is positive in a crystal.
A longitudinal wave $\mathbf u\propto\hat{\mathbf x}\cos qx$ stretches the medium along $\hat{\mathbf x}$ only, which both
changes its area and changes its shape, so elasticity theory gives
\[
mn\,c^2=K\quad\text{(superfluid)},\qquad mn\,c^2=K+\mu_s\quad\text{(crystal)},
\]
with $mn$ the mass density. Scale invariance fixes $K$ in every phase through its own $e_\infty$. At zero temperature
$P=n^2\,\partial(E/N)/\partial n$, and $E/N=e_\infty\pi\hbar^2n/m$ gives $P=e_\infty\pi\hbar^2n^2/m$ and
$K=2\pi e_\infty\hbar^2n^2/m$. For the shear modulus scale invariance fixes only the form,
$\mu_s=\tilde\mu\,\pi\hbar^2n^2/m$, with a dimensionless coefficient $\tilde\mu\ge0$ that depends on the crystal.
Together,
\[
\frac{s}{a_0}=\frac{1}{2\sqrt{\pi(2e_\infty+\tilde\mu)}},
\]
where $\tilde\mu=0$ gives the superfluid value $s=a_0/(2\sqrt{2\pi e_\infty})$ of Sec.~\ref{sec:analytic-wf}. On the
sphere the $f$-sum rule and the phonon frequency involve $l(l+1)/R^2$ instead of $q^2$, which raises the ratio at $l=1$
by $\sqrt2$: $S(q_1)/(q_1a_0)=1/\{2[\pi(e_\infty+\tilde\mu/2)]^{1/2}\}$. For the superfluid this gives $0.32$ for spinless
and $0.38$ for SU(2) semions, close to the values above. For the crystal, with its higher $e_\infty$ of $0.909$ and
$0.783$ (Table~\ref{tab:main}), it gives $0.30$ and $0.32$ at $\tilde\mu=0$, and less for $\tilde\mu>0$. The CB crystals
have $S(q_1)/(q_1a_0)=0.28$--$0.30$ (spinless) and $0.30$--$0.34$ [SU(2)] at $N=8$--24, also nearly independent of $N$
and comparable to these values. Since the competing
states differ in $e_\infty$ by less than a factor of 1.5, their slopes $s$ are similar, and $S(q)\simeq s\,q$ is a
common property of the compressible phases.

\FloatBarrier
\subsection{Spinless anyons at \texorpdfstring{$\theta=\pi/3$}{theta=pi/3}}\label{sec:res-pi3}
The shift sectors $\qbg=-5/3$, $-4/3$, $-1$ and $-1/3$ select the Moore--Read, CFL, PH-Pfaffian and anti-Pfaffian states, respectively, in the second Landau level for the composite fermions, and we initialize the CF liquid in each sector with the corresponding exact wavefunction. Table~\ref{tab:pi3} compares every ansatz with the CB crystal at the same sector and size. For $N\ge12$, the best liquid and crystal differ by less than $0.015$ per particle, with the liquid energy relative to the crystal ranging from $-0.013$ in the Moore--Read sector at $N=14$ to $+0.008$ in the PH-Pfaffian sector at $N=16$. The sign changes with both the sector and $N$; in the CFL sector, the trion Fermi liquid is lower by $0.010$ at $N=24$. CF networks initialized with random orbitals instead converge to non-uniform states above both candidates. The three-color metal and its 2SC descendants, in which colors 1 and 2 pair while color 3 forms either a closed-shell pocket or pairs with itself, give the lowest energies at $N\le9$. At $N=12$, they lie slightly below the crystal in the $\qbg=2/3$ and $5/3$ sectors, where no CF liquid is available, while the CF liquid remains lower in the anti-Pfaffian sector. Taken together, we are not confident to make claims on whether the liquid phase or the crystal phase has lower energy for this case, given how close their energy is. For the liquid phase, the CF and color constructions lead to the same trion liquid. In the CFL sector, the second-level CFs become charge-3 fermions that couple to no gauge field, the trion Fermi liquid of Sec.~\ref{sec:eft-trion}. In the color construction, the three-color metal is a fractionalized metal of three holon Fermi seas coupled to the SU(3) gauge field. Pairing colors 1 and 2 breaks the color gauge group, and the remaining pocket of color 3 carries the anyon number 3 of a trion and is neutral under the residual gauge group. The 2SC with an unpaired third color is therefore also a Fermi liquid of trions, with the same Fermi volume $n/3$. Pairing the trions gives a superfluid of six-anyon clusters, as  the Moore--Read, PH-Pfaffian or anti-Pfaffian state of the CFs in the second-Landau-level. Each shift sector admits only one of these CF states, so paired and unpaired trion liquids can be compared only across sectors, where the shift changes the finite-size energy.  We therefore cannot determine whether the trion Fermi liquid is unstable to pairing.

\begin{table}[h]
\caption{The table gives the energies of spinless anyons at $\theta=\pi/3$, with $E/N$ in units of $\Eunit$, organized by shift sector. For each CF liquid, we report the lowest energy among the variants initialized with the exact target state and using capped, split or uncapped contact coordinates. The symbols have the same meanings as in Table~\ref{tab:semions}.}
\label{tab:pi3}
\centering\footnotesize
\setlength{\tabcolsep}{4pt}
\begin{tabular}{lrrrr}
\toprule
\multicolumn{5}{l}{\emph{Moore--Read sector, $q_{bg}=-5/3$}} \\
$N$ & 8 & 14 & 20 & 26 \\
\midrule
CB network (crystal) & 0.7242 & 0.6880 & 0.6847 & 0.6823 \\
CB vortex lattice &  & 0.7982 & 0.7876 &  \\
CF liquid, exact Moore--Read start & 0.7383 & 0.6748 & 0.6849 &  \\
CF network, random start &  & $0.7016^{\ast}$ & $0.7040^{\ast}$ &  \\
\midrule
\multicolumn{5}{l}{\emph{CFL sector, $q_{bg}=-4/3$}} \\
$N$ & 9 & 15 & 21 & 24 \\
\midrule
CB network (crystal) & 0.6881 & 0.6733 & 0.6717 & 0.6684 \\
CB vortex lattice &  & 0.7734 & 0.7715 & 0.7644 \\
three-color metal & 0.6554 &  &  &  \\
2SC, third color a pocket & 0.6528 &  &  &  \\
CF liquid, exact CFL start & 0.6703 &  &  & 0.6582 \\
CF network, random start &  &  &  & $0.6835^{\ast}$ \\
\midrule
\multicolumn{4}{l}{\emph{PH-Pfaffian sector, $q_{bg}=-1$}} \\
$N$ & 10 & 16 & 22 \\
\midrule
CB network (crystal) & 0.6622 & 0.6569 & 0.6602 \\
CB vortex lattice &  & 0.7510 & 0.7542 \\
CF liquid, exact PH-Pfaffian start &  & 0.6644 &  \\
CF network, random start &  & $0.6743^{\ast}$ &  \\
\midrule
\multicolumn{5}{l}{\emph{anti-Pfaffian sector, $q_{bg}=-1/3$}} \\
$N$ & 6 & 12 & 18 & 24 \\
\midrule
CB network (crystal) & 0.6015 & 0.6236 & 0.6292 & 0.6399 \\
CB vortex lattice &  &  & 0.7154 & 0.7286 \\
three-color metal & 0.5928 &  &  &  \\
2SC, third color a pocket & 0.5931 &  &  &  \\
2SC, third color paired & 0.5927 & $0.6193^{\ast}$ &  &  \\
CF liquid, exact anti-Pfaffian start &  & 0.6178 & 0.6352 & $0.6446^{t}$ \\
CF network, random start &  &  & $0.6535^{\ast}$ & $0.6840^{t}$ \\
\midrule
\multicolumn{2}{l}{\emph{$q_{bg}=2/3$ (color states only)}} \\
$N$ & 12 \\
\midrule
CB network (crystal) & 0.5809 \\
three-color metal & 0.5717 \\
2SC, third color a pocket & 0.5692 \\
\midrule
\multicolumn{2}{l}{\emph{$q_{bg}=5/3$ (color states only)}} \\
$N$ & 12 \\
\midrule
CB network (crystal) & 0.5222 \\
2SC, third color paired & 0.5211 \\
\bottomrule
\end{tabular}
\end{table}

\clearpage
\subsection{SU(3) anyons at \texorpdfstring{$\theta=\pi/3$}{theta=pi/3}}\label{sec:res-su3}
\begin{figure}[h]
\centering
\includegraphics[width=\textwidth]{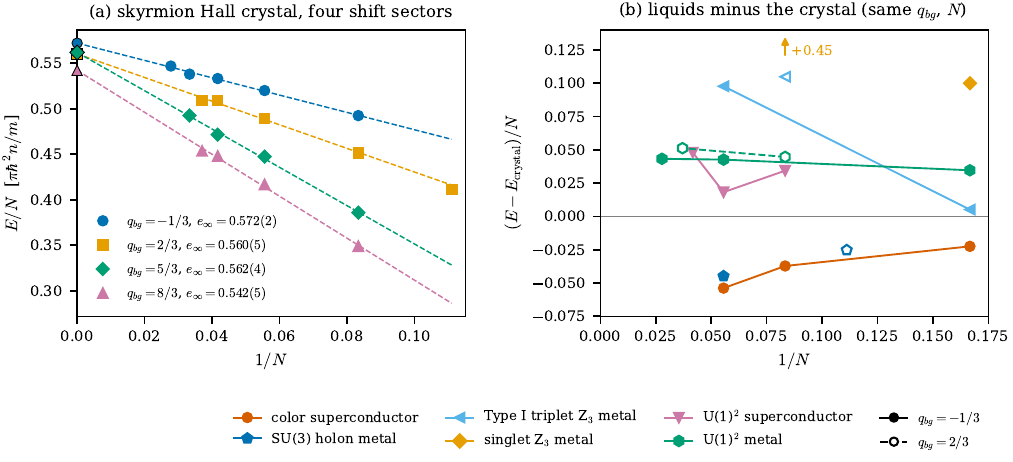}
\caption{We compare crystal and liquid energies for SU(3) anyons at $\theta=\pi/3$. Panel (a) shows the skyrmion Hall crystal in four shift sectors. Linear fits in $1/N$ are performed separately within each sector using $N\ge12$, and the corresponding extrapolated $e_\infty$ is shown. Panel (b) compares the energies of rotation-invariant liquids relative to the best crystal at the same sector and size, using filled symbols and solid lines for $\qbg=-1/3$ and open symbols and dashed lines for $\qbg=2/3$. The arrow marks the singlet Z$_3$ metal at $N=12$, whose energy lies above the plotted range.}
\label{fig:su3sectors}
\end{figure}

\paragraph{Energetics.}
The color superconductor has the lowest variational energy among the
states considered (Table~\ref{tab:su3}). We compare each liquid with the
crystal at the same $\qbg$ and $N$, since the shift changes the
finite-size correction [Fig.~\ref{fig:su3sectors}(a)]. In both chiral
sectors, $\qbg=-1/3$ and $5/3$, the color superconductor lies below the
crystal by an amount that grows with $N$, and its extrapolated energy
lies well below that of the crystal (Table~\ref{tab:main}). Because of
the shift, the $p+ip$ state at $\qbg=5/3$ lies below the $p-ip$ state at
$\qbg=-1/3$ at fixed $N$, but it extrapolates above it. Remarkably, the lowest energy requires no
optimization: the analytical wavefunction of Eq.~\eqref{eq:pfree}, which
has no variational parameters, lies slightly below the optimized color
superconductor at every size.

\paragraph{Flavor symmetry of the analytical state.}
As discussed in Sec.~\ref{sec:flavor-proj}, the same-color contact factor
weakly breaks the flavor symmetry of Eq.~\eqref{eq:pfree}. The Casimir
test of Sec.~\ref{sec:flavor-proj} shows that the state is an exact
singlet at $N=6$ and contains a non-singlet admixture of about 0.15\% at
$N=12$, in the 27-dimensional representation. The exact flavor singlet of Eq.~\eqref{eq:pfree-sym} has
the same energy at $N=6$ and a slightly higher one at $N=12$ and 18, by
at most 0.006 per particle.

\paragraph{Comparison with the optimized state.}
The analytical wavefunction has a high fidelity with the optimized color
superconductor, and the two extrapolate to the same $e_\infty$. The
fidelities in Table~\ref{tab:su3overlaps} show that the different
constructions of the color superconductor describe the same phase. The
color state with only a two-body Jastrow factor and the analytical
wavefunction, with the contact correction in all factors or in the pair
function only, have $-\ln F/N$ more than an order of magnitude smaller
than the competing phases at $N\ge12$. The exception is the SU(3) holon
metal at $N=18$, where two holons per species fill the lowest shell, so
that lowest-shell pairing reproduces the closed-shell metal.

\begin{table}[h]
\caption{The table gives the energies of SU(3) anyons at $\theta=\pi/3$, with $E/N$ in units of $\Eunit$, organized by shift sector. The paired states require $\qbg=-1/3$ or $5/3$, while states with Fermi pockets are evaluated at closed shells, as described in the text. Each crystal entry is the best CB network at that size; at $\qbg=-1/3$ and $N=12$ this is the $K=16$ network, whereas the $1/N$ fits use the $K=8$ network with energy 0.4925. The symbols have the same meanings as in Table~\ref{tab:semions}.}
\label{tab:su3}
\centering\footnotesize
\setlength{\tabcolsep}{4pt}
\begin{tabular}{lrrrrrrrr}
\toprule
\multicolumn{9}{l}{\emph{$q_{bg}=-1/3$}} \\
$N$ & 6 & 9 & 12 & 18 & 24 & 30 & 36 & 42 \\
\midrule
skyrmion Hall crystal (CB network) & 0.4256 & 0.4925 & 0.4888 & 0.5200 & 0.5332 & 0.5380 & 0.5469 &  \\
color superconductor (optimized) & 0.4031 &  & 0.4515 & 0.4662 &  &  &  &  \\
color superconductor, two-body Jastrow only & 0.4204 &  & 0.4837 & $0.5011^{t}$ &  &  &  &  \\
color superconductor (analytical) & 0.4014 &  & 0.4489 & 0.4637 & 0.4766 &  &  &  \\
analytical, pair function only & 0.4249 &  & 0.4937 &  &  &  &  &  \\
SU(3) holon metal &  &  &  & 0.4752 &  &  &  &  \\
Type I triplet Z$_3$ metal (color sum) & 0.4307 &  &  & 0.6178 &  &  &  &  \\
singlet Z$_3$ metal & 0.5256 &  & 0.9431 &  &  & $0.8549^{\ast}$ &  & $1.0791^{\ast}$ \\
U(1)$^2$ metal & 0.4602 &  &  & 0.5627 &  &  & 0.5902 &  \\
U(1)$^2$ superconductor &  &  & 0.5232 & 0.5382 & 0.5813 &  &  &  \\
general CF network (MP3), random start & 0.4481 &  & $0.5506^{\ast}$ & $0.6030^{\ast}$ & $0.6151^{\ast}$ &  &  &  \\
general CF network (attention), random start & 0.4349 &  & $0.5267^{\ast}$ &  &  &  &  &  \\
\midrule
\multicolumn{6}{l}{\emph{$q_{bg}=2/3$}} \\
$N$ & 9 & 12 & 18 & 24 & 27 \\
\midrule
skyrmion Hall crystal (CB network) & 0.4112 & 0.4513 & 0.4895 & 0.5089 & 0.5090 \\
SU(3) holon metal & 0.3860 &  &  &  &  \\
Type I triplet Z$_3$ metal (color sum) &  & 0.5563 &  &  &  \\
singlet Z$_3$ metal & $0.4653^{\ast}$ &  &  &  &  \\
U(1)$^2$ metal &  & 0.4960 &  &  & 0.5603 \\
general CF network (MP3), random start & $0.5046^{\ast}$ & $0.5376^{\ast}$ &  &  &  \\
general CF network (attention), random start & $0.4990^{\ast}$ & $0.5240^{\ast}$ &  &  &  \\
\midrule
\multicolumn{6}{l}{\emph{$q_{bg}=5/3$}} \\
$N$ & 6 & 12 & 18 & 24 & 30 \\
\midrule
skyrmion Hall crystal (CB network) & $0.2778^{t}$ & 0.3859 & 0.4474 & 0.4716 & 0.4926 \\
color superconductor (optimized) &  & 0.3823 & 0.4266 &  &  \\
SU(3) holon metal &  &  & 0.4270 &  &  \\
singlet Z$_3$ metal &  & 0.9388 &  &  &  \\
U(1)$^2$ superconductor &  & 0.4287 & 0.4933 & 0.5285 &  \\
\midrule
\multicolumn{5}{l}{\emph{$q_{bg}=8/3$}} \\
$N$ & 12 & 18 & 24 & 27 \\
\midrule
skyrmion Hall crystal (CB network) & 0.3489 & 0.4167 & 0.4480 & 0.4537 \\
Type II triplet Z$_3$ metal (product) & $0.5865^{\ast}$ &  &  & $0.6572^{\ast}$ \\
\bottomrule
\end{tabular}
\end{table}

\begin{table}[h]
\caption{For SU(3) anyons at $\qbg=-1/3$, the table gives the fidelity $F$ of each state with the optimized color superconductor of the same size. The one-sided and two-sided estimates agree within 0.02. The three color-superconductor rows compare the color superconductor with only a two-body Jastrow factor and the analytical wavefunction of Eq.~\eqref{eq:pfree}, with the contact correction in all factors or in the pair function only. The last of these has no $J_{ij}$.}
\label{tab:su3overlaps}
\centering\footnotesize
\begin{tabular}{lccc}
\toprule
state & $N=6$ & $N=12$ & $N=18$ \\
\midrule
skyrmion Hall crystal & 0.462(3) & 0.028(1) & 0.0017(12) \\
CB network with moving zeros & 0.710(3) &  &  \\
color superconductor, two-body Jastrow only & 0.963(1) & 0.919(1) &  \\
color superconductor (analytical) & 0.975(1) & 0.951(1) & 0.893(2) \\
analytical, pair function only & 0.932(1) & 0.822(2) & 0.680(5) \\
SU(3) holon metal &  &  & 0.691(5) \\
Type I triplet Z$_3$ metal & 0.889(1) &  &  \\
singlet Z$_3$ metal & 0.191(4) & 0.0000(3) &  \\
U(1)$^2$ metal & 0.210(3) &  & 0.0001(3) \\
U(1)$^2$ superconductor &  & 0.0001(2) & 0.0000(3) \\
general CF network (MP3) & 0.855(2) & 0.0021(4) & 0.0001(2) \\
general CF network (attention) & 0.872(2) & 0.0022(5) &  \\
\bottomrule
\end{tabular}
\end{table}

\subsection{Skyrmion Hall crystal}\label{sec:res-sky}
\begin{figure}[h]
\centering
\includegraphics[width=\textwidth]{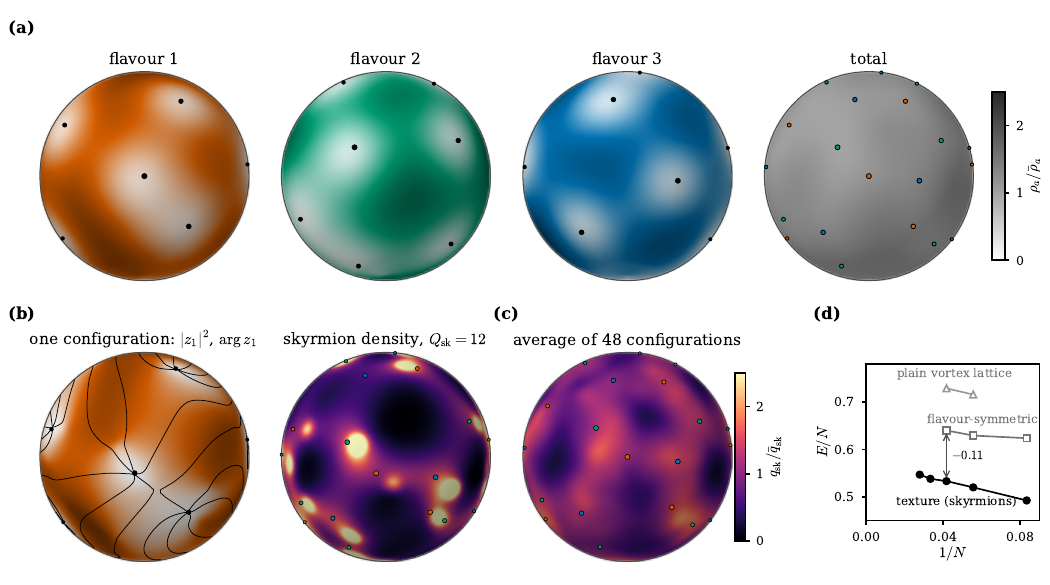}
\caption{The skyrmion Hall crystal at $N=36$ and $\qbg=-1/3$, with $N_v=12$ vortices per flavor, is shown on the front hemisphere. (a) The flavor densities $\rho_a/\bar\rho_a$ and total density are obtained from $2.6\times10^5$ samples. Black dots mark each flavor's vortex lattice sites, identified from 48 configurations; all 36 vortex sites are shown on the total-density plot. Each flavor density vanishes at its own vortices, while the total remains nearly uniform. (b) A single configuration is probed by moving one anyon through the wavefunction. The left panel shows the flavor-1 weight $|z_1|^2$ and contours of $\arg z_1$ at intervals of $60^\circ$, radiating from the vortices (dots), while the right panel shows the skyrmion density of $[z_1\!:\!z_2\!:\!z_3]$ in units of its mean $N_v/4\pi R^2$, together with the vortices of all three flavors. Its integral gives $|Q_{\rm sk}|=12$. (c) The skyrmion density is averaged over 48 configurations and shown with the lattice sites. (d) The energy $E/N$ at $\qbg=-1/3$ is compared for the state with a flavor texture (the SU(3) CB network with flavor vortices in Table~\ref{tab:su3}), the flavor-symmetric crystal (equivalent to the spinless CB network in Table~\ref{tab:pi3}), and the plain vortex lattice.}
\label{fig:skyrmion}
\end{figure}
Figure~\ref{fig:skyrmion} shows the crystal at $N=36$. We locate its vortices by moving one particle of a given flavor over a grid while holding the remaining particles at a configuration sampled from $|\Psi|^2$. The vortices are not fixed network parameters, since the backflow and environment-dependent coefficients in Eq.~\eqref{eq:cbnet} allow them to move with the other particles. Across 48 sampled configurations, the vortices of each flavor cluster around $N_v$ lattice sites, with rms displacements of 0.2--0.4$a_0$ compared with site spacings of 1.2--1.5$a_0$ for $N=12$--36. This separation of scales defines a well-formed vortex lattice whose vortices fluctuate around their sites, much as particles do in a quantum crystal. The three sublattices interleave, allowing the other two flavors to fill each vortex core, as in interlaced vortex lattices of two-component condensates \cite{Mueller2002}. Each flavor density vanishes at its own sites and reaches about twice its mean between them, while the total density varies by only about $\pm10\%$, so the dominant modulation is in the flavor texture.

\paragraph{Extracting the texture from the wavefunction.}
To extract the flavor texture of the correlated variational state,
we sample a configuration from $|\Psi|^2$ and move one flavor-1 anyon
to a probe position $\mathbf r$, obtaining a CB amplitude $z_1(\mathbf r)$.
We obtain $z_2$ and $z_3$ by exchanging the probe's flavor with that of a
fixed reference anyon of the corresponding flavor.
The three amplitudes have identical particle positions and total flavor
populations. Common factors, including the statistical phase and
flavor-independent Jastrow factors, therefore cancel in the ratios
$z_1:z_2:z_3$. We thus regard $(z_1,z_2,z_3)$ only up to an overall
complex factor. This defines a point $[z_1:z_2:z_3]$ of the complex
projective plane CP$^2$ in homogeneous coordinates, and
$\mathbf r\mapsto[z_1:z_2:z_3]$ maps the sphere to CP$^2$. Because the
reference anyons carry different flavors in the three amplitudes, $z_2$
and $z_3$ acquire scale factors relative to $z_1$, which bias the flavor
weights. We remove this bias by using
$Z_b=\sqrt{w_b}z_b/(\sum_cw_c|z_c|^2)^{1/2}$, with positive constants
$w_b$ chosen so that the spatial average of each $|Z_b|^2$ is $1/3$. The
$w_b$ do not depend on $\mathbf r$, but we determine them separately for
each sampled configuration, since the scale factors depend on the
positions of the reference anyons.
These nonsingular rescalings preserve the topological charge.
For the correlated state, this defines a conditional texture for each
sampled configuration. Its averaged flavor weights are compared with
the measured densities below.

Numerically, the Berry flux through each grid plaquette is the phase of
the cyclic overlap product
$(\mathbf Z_1^\dagger\mathbf Z_2)(\mathbf Z_2^\dagger\mathbf Z_3)
(\mathbf Z_3^\dagger\mathbf Z_4)(\mathbf Z_4^\dagger\mathbf Z_1)$,
where the subscripts label its corners in order.
Local phase choices cancel in this product.
Summing the plaquette phases and dividing by $2\pi$ gives $Q_{\rm sk}$;
we report its magnitude when comparing it with $N_v$. The CP$^2$ map gives $|Q_{\rm sk}|=N_v$ in 46 of the 48 configurations and $N_v-1$ in the remaining two, with an average skyrmion charge of $1/3$ per vortex cell. Averaged over the configurations, the composition $|Z_b|^2$ of the texture reproduces the flavor fraction $\rho_b/\rho$ measured from the samples, the density of flavor $b$ divided by the total density: their Pearson correlation coefficient over the sphere is 0.99 for each flavor. The averaged skyrmion density peaks where the total density is lowest.  Removing the texture makes the crystal flavor symmetric and raises its energy to that of spinless anyons, by 0.11--0.13 per particle at $N=12$--24 [Fig.~\ref{fig:skyrmion}(d)]. This SU(3) flavor texture is analogous to the spin textures of quantum Hall skyrmion crystals \cite{Sondhi1993,Brey1995} and spinor Bose condensates \cite{OhmiMachida1998,Ho1998,Mueller2002}. In the doped FCI of twisted MoTe$_2$, the SU(3) flavor is a valley index arising from the projective action of translations on the charge-$e/3$ anyons, which produces a threefold-degenerate band with three valleys \cite{Shi2025a,Zhang2026a,Han2026}. The flavor texture therefore breaks moir\'e translation symmetry, and local probes should detect the skyrmion Hall crystal as charge order on the moir\'e scale.

%=======================================================================================================================
\end{document}